\documentclass[reprint,nofootinbib,aps,prd,superscriptaddress,amsmath,amssymb]{revtex4-2}

\usepackage{graphicx}
\usepackage{dcolumn}
\usepackage{bm}
\usepackage{breqn}
\usepackage{color}
\usepackage{subcaption}
\usepackage{booktabs}
\usepackage[colorlinks=true,linkcolor=red,citecolor=blue]{hyperref}

\newcommand{\episode}{\varepsilon}

\begin{document}

\title{External magnetic-field effects on dipolar-particle orbits and critical collisions in Kerr--Bertotti--Robinson spacetime}

\author{Haowei Chen}
\email{chenhaowei@zjut.edu.cn}
\author{Hengyu Xu}
\email{xuhengyu0501@outlook.com}
\author{Shao-Jun Zhang}
\email{sjzhang@zjut.edu.cn (corresponding author)}
\affiliation{Institute for Theoretical Physics and Cosmology$,$ Zhejiang University of Technology$,$ Hangzhou 310032$,$ China}
\affiliation{School of Physics and Optical Engineering$,$ Zhejiang University of Technology$,$ Hangzhou 310032$,$ China}
\date{\today}

\begin{abstract}
Strong magnetic fields affect particle dynamics through two
distinct channels: gravitational backreaction deforms the spacetime, while
direct coupling to an intrinsic magnetic moment depends on the relative
orientation of the field and the dipole. We disentangle these effects by
studying equatorial orbits and near-horizon collisions of electrically neutral
magnetized particles in the exact Kerr--Bertotti--Robinson spacetime. The
field-induced geometric deformation shifts turning points and
circular-orbit domains and can eliminate a finite effective-potential well
together with its bound orbits. Even without direct dipole coupling, it can
also offset the Kerr periapsis advance and produce a finite-radius
zero-precession orbit. Direct dipole coupling breaks the symmetry under
magnetic-field reversal and shifts the radius, energy, and angular momentum of
the innermost stable circular orbit in an orientation-dependent manner. The
formal ultrarelativistic endpoints of these orbit branches, however, remain
fixed by the background geometry and approach circular null orbits. In the
Ba\~nados--Silk--West mechanism, both magnetic effects modify finite-radius
potential barriers and hence the ability of a critical particle to reach the
near-horizon collision region. In the representative nonzero-field cases
examined here, an exactly critical particle released from infinity is blocked
before reaching an extremal horizon, although a locally admissible collision
between critical and usual particles can still produce unbounded
center-of-mass energy. Finite dipole coupling shifts the barriers but does not
change the leading near-horizon divergence. Near a nonextremal horizon, an
exactly critical particle is excluded and the collision energy remains finite.
\end{abstract}


\maketitle

\section{Introduction}

Astrophysical black holes are commonly surrounded by magnetized
plasma, which regulates accretion and can power relativistic outflows and
jets~\cite{Stuchlik:2020Cosmic,Blandford:1977BZ}. When the electromagnetic
stress--energy is negligible, the field may be treated on a fixed Kerr
background, as in Wald's uniform-field construction~\cite{Wald:1974UniformField}.
If the field backreacts appreciably, the metric and Maxwell field must instead
be determined self-consistently. Standard examples include the Melvin universe
and the magnetized Schwarzschild and Kerr solutions~
\cite{Melvin:1964Geons,Ernst:1976MagneticUniverse,ErnstWild:1976KerrMagnetic}.
Rotating Melvin-type solutions may, however, develop ergoregions extending
arbitrarily far from the black hole~\cite{Gibbons:2013Ergoregions}.

Starting from the uniform Einstein--Maxwell field of the
Bertotti--Robinson universe~
\cite{Bertotti:1959Uniform,Robinson:1959MaxwellEinstein}, Podolsk\'y and
Ovcharenko recently obtained the Kerr--Bertotti--Robinson (Kerr--BR) black
hole~\cite{Podolsky:2025KerrBR,Ovcharenko:2026KerrNewmanBR}; the solution was later embedded in a broader
family with nonaligned electromagnetic fields~
\cite{Ovcharenko:2025Nonaligned}. Kerr--BR is an exact electrovacuum spacetime
that reduces to Kerr at zero field, is of Petrov type D, has bounded
ergoregions, and approaches a uniform Bertotti--Robinson-type field toward its
outer radial end~\cite{Podolsky:2025KerrBR}. It therefore provides a controlled
setting in which the external field affects particle motion both through the
Maxwell interaction and through its backreaction on the geometry.

Recent studies have examined geodesic integrability, equatorial and spherical
motion, ISCOs, inspirals, and quasiperiodic oscillations in this geometry~
\cite{Wang:2026KerrBRGeodesics,Wang:2025KerrBRISCO,Mustafa:2026KerrBRDynamics,
Wang:2026KBRSpherical,Lu:2026KBRIntegrability}. Optical observables,
photon-ring critical behavior, and gravitational-wave signatures have also
been investigated~
\cite{Zeng:2025KerrBROptics,Wan:2026KBRPhotonRings,Li:2026KerrBRGW}, together
with magnetic reconnection and the magnetic Penrose process~
\cite{Zeng:2025KerrBRReconnection,Mirkhaydarov:2026KerrBRMPP}. Related work has
addressed periodic and nonintegrable motion in the static limit, spinning
test particles in the rotating spacetime, near-horizon symmetry, flux
expulsion, thermodynamics, and global stability~
\cite{Xamidov:2026SchwarzschildBRPeriodic,Xu:2026SBRChaos,
Zhang:2026KerrBRSpin,Siahaan:2026KerrBRCFT,Siahaan:2026KBRMeissner,
Hu:2026KerrBRThermodynamics,Zhou:2026KerrBRNotes}. A treatment that follows
intrinsic magnetic-dipole motion from finite-radius orbital structure to
critical near-horizon collisions is nevertheless still lacking.

Neutral particles with intrinsic magnetic dipole moments are particularly
useful for separating the two roles of the field. They experience no Lorentz
force associated with a net charge; their direct electromagnetic response is
instead governed by the local dipole--field interaction. We use the covariant
scalar-coupling prescription of Ref.~\cite{Preti:2004Polarized}, which has been
applied to effective potentials, circular motion, marginal stability, and
particle acceleration in Schwarzschild and Kerr backgrounds~
\cite{DeFelice:2003Schwarzschild,DeFelice:2004Kerr,
Abdujabbarov:2014MagnetizedAcceleration,Uktamov:2024MagnetizedSchwarzschild}.
Such coupling can shift the ISCO and mimic some effects of geometric
deformations~\cite{Narzilloev:2021QuasiSchwarzschild}. In Kerr--BR, setting
$\mu=0$ retains the backreacted geometry, whereas reversing the field at fixed
$|B|$ reverses the direct coupling while leaving the metric unchanged. These
comparisons isolate the geometric and dipolar contributions.

The radial effective potential provides a common description of finite-radius
orbits and near-horizon accessibility. Its wells and barriers determine
turning points, bound and scattering trajectories, circular orbits, and the
ISCO. Closed strong-field orbits provide an additional diagnostic through
their zoom--whirl structure and periapsis precession. Their rational-frequency
classification was developed in
Refs.~\cite{Glampedakis:2002ZoomWhirl,Levin:2008Periodic} and has been applied
to magnetic dipoles around a magnetized Kerr black hole~
\cite{Jumaniyozov:2026MagnetizedKerrPeriodic}. In Kerr--BR, the same potential
also determines whether a BSW critical particle can reach a proposed collision
radius, thereby linking the orbital and collision analyses.

Ba\~nados, Silk, and West showed that a critical particle colliding with a
usual particle near an extremal rotating black hole can have an unbounded
local center-of-mass energy in the horizon limit~\cite{Banados:2009pr}.
Subsequent studies clarified the fine tuning, general kinematics, and
astrophysical limitations of this mechanism~
\cite{Berti:2009BSWComment,Jacobson:2009BSWLimits,
Harada:2011GeneralGeodesic,Harada:2014vka}. Criticality alone is insufficient:
the trajectory must remain future directed and radially admissible~
\cite{Zaslavskii:2010Universal,Hejda:2017KinematicRestrictions}. The mechanism
can survive regular finite forces~\cite{Ovcharenko:2023FiniteForces}, whereas
an exactly critical particle is excluded from a neighborhood of a
nonextremal horizon~\cite{Zaslavskii:2020NonextremalAccelerator}.
Electromagnetic fields and magnetic interactions can further alter radial
accessibility~
\cite{Frolov:2011ea,Igata:2012js,Zaslavskii:2014eda,
Jumaniyozov:2024MagnetizedKerr}. Moreover, divergent local collision energy
does not by itself imply efficient extraction to infinity, which depends on
the production and propagation of the collision products~
\cite{Piran:1977CollisionalPenrose,Zaslavskii:2012BSWPenrose,
Bejger:2012CollisionalPenrose,Schnittman:2014RevisedLimit}.

In this work, we consider equatorial motion with the dipole fixed normal to the
equatorial plane in the frame of zero-angular-momentum observers (ZAMOs). We derive the
radial effective potential on the future-directed branch and use it to study
turning points, bound and scattering motion, closed orbits, circular-orbit
domains, and ISCOs. The closed orbits are defined using the regularized
$2\pi$-periodic azimuthal angle and classified by their zoom--whirl structure.
Comparisons among $\mu=0$, $B>0$, and $B<0$ separate geometric backreaction
from the direct dipole coupling. The results show that the backreacted geometry
can qualitatively alter finite-radius precession even without direct coupling,
while nonzero $\mu$ breaks the $B\to-B$ symmetry and shifts the orbital and
ISCO branches.

We then construct the local effective center-of-mass energy and test critical
motion against the same radial admissibility condition. On a locally allowed
extremal branch, the critical--usual divergence persists, and a fixed finite
dipole moment changes finite-radius barriers but not the leading near-horizon
exponent. For a nonextremal horizon, an exactly critical particle is excluded
from the near-horizon region; any exterior allowed branch ends at a finite
turning point with finite collision energy. These are local test-particle
statements and do not establish the energy received in an outer observation
region.

The remainder of the paper is organized as follows. Section~II formulates the
magnetic-dipole dynamics in Kerr--BR spacetime. Section~III presents the
orbital analysis, and Section~IV examines critical collisions near extremal
and nonextremal horizons. Section~V summarizes the results. The appendices
collect the ZAMO-frame field components and numerical data. We use geometrized
units with $G=c=4\pi\epsilon_0=1$, where $G$ is Newton's gravitational
constant, $c$ is the speed of light in vacuum, and $\epsilon_0$ is the vacuum
permittivity.
In the numerical analysis, we set the black-hole mass
$M=1$, so all other dimensionful quantities are expressed in the appropriate
powers of $M$.

\section{Motion of Particles with Magnetic Dipole Moment around Kerr–Bertotti–Robinson black holes}

\begin{figure}[!htbp]
    \centering
    \includegraphics[width=0.9\linewidth]{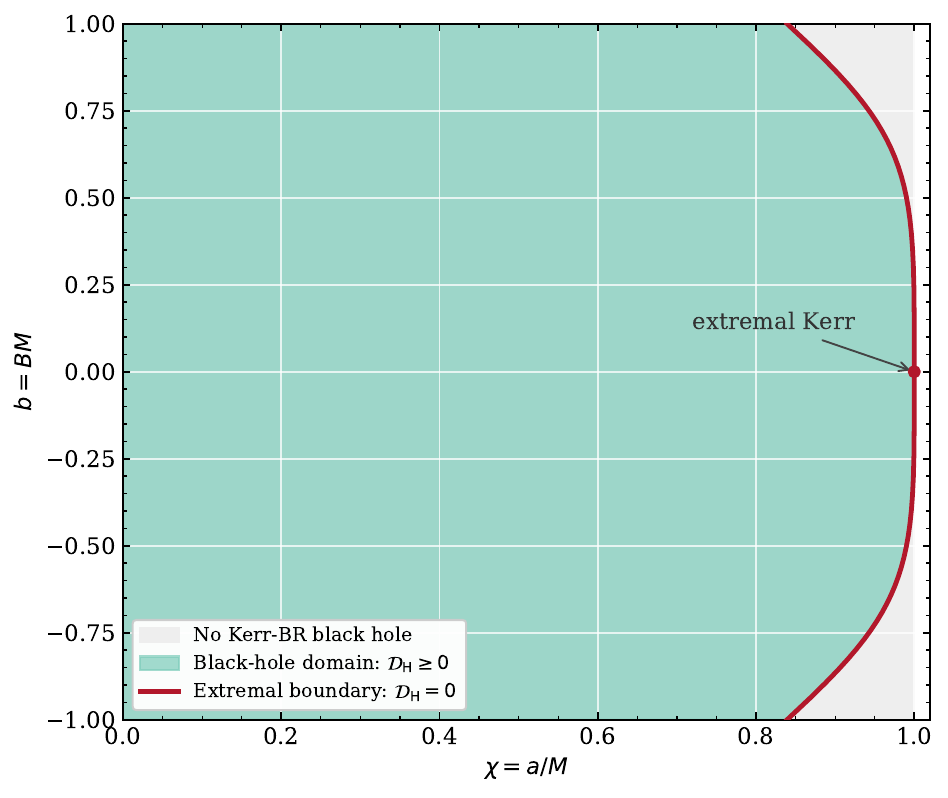}
    \caption{Horizon-existence domain in the dimensionless $(\chi,b)$-plane.
    The shaded region satisfies Eq.~(\ref{dimensionless-BH-domain}); the gray
    region is excluded, and the red curves mark the extremal boundary
    $\mathcal{D}_{\rm H}=0$. Only the range $|b|\leq1$ is displayed.}
    \label{fig:BH-parameter-space}
\end{figure}

In this section, we study the motion of a test particle with an intrinsic magnetic dipole moment around the Kerr-Bertotti-Robinson (Kerr-BR) black hole. The metric of the Kerr-BR black hole is \cite{Podolsky:2025KerrBR}
\begin{align}
ds^2={}&\frac{1}{\Omega^2}\Bigg[
-\frac{Q}{\rho^2}
\bigl(dt-a\sin^2\theta\,d\varphi\bigr)^2
+\frac{\rho^2}{Q}\,dr^2 \notag\\
&+\frac{\rho^2}{P}\,d\theta^2
+\frac{P}{\rho^2}\sin^2\theta \notag\\
&\quad\times
\bigl(a\,dt-(r^2+a^2)d\varphi\bigr)^2
\Bigg],
\label{KerrBR}
\end{align}
where
\begin{align}
\rho^2 &= r^2 + a^2 \cos^2\theta , \\
P &= 1 + B^2\left(\frac{M^2 I_2}{I_1^2} - a^2\right)\cos^2\theta ,
\label{P-function}\\[0.5em]
Q &= (1 + B^2 r^2)\,\Delta , \\[0.5em]
\Omega^2 &= (1 + B^2 r^2) - B^2 \Delta \cos^2\theta , \\[0.5em]
\Delta
&= \left(1 - \frac{B^2 M^2 I_2}{I_1^2}\right) r^2
- \frac{2 M I_2}{I_1}\, r
+ a^2 , \\[0.5em]
I_1 &= 1 - \frac{1}{2} B^2 a^2 ,
\qquad
I_2 = 1 - B^2 a^2 .
\end{align}
Here $M$ and $a$ are the black hole mass and spin parameters, respectively. The magnetic field parameter $B$ denotes the strength of the external magnetic field. This metric is an exact electrovacuum solution to the Einstein--Maxwell equations and describes a rotating black hole immersed in an external, asymptotically uniform Bertotti–Robinson magnetic field aligned ($B>0$) or anti-aligned ($B<0$) with the rotation axis. In the zero-field limit $B\to0$, the solution reduces to the Kerr spacetime. Setting $M\to0$ with $B\neq0$ instead removes the black hole and yields the standard Bertotti--Robinson universe supported by a uniform magnetic field~\cite{Bertotti:1959Uniform,
Robinson:1959MaxwellEinstein,Podolsky:2025KerrBR}. The horizons of the black hole are determined by the roots of the equation $\Delta=0$, which admits two solutions in generic,
\begin{align}
    {r}_{ \pm  } = \frac{M{I}_{2} \pm  \sqrt{{M}^{2}{I}_{2} - {a}^{2}{I}_{1}^{2}}}{{I}_{1}^{2} - {B}^{2}{M}^{2}{I}_{2}}{I}_{1},
\end{align}
localizing the outer and inner black hole horizons, respectively. For the metric to describe a
black hole, $r_\pm$ must be real. This requires
\begin{align}
    \mathcal{D}_{\rm H}\equiv M^2 I_2-a^2 I_1^2 \ge 0,
    \label{BH-domain-condition}
\end{align}
with equality defining the extremal case $r_+=r_-$. Introducing the
dimensionless parameters
\begin{align}
    \chi\equiv \frac{a}{M},\qquad b\equiv B M,
\end{align}
the above horizon-existence condition (\ref{BH-domain-condition}) becomes
\begin{align}
    1-b^2\chi^2
    -\chi^2\left(1-\frac{1}{2}b^2\chi^2\right)^2
    \ge 0.
    \label{dimensionless-BH-domain}
\end{align}
For $b=0$, Eq.~(\ref{dimensionless-BH-domain}) reduces to the Kerr
bound $\chi\leq1$. For $\chi=0$, the geometry becomes the
Schwarzschild--BR solution, whose horizon radius
$r_h=2M/(1-b^2)$ requires $|b|<1$. Figure~\ref{fig:BH-parameter-space}
shows the black-hole domain within $|b|\leq1$, which is symmetric under
$b\to-b$.

For a fixed nonzero spin $0<\chi<1$,
Eq.~(\ref{dimensionless-BH-domain}) gives
$|b|\leq b_{\rm ext}(\chi)$, where
\begin{align}
    b_{\rm ext}^2
    =
    \frac{2}{\chi^4}
    \left(1-\sqrt{1-\chi^2}\right)\sqrt{1-\chi^2}.
    \label{Bext-condition}
\end{align}
This bound decreases monotonically with $\chi$ and satisfies
\begin{align}
    b_{\rm ext}^2
    &=
    \chi^{-2}-\frac14+\mathcal O(\chi^2),
    &&\chi\to0^+,\notag\\
    b_{\rm ext}
    &\sim\sqrt{2}\,(1-\chi^2)^{1/4},
    &&\chi\to1^-.
\end{align}
Thus, $b_{\rm ext}$ diverges along the nonzero-spin extremal curve as $\chi\to0^+$ and
terminates at the extremal Kerr point $(\chi,b)=(1,0)$; the separate
static limit remains restricted by $|b|<1$.

\subsection{The electromagnetic field}
The electromagnetic field of the Kerr--BR solution is encoded in a complex
one-form potential $A$ which takes the form,
\begin{equation}
\begin{aligned}
A={}&e^{i\xi}\frac{1}{2B}\Bigg[
\Omega_{,r}\frac{a\,dt-(r^2+a^2)d\varphi}
{r+i a\cos\theta}
\\
&+\frac{i\Omega_{,\theta}}{\sin\theta}
\frac{dt-a\sin^2\theta\,d\varphi}
{r+i a\cos\theta}
+(\Omega-1)d\varphi
\Bigg].
\end{aligned}
\end{equation}
Here $\xi$ parameterizes the electric--magnetic
duality rotation, and $\Omega$ denotes the positive square root of the metric function
$\Omega^2$, and $\Omega_{,r}\equiv\partial_r\Omega$,
$\Omega_{,\theta}\equiv\partial_\theta\Omega$. In this work, we choose $\xi=0$. This choice fixes the magnetic duality
sector: In the Bertotti--Robinson limit $M=0$, the ZAMO-measured electric field vanishes, so the $\xi=0$ field is purely magnetic in that frame (see Appendix~\ref{app:magnetic-field}), whereas $\xi=\pi/2$ gives the dual purely electric field. For a
rotating Kerr--BR black hole with $M\neq0$, however, the ZAMO decomposition
generally contains nonvanishing electric-field components.
Taking the physical real potential $A^{\rm real}=2\,{\rm Re}\,A$ gives
\begin{align}
A_t &=
\frac{
a r\,\Omega_{,r}
+a\,\Omega_{,\theta}\cot\theta
}{
B\left(r^2+a^2\cos^2\theta\right)
},
\\[0.8em]
A_\varphi &=
\frac{1}{
B\left(r^2+a^2\cos^2\theta\right)
}
\Big[
(\Omega-1)\left(r^2+a^2\cos^2\theta\right)
\nonumber\\
&\qquad
-r(r^2+a^2)\Omega_{,r}
-a^2\Omega_{,\theta}\cos\theta\sin\theta
\Big].
\label{Aphi}
\end{align}

For an observer with four-velocity $u^\mu$, the electric and
magnetic fields measured in the spatial frame $e_{(i)}^\mu$ are defined by

\begin{align}
    {E}^{(i)} \equiv  {F}^{\mu \nu }{u}_{\mu }{e}_{\nu }^{\left( i\right) },\;{B}^{(i)} \equiv  \frac{1}{2}{\epsilon }^{\mu \nu \rho \sigma }{F}_{\rho \sigma }{u}_{\mu }{e}_{\nu }^{\left( i\right) },
    \label{electromagnetic}
\end{align}
where $F_{\mu\nu} \equiv \partial_\mu A_\nu - \partial_\nu A_\mu$ and
$e_\nu^{(i)}=g_{\nu\lambda}e_{(i)}^\lambda$. We adopt the orientation
$\epsilon^{tr\theta\varphi}=-1/\sqrt{-g}$, consistently with the
convention employed in the calculation of the field components.

In a stationary and axisymmetric spacetime, such as the Kerr-BR black hole considered here, a natural class of fiducial observers is provided by the zero-angular-momentum-observers (ZAMOs). These observers are defined by the condition that their angular momentum with respect to the axial Killing vector vanishes, $u_\varphi = g_{\varphi t}u^t + g_{\varphi \varphi}u^\varphi = 0$, which, assuming no radial or poloidal motion, leads to a four-velocity of the form
\begin{align}
    u^\mu=N^{-1}\left(k_{(t)}^{\mu}
    +\omega k_{(\varphi)}^{\mu}\right)=\left(\frac{1}{N},0,0,\frac{\omega}{N}\right).
    \label{ZAMO}
\end{align}
Here $N=\sqrt{PQ\rho^2/(R\Omega^2)}$ is the lapse function,
$R=P(r^2+a^2)^2-Qa^2\sin^2\theta$, and
$\omega=-g_{t\varphi}/g_{\varphi\varphi}
   =a[P(r^2+a^2)-Q]/R$ is the frame-dragging angular velocity.
Following Ref.~\cite{Podolsky:2025KerrBR}, we use the spatial
orthonormal frame vectors
\begin{equation}
\begin{aligned}
e_{(r)} &=
\frac{\Omega\sqrt{Q}}{\rho}\,\partial_r,
\qquad
e_{(\theta)} =
-\frac{\Omega\sqrt{P}}{\rho}\,\partial_\theta,
\\
e_{(\varphi)} &=
-\frac{\Omega}{\rho\sqrt{P}\sin\theta}
\left(
a\sin^2\theta\,\partial_t+\partial_\varphi
\right).
\end{aligned}
\label{polar-tetrad-vector}
\end{equation}
So for ZAMOs, the observed components of the electromagnetic field can be calculated with Eqs.~(\ref{electromagnetic}) (\ref{ZAMO}) (\ref{polar-tetrad-vector}). The explicit full expressions of these components are rather long and we put them in Appendix~\ref{app:magnetic-field}. For equatorial motion with the dipole moment aligned in the polar direction we consider in this work, only the $\theta$-component of the magnetic field enters the interaction whose form is
\begin{align}
    B^{\hat{\theta}}
    =
    \Omega^2\sqrt{\frac{Q}{R}}\,
    \csc\theta\,\partial_r A_\varphi .
    \label{Btheta-compact}
\end{align}
Here we use the hatted symbols to denote quantities observed in ZAMO-frame. On the equatorial plane $\theta=\pi/2$, this reduces further to
\begin{equation}
\begin{aligned}
B^{\hat{\theta}}
&=
-Br(1-a^2B^2)
\left.
\sqrt{\frac{\Delta}{R}}
\right|_{\theta=\pi/2},
\end{aligned}
\label{Btheta-equatorial-compact}
\end{equation}
here, on the equatorial plane,
\[
R=(r^2+a^2)^2-a^2(1+B^2r^2)\Delta .
\]
\subsection{Equations of motion}

The dynamics of a test particle with an intrinsic magnetic dipole moment is governed by the following Lagrangian~\cite{Preti:2004Polarized}
\begin{align}
    {\cal L}  = \frac{1}{2}\left( {m + U}\right) {g}_{\mu \nu }\dot{x}^\mu \dot{x}^\nu - \frac{1}{2} U,
    \label{Lagrangian}
\end{align}
where $m$ denotes the particle's rest mass, and the dot represents the derivative with respect to the proper time $\tau$, i.e., $\dot{x}^\mu \equiv d x^\mu /d\tau$. The magnetic interaction energy $U$ takes the form

\begin{align}
    U=-\frac{1}{2}F^{\alpha\beta}P_{\alpha\beta},
    \qquad
    P_{\alpha\beta}=\epsilon_{\alpha\beta\rho\sigma}u^\rho\mu^\sigma-2u_{[\alpha}d_{\beta]},
    \label{interaction-polarization}
\end{align}
where $P_{\alpha\beta}$ is the antisymmetric polarization (dipole-moment) tensor. The dipole vectors $d^\alpha$ and $\mu^\alpha$ denote the electric and magnetic
dipole-moment four-vectors, respectively. Relative to the ZAMO
four-velocity $u^\alpha$, they satisfy
$u_\alpha d^\alpha=u_\alpha\mu^\alpha=0$.
In the adopted ZAMO-frame decomposition, we restrict attention
to the purely magnetic case and hence set $d^\alpha=0$. Additionally, we consider the magnetic moment to be perpendicular to the equatorial plane in the ZAMO-frame, taking $\mu>0$ to denote its magnitude,
we choose $\mu^{\hat\alpha}=(0,0,-\mu,0)$, i.e., the dipole points
along $-e_{(\theta)}$. This selects one of the two orientations normal
to the equatorial plane. With our conventions, it is aligned with the
local magnetic field for $B>0$ and anti-aligned for $B<0$.

The magnetic interaction energy then reduces to
\begin{align}
    U=-\mu_{\hat\alpha}B^{\hat\alpha}=\mu B^{\hat\theta}.
    \label{U}
\end{align}
For this velocity-independent coupling, the conjugate momentum is
$p_\mu=(m+U)g_{\mu\nu}\dot x^\nu$. The normalization
$g_{\mu\nu}\dot x^\mu\dot x^\nu=-1$ therefore gives
$g^{\mu\nu}p_\mu p_\nu=-(m+U)^2$.
Since $t$ and $\varphi$ are cyclic coordinates, their conjugate momenta are conserved:
\begin{align}
    p_t \equiv \frac{\partial {\cal L}}{\partial \dot{t}} = (m+U)\left(g_{tt}\dot t + g_{t\varphi}\dot\varphi\right) \equiv -E,\label{Energy}\\
    p_\varphi \equiv \frac{\partial {\cal L}}{\partial \dot{\varphi}} = (m+U)\left(g_{t\varphi}\dot t + g_{\varphi\varphi}\dot\varphi\right) \equiv L,
    \label{AngularMomentum}
\end{align}
Here $E$ and $L$ are the conserved energy and angular momentum, respectively.
For equatorial motion, we have $p_\theta=0$. Then Eqs.~(\ref{Energy}),(\ref{AngularMomentum}) give the radial equation
\begin{align}
    (1+\beta)^2g_{rr}\dot{r}^2=-g^{tt}\episode^2-g^{\varphi\varphi} l^2+2g^{t\varphi}\episode l-(1+\beta)^2,
    \label{motionEq}
\end{align}
where $\episode \equiv E/m$ and $l \equiv L/m$ denote the specific energy and the specific angular momentum of the particle, respectively. Furthermore, we introduce the magnetic coupling parameter $\beta\equiv U/m=\mu B^{\hat{\theta}}/m$, which characterizes the strength of the interaction between the particle’s intrinsic magnetic dipole moment and the external magnetic field. Throughout this work, we restrict to $1+\beta>0$, which ensures that $p^\mu$ and the particle four-velocity $\dot{x}^\mu$ have the same time orientation.

\subsection{Effective potential}

Using Eqs.~(\ref{Energy}), (\ref{AngularMomentum}), and (\ref{motionEq}), the radial equation, regarded as a quadratic equation for the specific energy $\episode$, can be rewritten as
\begin{align}
    \alpha \episode^2 + 2\delta \episode-\eta=0,
    \label{energy-quadratic}
\end{align}
where $\alpha=-g^{tt},
\delta=g^{t\varphi}l$ and
$\eta=g^{\varphi\varphi}l^2+(1+\beta)^2
+g_{rr}\dot{r}^2(1+\beta)^2$. Solving Eq.~(\ref{energy-quadratic}) for $\episode$ gives the two algebraic branches $\episode_\pm=(-\delta\pm\sqrt{\delta^2+\alpha\eta})/\alpha$. Since
$ (1+\beta)\dot t=\alpha\episode+\delta $, the two branches satisfy
\begin{align}
(1+\beta)\dot t
=\begin{cases}
+\sqrt{\delta^2+\alpha\eta},&\episode=\episode_+,\\
-\sqrt{\delta^2+\alpha\eta},&\episode=\episode_-.
\end{cases}
\end{align}
On the branch $1+\beta>0$, $\episode_+$ is future directed, whereas
$\episode_-$ is past directed. We therefore define the effective potential as
the value of $\episode_+$ at a radial turning point:
\begin{equation}
\begin{aligned}
V_{\rm eff}(r)
&\equiv \episode(\dot{r}=0)
=\frac{g^{t\varphi}l}{g^{tt}}
\\
&\quad
+\frac{
\sqrt{
l^2\left[(g^{t\varphi})^2-g^{tt}g^{\varphi\varphi}\right]
-g^{tt}(1+\beta)^2
}
}{
-g^{tt}
}.
\end{aligned}
\label{Veff}
\end{equation}
The turning points ($\dot{r}=0$) and extrema of $V_{\rm eff}$ determine the allowed radial regions and the possible circular orbits, respectively.

\begin{figure*}[!htbp]
    \centering
    \begin{subfigure}[b]{0.4\linewidth}
        \centering
        \includegraphics[width=\linewidth]{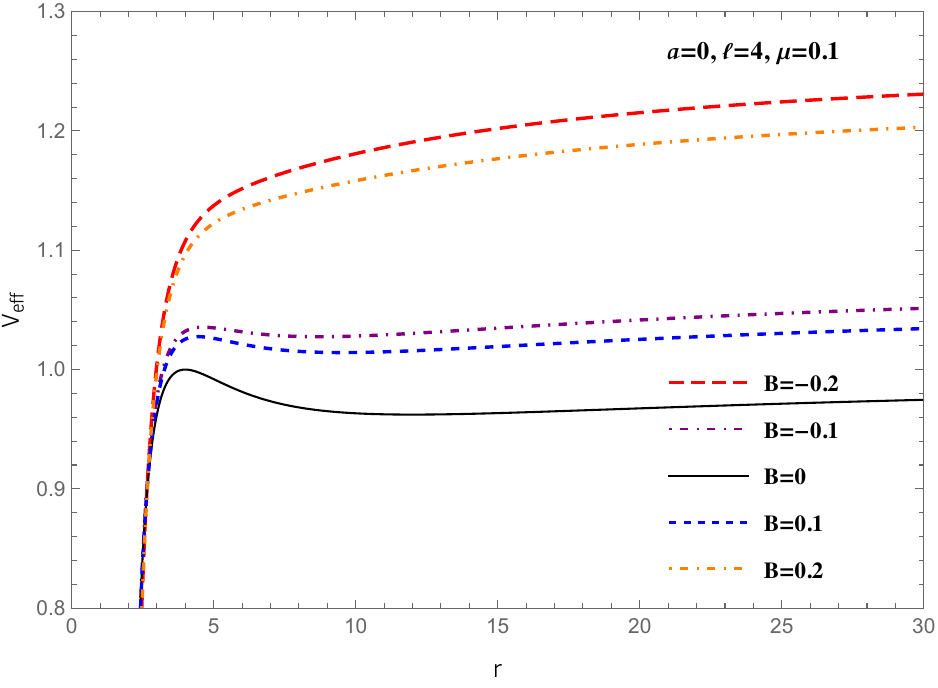}
        \caption{Varying $B$ for $a=0$, $l=4$, and $\mu=0.1$.}
        \label{fig:veff-B}
    \end{subfigure}
    \hspace{0.04\linewidth}
    \begin{subfigure}[b]{0.4\linewidth}
        \centering
        \includegraphics[width=\linewidth]{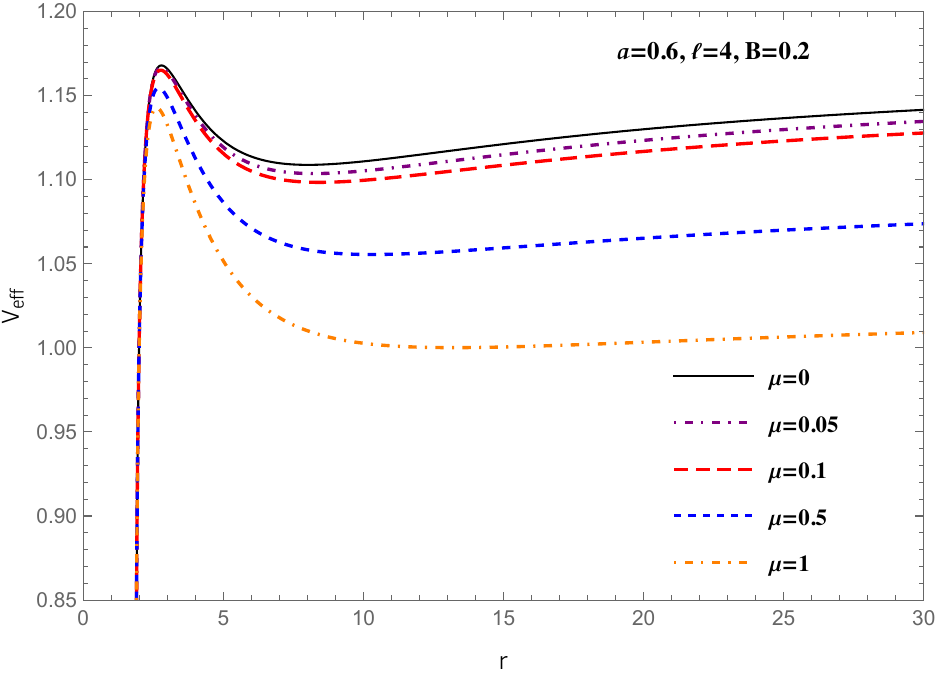}
        \caption{Varying $\mu$ for $a=0.6$, $B=0.2$, and $l=4$.}
        \label{fig:veff-mu}
    \end{subfigure}

    \vspace{0.15cm}

    \begin{subfigure}[b]{0.4\linewidth}
        \centering
        \includegraphics[width=\linewidth]{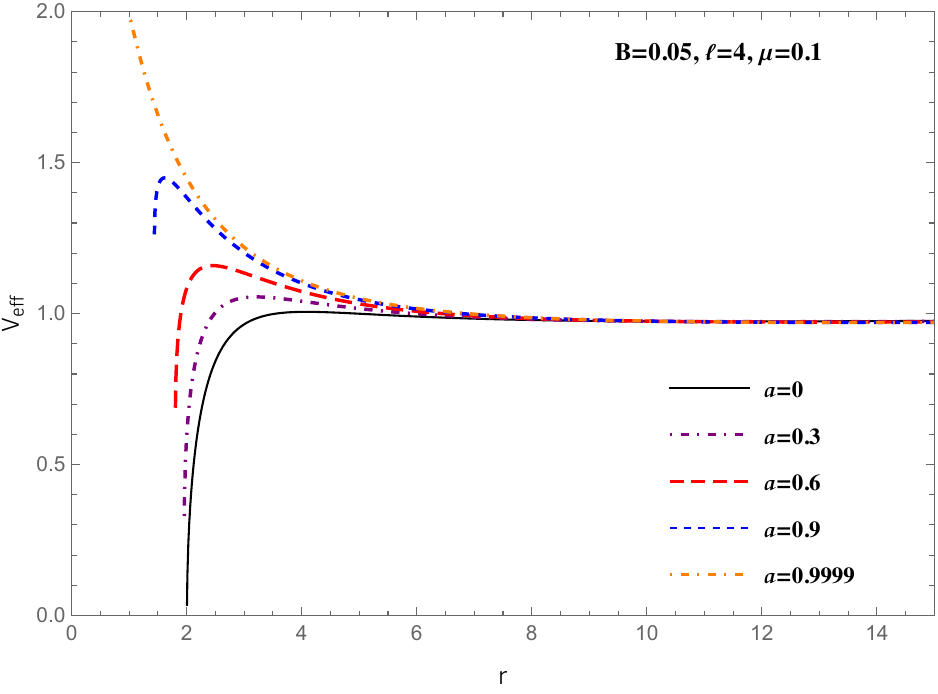}
        \caption{Varying $a$ for $B=0.05$, $l=4$, and $\mu=0.1$.}
        \label{fig:veff-a}
    \end{subfigure}
    \hspace{0.04\linewidth}
    \begin{subfigure}[b]{0.4\linewidth}
        \centering
        \includegraphics[width=\linewidth]{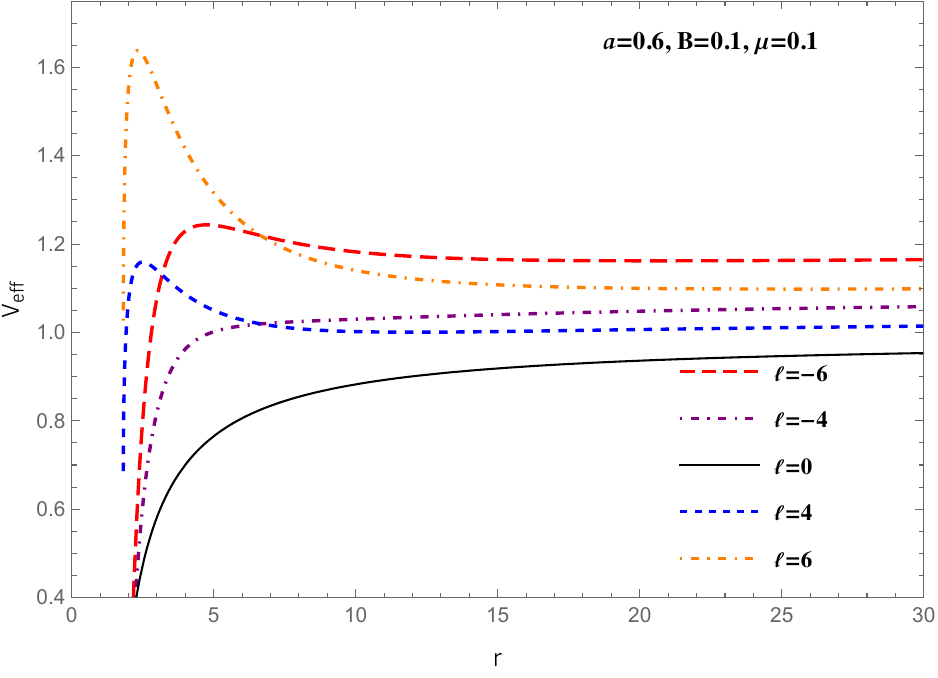}
        \caption{Varying $l$ for $a=0.6$, $B=0.1$, and $\mu=0.1$.}
        \label{fig:veff-l}
    \end{subfigure}

    \caption{Effective potential $V_{\rm eff}$ for a magnetized particle in the equatorial plane of a Kerr--BR black hole.}
    \label{VeffNER}
\end{figure*}

Fig.~\ref{VeffNER} compares the effects of $B$, $\mu$, $a$, and $l$ on $V_{\rm eff}$. In Fig.~\ref{fig:veff-B}, the change in the potential is not simply proportional to $B$, and reversing the sign of $B$ produces different curves: for the same $|B|$, the negative-$B$ branch lies higher at large radii. This difference results from the combined effects of the $B$-dependent geometry and the dipole coupling $\beta=\mu B^{\hat\theta}/m$. In Fig.~\ref{fig:veff-mu}, increasing $\mu$ lowers the potential outside the inner maximum and deepens the outer well, while the inner peak changes less significantly. Fig.~\ref{fig:veff-a} shows that increasing $a$ mainly modifies the near-horizon region by moving the inner peak inward and raising the barrier. Finally, Fig.~\ref{fig:veff-l} shows that increasing $|l|$ strengthens the centrifugal barrier, while the difference between positive and negative $l$ is caused by frame dragging.

\section{orbits}

In this section, we systematically investigate the structure of orbits of the magnetized particle, and analyze the influence of the magnetic field on the orbital properties.

For equatorial motion, Eqs.~(\ref{motionEq}), (\ref{Energy}), and (\ref{AngularMomentum}) give
\begin{align}
    \dot{r}^2 = \frac{-g^{tt}\episode^2 - g^{\varphi\varphi}l^2 + 2g^{t\varphi}\episode l - (1+\beta)^2}{(1+\beta)^2 g_{rr}},
    \label{rdot}
\end{align}
\begin{align}
    \dot{\varphi}=-\frac{g_{t\varphi}\episode+g_{tt}l}{(1+\beta)(g^{2}_{t\varphi}-g_{tt}g_{\varphi\varphi})}=\frac{g^{\varphi\varphi}l-g^{t\varphi}\episode}{1+\beta}.
\end{align}
With them, one can obtain the orbital equation 
\begin{align}
    \left(\frac{dr}{d\varphi}\right)^2 = \frac{\dot{r}^2}{\dot{\varphi}^2} =\frac{-g^{tt} \episode^2 - g^{\varphi\varphi} l^2 + 2g^{t\varphi} \episode l - (1 + \beta)^2}{g_{rr} (g^{\varphi\varphi} l - g^{t\varphi} \episode)^2}
    \label{orbiteq}
\end{align}

\subsection{Axial regularity and azimuthal normalization}
As pointed out in Ref.~\cite{Podolsky:2025KerrBR}, the metric (\ref{KerrBR}) may have conical singularity on the symmetry axis ($\theta=0, \pi$). To remove such kind of singularity, the period of the original azimuthal coordinate $\varphi$ should be normalized as
\begin{align}
    \varphi\sim\varphi+2\pi\mathcal C,
    \label{regular-azimuthal-period}
\end{align}
where the conicity parameter $\mathcal C$ is
\begin{align}
    \mathcal C
    &=
    \left.\frac{1}{P}\right|_{\theta=0,\pi}
    =
    \left[
    1+B^2\left(\frac{M^2I_2}{I_1^2}-a^2\right)
    \right]^{-1}.
    \label{conicity-factor}
\end{align}

For quantities that count complete revolutions, it is convenient to introduce
the standard-period angle
\begin{align}
    \Phi\equiv\frac{\varphi}{\mathcal C},
    \qquad
    \Phi\sim\Phi+2\pi .
    \label{standard-period-angle}
\end{align}
For Cartesian-type visualizations of equatorial trajectories, the normalized
angle must be used:
\begin{align}
    x=r\cos\Phi
    =r\cos\left(\frac{\varphi}{\mathcal C}\right),
    \qquad
    y=r\sin\Phi
    =r\sin\left(\frac{\varphi}{\mathcal C}\right).
    \label{regular-orbit-plotting}
\end{align}

\subsection{Bound and unbound orbits}

\begin{figure*}[!htbp]
    \centering
    \begin{subfigure}[t]{0.395\linewidth}
        \vspace{0pt}
        \centering
        \includegraphics[width=\linewidth,height=0.19\textheight,keepaspectratio]{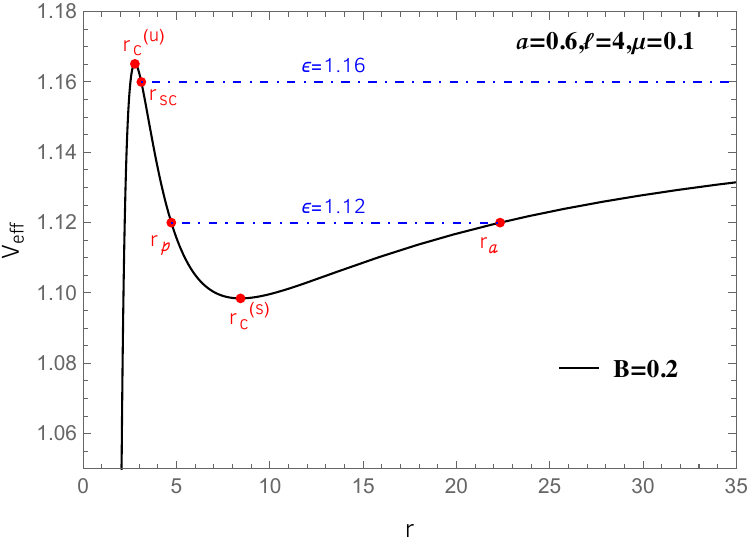}
        \caption{Potential and characteristic radii.}
        \label{fig:orbit-potential}
    \end{subfigure}\hfill
    \begin{subfigure}[t]{0.295\linewidth}
        \vspace{0pt}
        \centering
        \includegraphics[width=\linewidth,height=0.19\textheight,keepaspectratio]{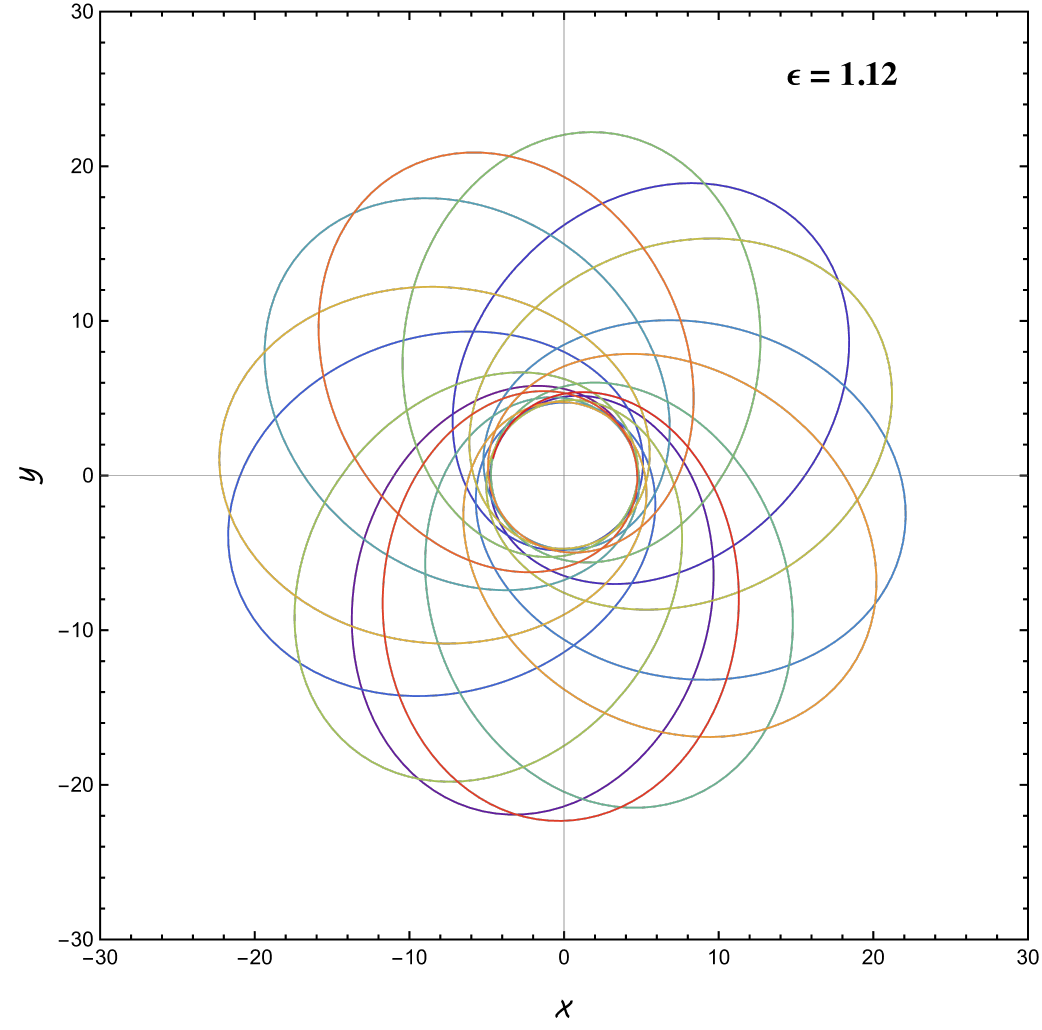}
        \caption{Bound orbit, $\episode=1.12$.}
        \label{fig:orbit-bound}
    \end{subfigure}
\begin{subfigure}[t]{0.295\linewidth}
        \vspace{0pt}
        \centering
        \includegraphics[width=\linewidth,height=0.19\textheight,keepaspectratio]{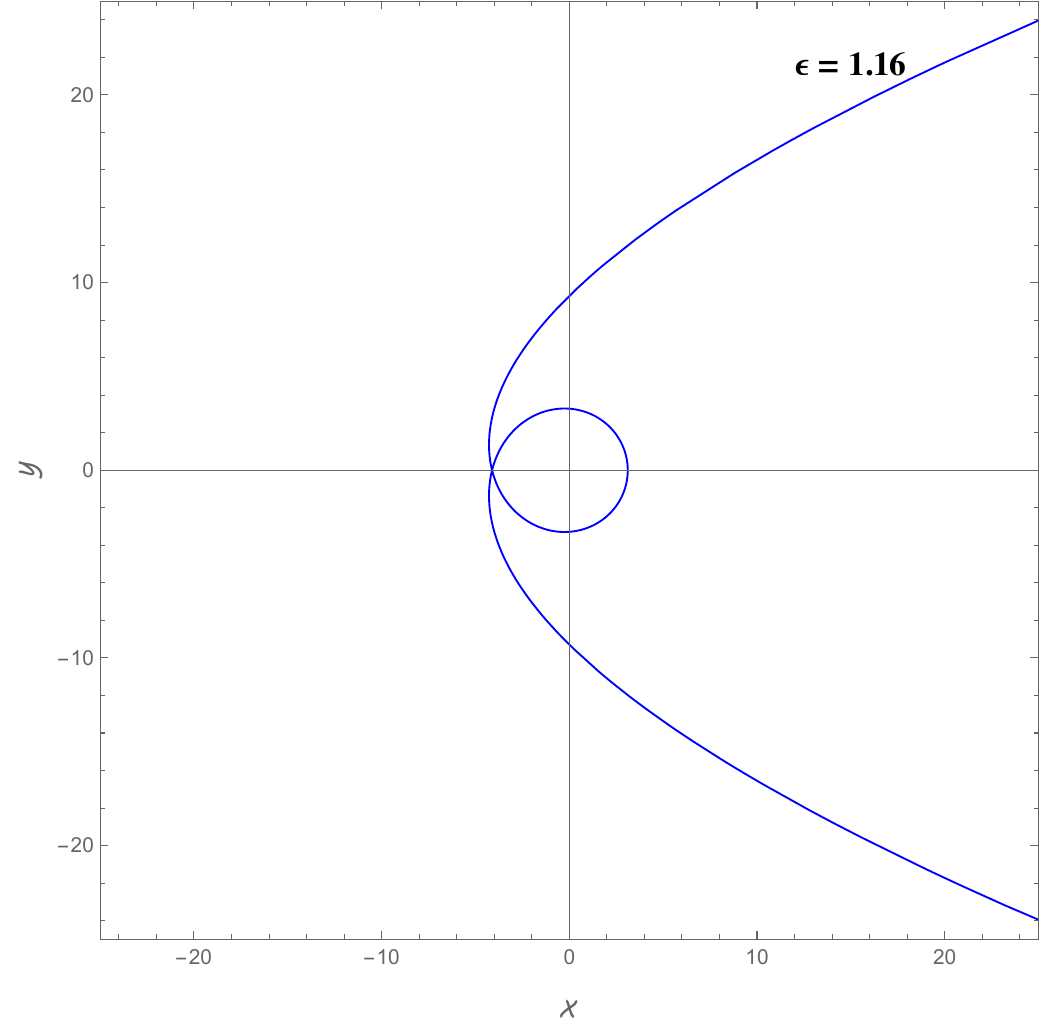}
        \caption{Scattering orbit, $\episode=1.16$.}
        \label{fig:orbit-scattering}
    \end{subfigure}\hfill
    \caption{Effective potential and representative equatorial trajectories for $a=0.6$, $B=0.2$, $l=4$, and $\mu=0.1$. The notation for the characteristic radii is summarized in Table~\ref{tab:characteristic-radii}. Panels (b) and (c) use the regular plotting coordinates $(x,y)=(r\cos\Phi,r\sin\Phi)$, with $\Phi=\varphi/\mathcal C$, as defined in Eq.~(\ref{regular-orbit-plotting}).}
    \label{fig:orbit-classification}
\end{figure*}

To connect the effective potential with the trajectories, we use the
$\mu=0.1$ curve of Fig.~\ref{fig:veff-mu} with $a=0.6$, $B=0.2$, and
$l=4$, reproduced in Fig.~\ref{fig:orbit-potential}.  At fixed specific
energy, each connected interval satisfying
$\episode\geq V_{\rm eff}(r)$ defines an allowed radial region.  A radial
turning point $r_t$ satisfies $\dot r=0$, or equivalently
$\episode=V_{\rm eff}(r_t)$.
The local minimum
$r_{\rm c}^{(\rm s)}$ is a stable circular radius, whereas the local maximum
$r_{\rm c}^{(\rm u)}$ is an unstable circular radius.  The notation for all
characteristic radii used in this subsection is summarized in
Table~\ref{tab:characteristic-radii}.

For the outer bound region, the smaller and larger turning
radii are identified as the periastron $r_p$ and apastron $r_a$,
respectively.  At $\episode=1.12$, the motion is confined to
$r_p\leq r\leq r_a$, and the resulting trajectory is the precessing rosette
shown in Fig.~\ref{fig:orbit-bound}.

For $\episode=1.16$, unbound scattering satisfies
\begin{align}
    V_{\rm eff}(\infty)<\episode
    <V_{\rm eff}\!\left(r_{\rm c}^{(\rm u)}\right).
\end{align}
The incoming particle reaches the closest-approach turning point $r_{\rm sc}$
and returns to infinity, as shown in Fig.~\ref{fig:orbit-scattering}.  As
$\episode\to V_{\rm eff}(r_{\rm c}^{(\rm u)})^{-}$,
$r_{\rm sc}\to r_{\rm c}^{(\rm u)}$; crossing the barrier instead produces
plunging motion.  Any additional near-horizon intersections belong to
separate horizon-connected allowed components and are not turning points of
the trajectories displayed in panels (b) and (c).

\begin{table*}[!htbp]
    \centering
    \small
    \setlength{\tabcolsep}{7pt}
    \renewcommand{\arraystretch}{1.12}
    \begin{tabular}{ccc}
        \toprule
        Symbol & Physical role & Defining condition \\
        \midrule
        $r_t$ & Radial turning radius
            & $\dot r=0$, or $\episode=V_{\rm eff}(r_t)$ \\
        $r_p$ & Periastron of a bound orbit
            & $\episode=V_{\rm eff}(r_p)$ \\
        $r_a$ & Apastron of a bound orbit
            & $\episode=V_{\rm eff}(r_a)$ \\
        $r_{\rm sc}$ & Closest approach of the scattering orbit
            & $\episode=V_{\rm eff}(r_{\rm sc})$ \\
        $r_{\rm c}^{(\rm s)}$ & Stable circular radius
            & $V_{\rm eff}'(r_{\rm c}^{(\rm s)})=0$,
              $V_{\rm eff}''(r_{\rm c}^{(\rm s)})>0$ \\
        $r_{\rm c}^{(\rm u)}$ & Unstable circular radius
            & $V_{\rm eff}'(r_{\rm c}^{(\rm u)})=0$,
              $V_{\rm eff}''(r_{\rm c}^{(\rm u)})<0$ \\
        \bottomrule
    \end{tabular}
    \caption{Characteristic radii appearing in the text.}
    \label{tab:characteristic-radii}
\end{table*}

\begin{figure*}[!htbp]
    \centering

    \begin{subfigure}[b]{0.46\linewidth}
        \centering
        \includegraphics[width=\linewidth]{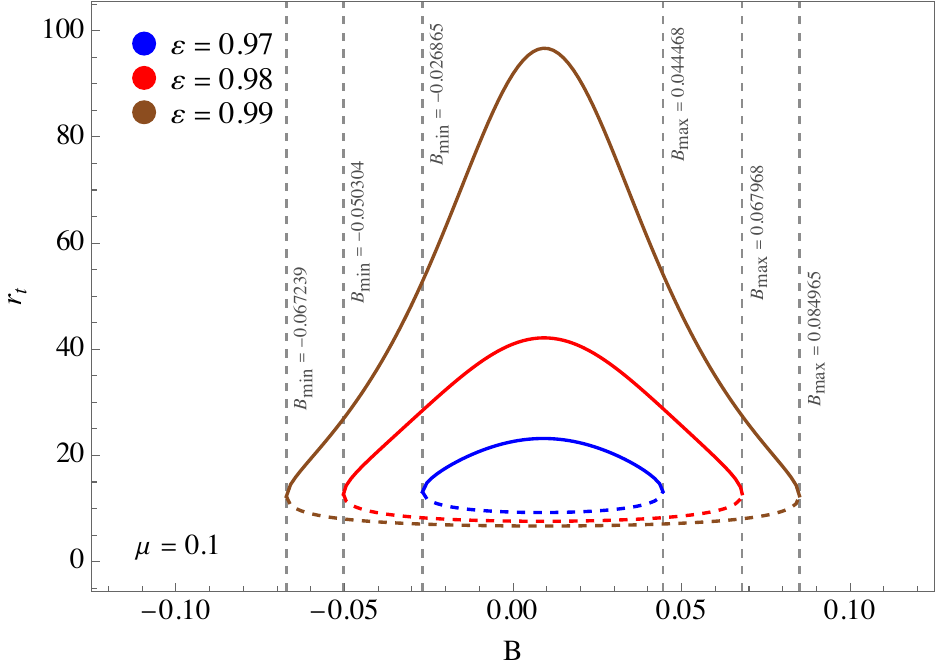}
        \caption{}
        \label{rtB}
    \end{subfigure}
    \quad
    \begin{subfigure}[b]{0.46\linewidth}
        \centering
        \includegraphics[width=\linewidth]{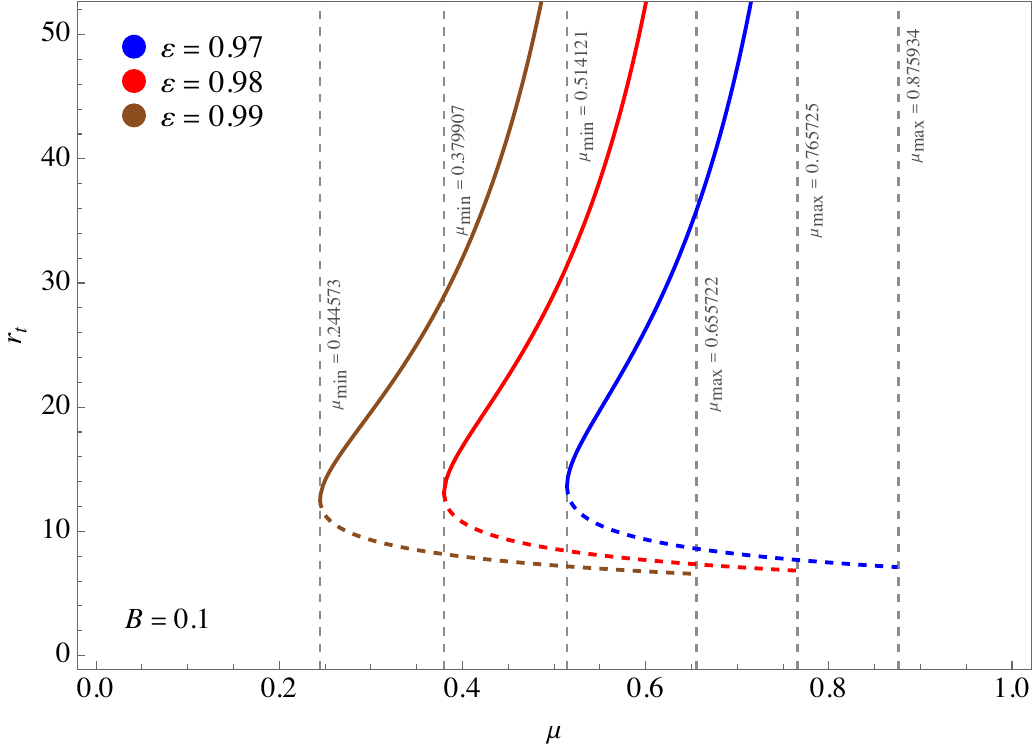}
        \caption{}
        \label{rtmu}
    \end{subfigure}

    \caption{Turning-point radii of the outer bound region,
    with $r_p$ shown by dashed curves and $r_a$ by solid curves, as functions
    of $B$ in panel (a) and $\mu$ in panel (b), at $a=0.6$ and $l=4$.
    Panel (a) fixes $\mu=0.1$, whereas panel (b) fixes $B=0.1$.
    The colors correspond to $\episode=0.97$, $0.98$, and $0.99$, and the
    vertical gray dashed lines mark the corresponding critical values.
    In panel (b), the solid curves
    leave the plotting range as $r_a\to\infty$ when
    $\mu\to\mu_{\max}$.}
    \label{rl}
\end{figure*}

For each fixed specific energy $\episode$, two distinct finite turning
points, $r_p$ and $r_a$, exist for $B_{\min}<B<B_{\max}$ in the
$B$ scan and for $\mu_{\min}<\mu<\mu_{\max}$ in the $\mu$ scan,
thereby defining the parameter ranges for finite noncircular bound
orbits. At $B_{\min}$, $B_{\max}$, and $\mu_{\min}$, the two turning
points coalesce into a stable circular orbit, $r_p=r_a$, as shown in
Fig.~\ref{rtB}. The left panel also shows that a larger magnetic-field
magnitude requires a higher $\episode$ to sustain a bound orbit.
When the magnetic field becomes sufficiently strong, the
effective-potential well disappears and bound orbits can no longer
form, as illustrated in Fig.~\ref{fig:veff-B}. In Fig.~\ref{rtmu},
increasing $\mu$ enlarges the radial excursion from $r_p$ to $r_a$.
At the asymptotic threshold $\mu_{\max}$, defined by
$V_{\rm eff}(\infty)=\episode$, $r_a\to\infty$ while $r_p$ remains finite,
and the orbit approaches the marginally unbound limit. The corresponding
critical values are marked by the vertical gray dashed lines in Fig.~\ref{rl}.

Radial confinement does not imply that an orbit closes after a
finite number of azimuthal revolutions. Closed periodic orbits form the subset of bound trajectories for which the
radial and azimuthal frequencies are commensurate.  We classify them using
the Levin--Perez-Giz framework~\cite{Levin:2008Periodic}, in which zoom--whirl trajectories alternate
between extended radial excursions and rapid revolutions near the periastron.

Introducing $u=1/r$, Eq.~(\ref{orbiteq}) becomes

\begin{equation}
    \left(\frac{du}{d\varphi}\right)^2 = u^4 \left[ \frac{-g^{tt} \episode^2 - g^{\varphi\varphi} l^2 + 2g^{t\varphi} \episode l - (1 + \beta)^2}{g_{rr} (g^{\varphi\varphi} l - g^{t\varphi} \episode)^2} \right].
    \label{ueq}
\end{equation}

Using the turning radii defined above, we define the eccentricity $e$ and the
semi-latus rectum $\lambda$ through
\begin{align}
    u_1=\frac{1-e}{\lambda},
    \qquad
    u_2=\frac{1+e}{\lambda},
    \label{bound-turning-parameterization}
\end{align}
where $u_1=1/r_a$ and $u_2=1/r_p$. For prescribed $(e,\lambda)$, the corresponding values of $\episode$ and $l$ are determined by the two turning-point conditions
$\episode=V_{\rm eff}(r_p)=V_{\rm eff}(r_a)$.

The accumulated azimuthal angle in the original coordinate
$\varphi$ during one complete radial oscillation is then directly obtained
by integrating Eq.~(\ref{ueq}):
\begin{widetext}
\begin{align}
    \Delta\varphi = 2\int_{u_1}^{u_2} \frac{1}{u^2} \left[ \frac{-g^{tt} \episode^2 - g^{\varphi\varphi} l^2 + 2g^{t\varphi} \episode l - (1 + \beta)^2}{g_{rr} (g^{\varphi\varphi} l - g^{t\varphi} \episode)^2} \right]_{r=1/u}^{-1/2} du.
\end{align}
\end{widetext}
The corresponding advance in the standard $2\pi$-periodic angle
$\Phi=\varphi/\mathcal C$ is
$\Delta\Phi\equiv\Delta\varphi/\mathcal C$, where $\mathcal C$ is the
conicity factor required by axial regularity and defined in
Eq.~(\ref{conicity-factor}). The orbital precession parameter can then be defined as
\begin{align}
    q
    =w+\frac{v}{z}
    =\frac{\Delta\Phi}{2\pi}-1
    =\frac{\Delta\varphi}{2\pi\mathcal C}-1,
    \label{regular-precession-parameter}
\end{align}
where $z$ denotes the number of zooms, $w$ is the number of whirls near the periastron, and $v$ determines the order in which successive leaves are traced. Periodic orbits occur when $q$ takes a rational value.

\begin{figure*}[!htbp]
    \centering

    \begin{minipage}[c]{0.32\linewidth}
        \centering
        \includegraphics[width=\linewidth]{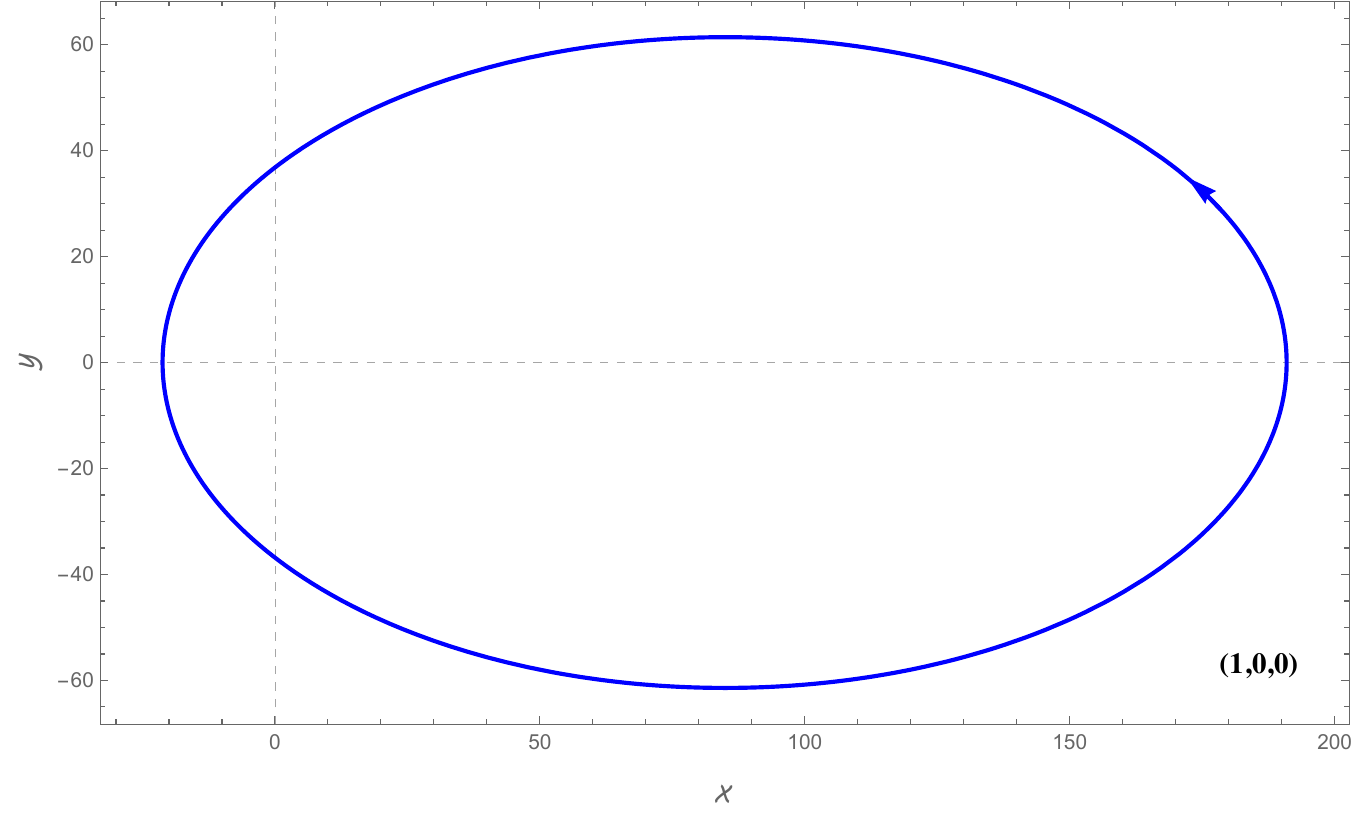}
    \end{minipage}\hfill
    \begin{minipage}[c]{0.32\linewidth}
        \centering
        \includegraphics[width=\linewidth]{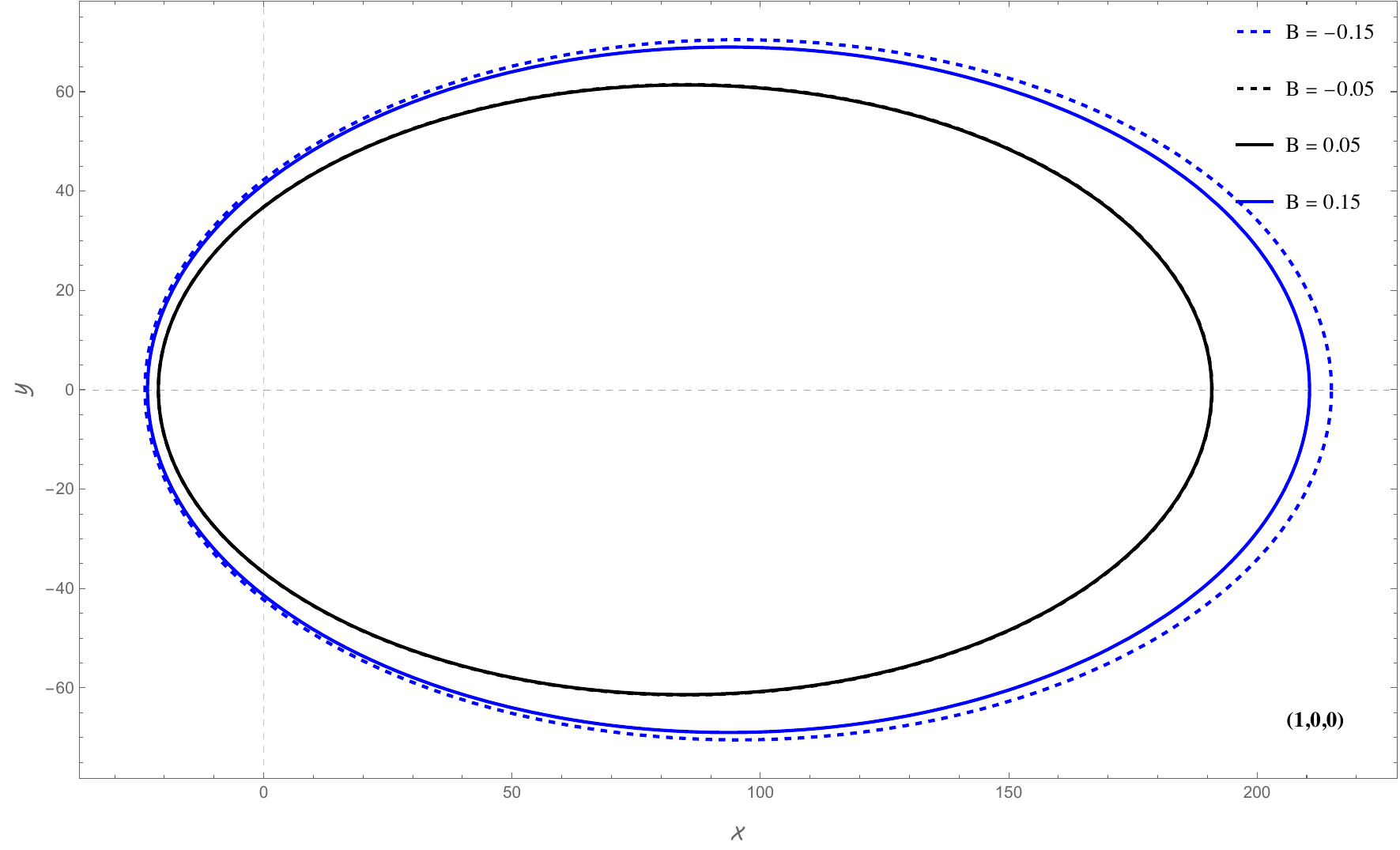}
    \end{minipage}\hfill
    \begin{minipage}[c]{0.32\linewidth}
        \centering
        \includegraphics[width=\linewidth]{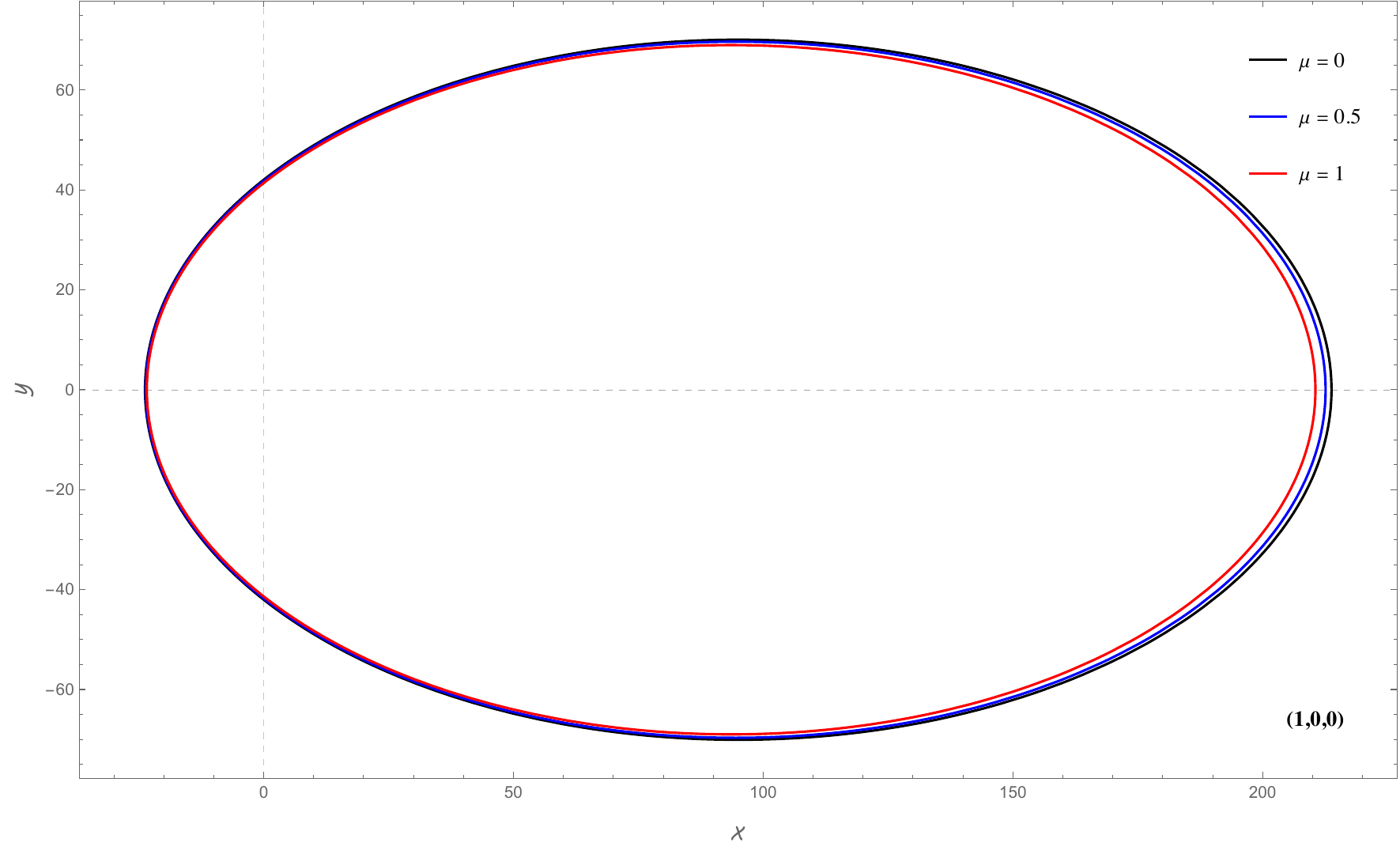}
    \end{minipage}

    \medskip

    \begin{minipage}[c]{0.32\linewidth}
        \centering
        \includegraphics[width=\linewidth]{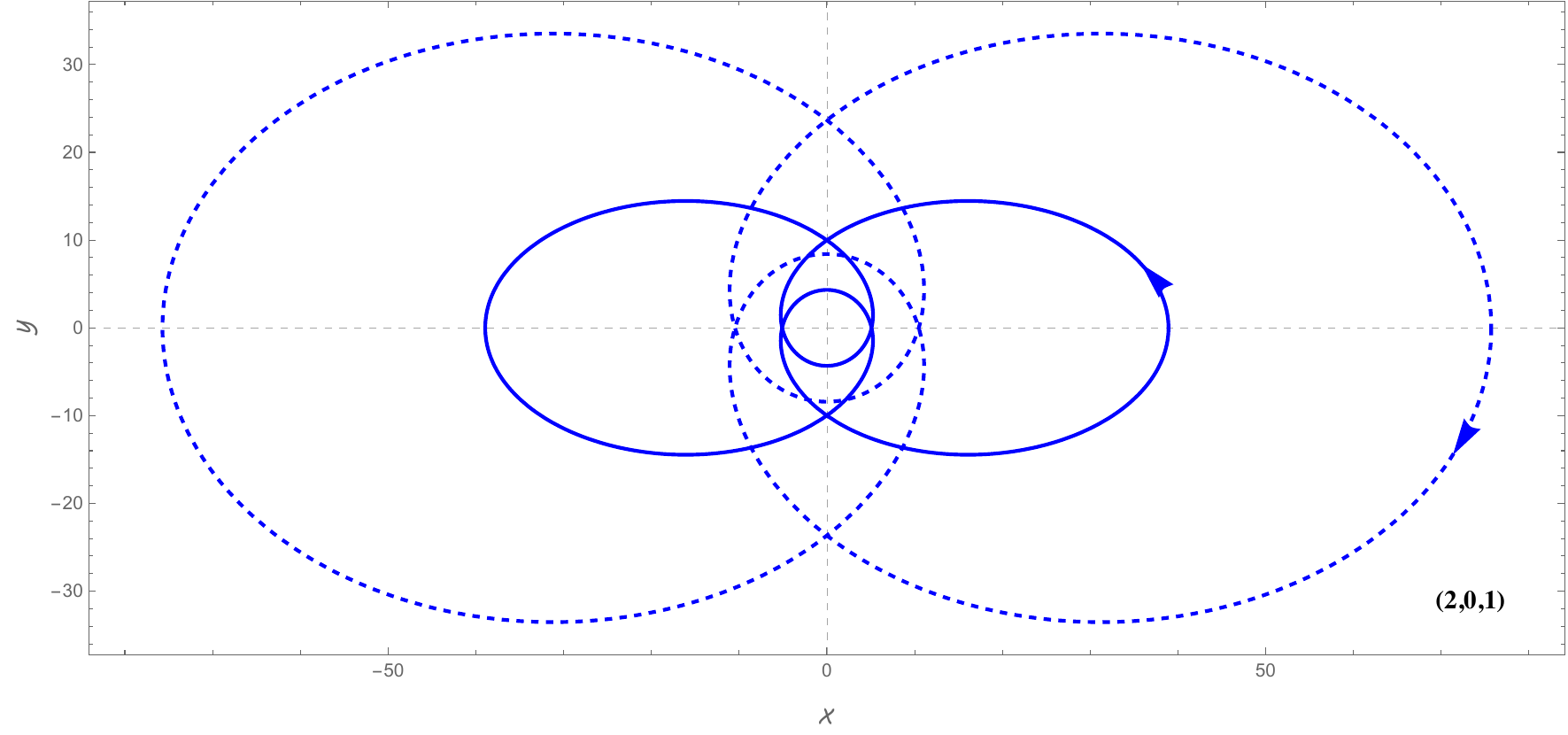}
    \end{minipage}\hfill
    \begin{minipage}[c]{0.32\linewidth}
        \centering
        \includegraphics[width=\linewidth]{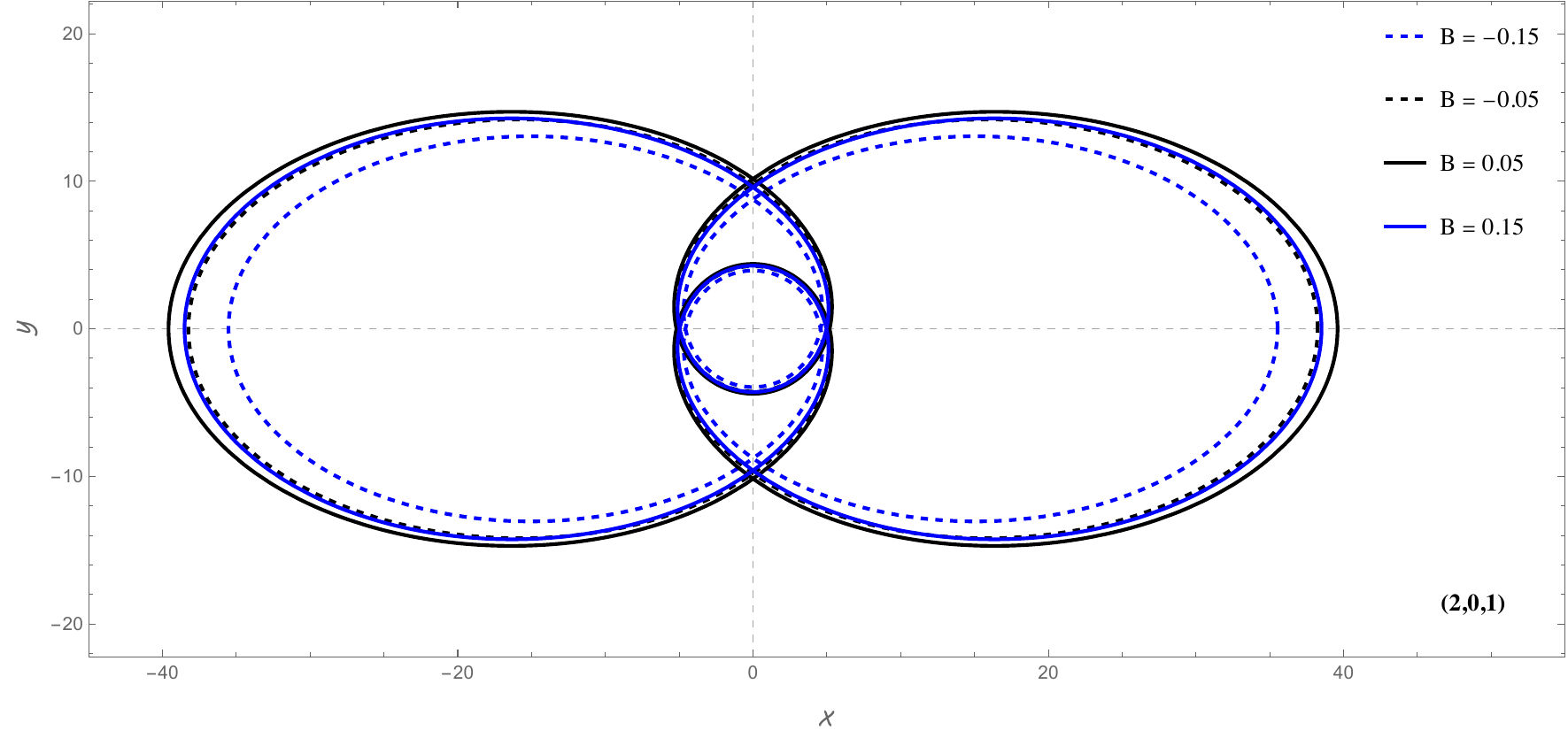}
    \end{minipage}\hfill
    \begin{minipage}[c]{0.32\linewidth}
        \centering
        \includegraphics[width=\linewidth]{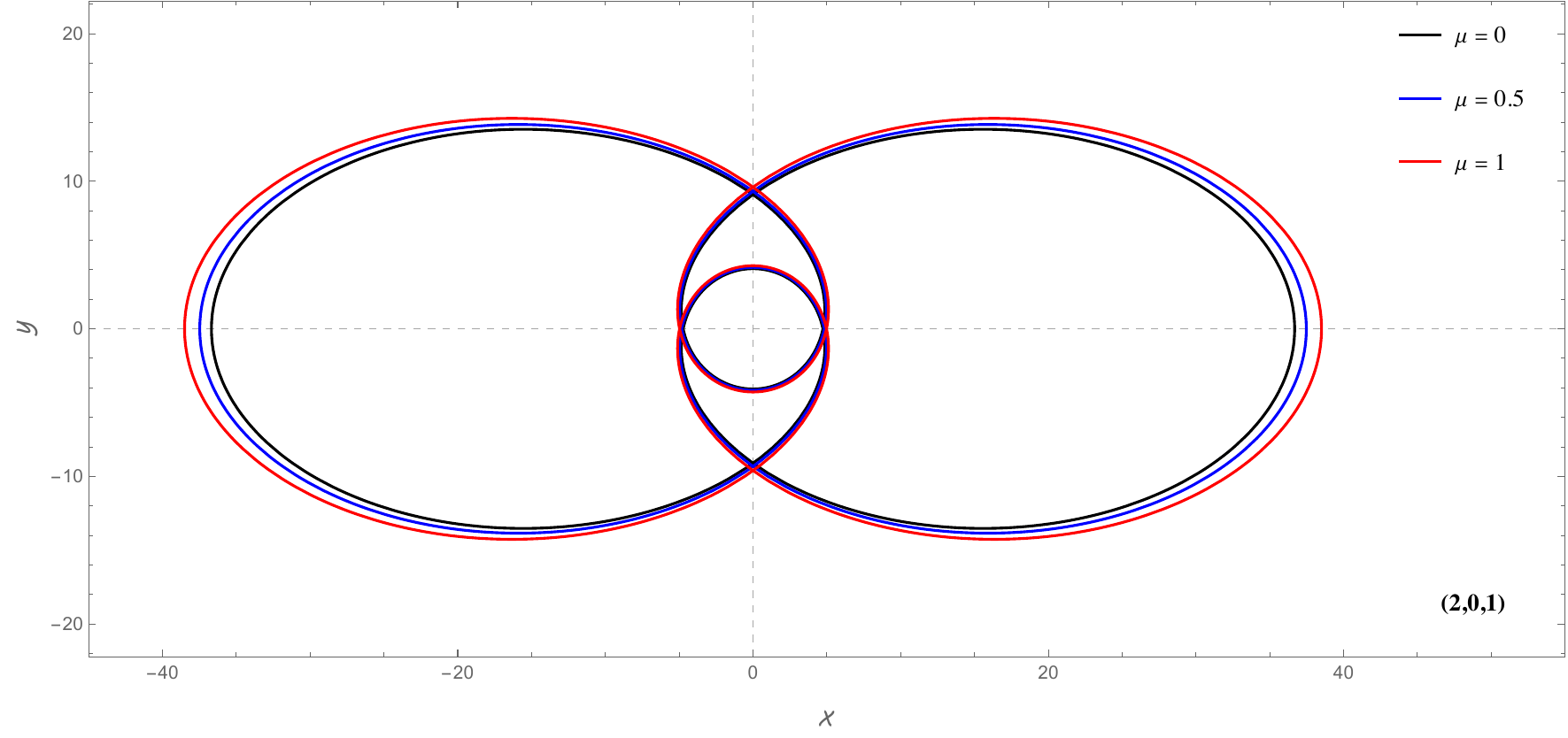}
    \end{minipage}

    \medskip

    \begin{minipage}[c]{0.32\linewidth}
        \centering
        \includegraphics[width=\linewidth]{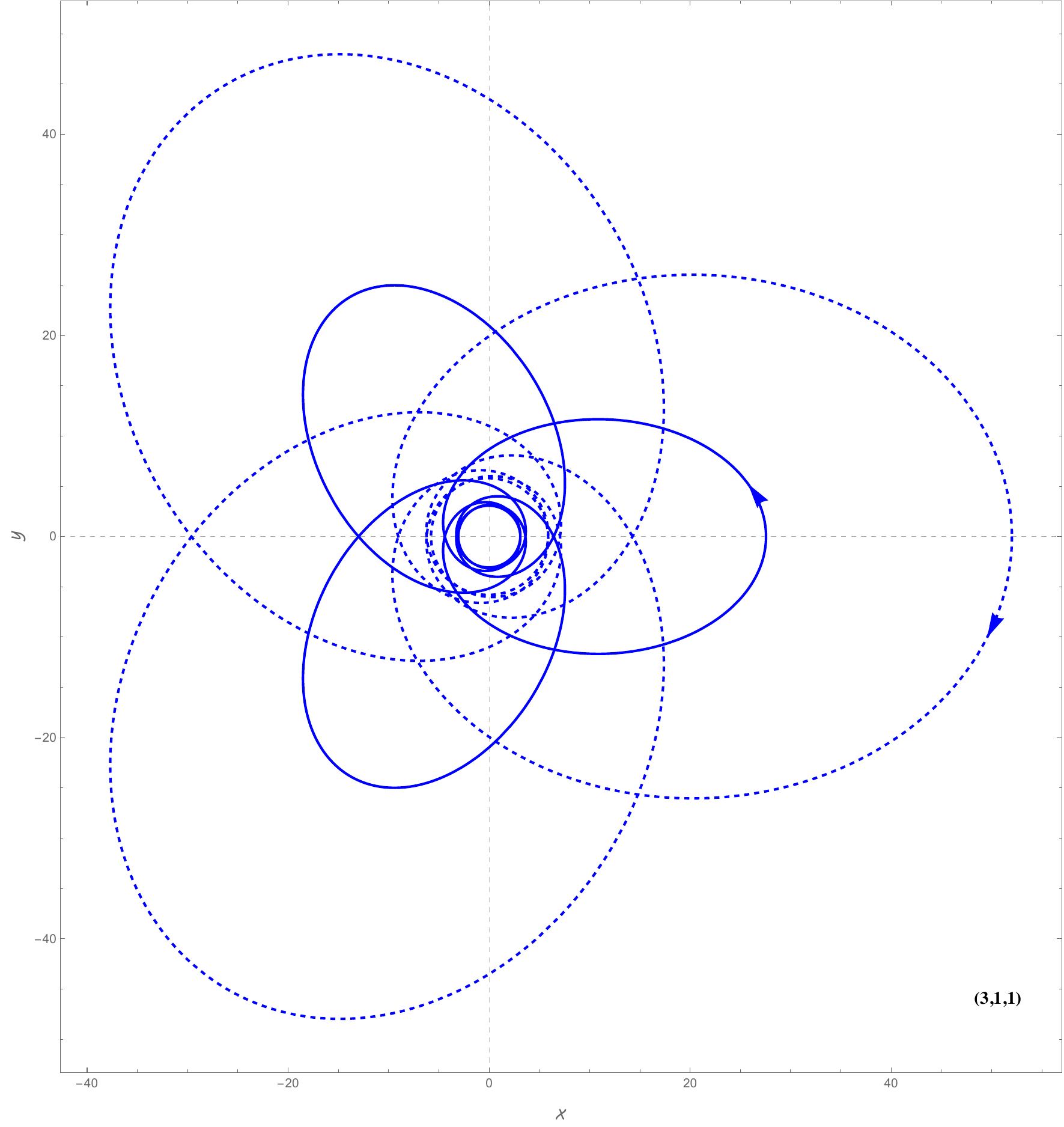}
    \end{minipage}\hfill
    \begin{minipage}[c]{0.32\linewidth}
        \centering
        \includegraphics[width=\linewidth]{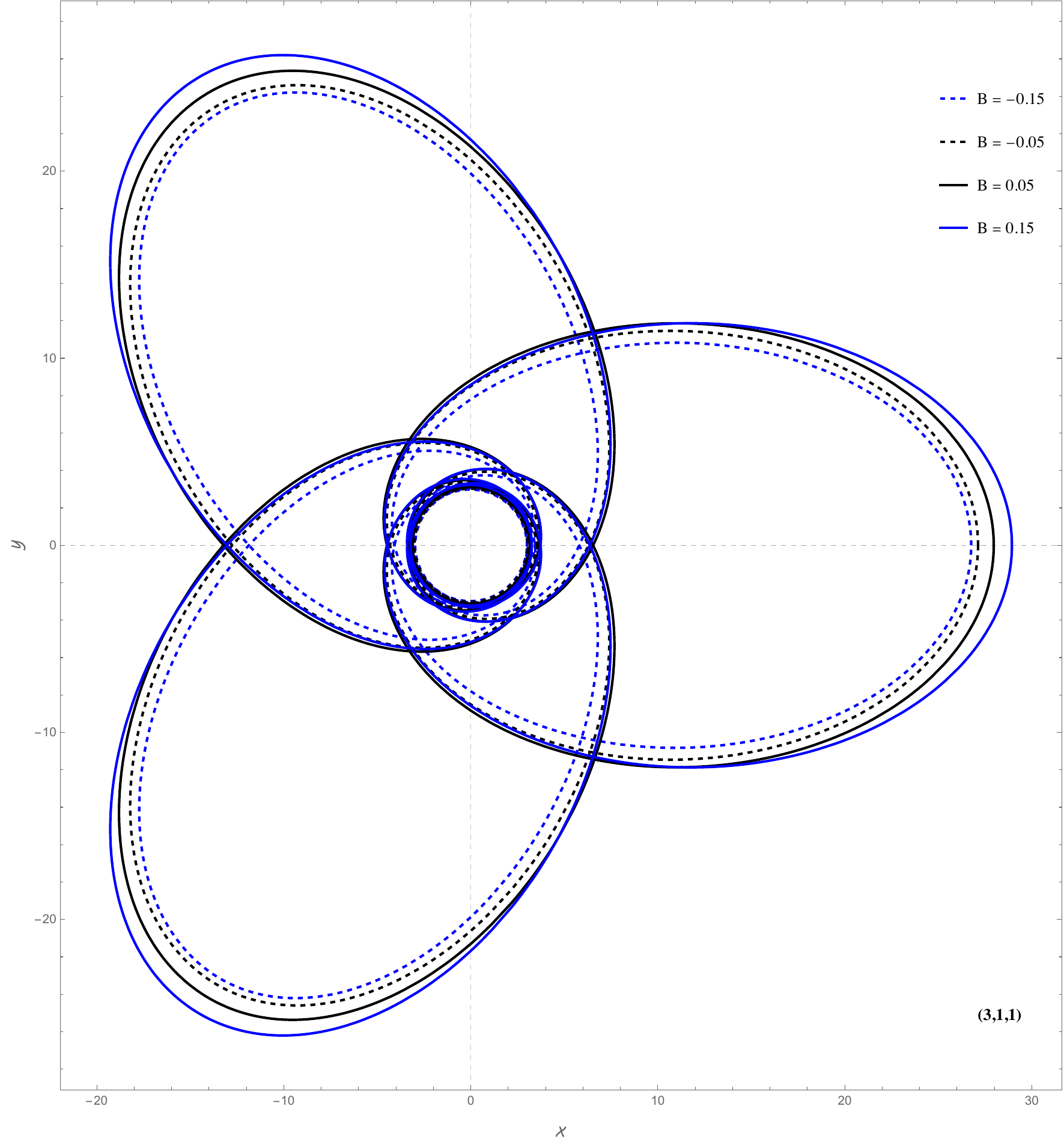}
    \end{minipage}\hfill
    \begin{minipage}[c]{0.32\linewidth}
        \centering
        \includegraphics[width=\linewidth]{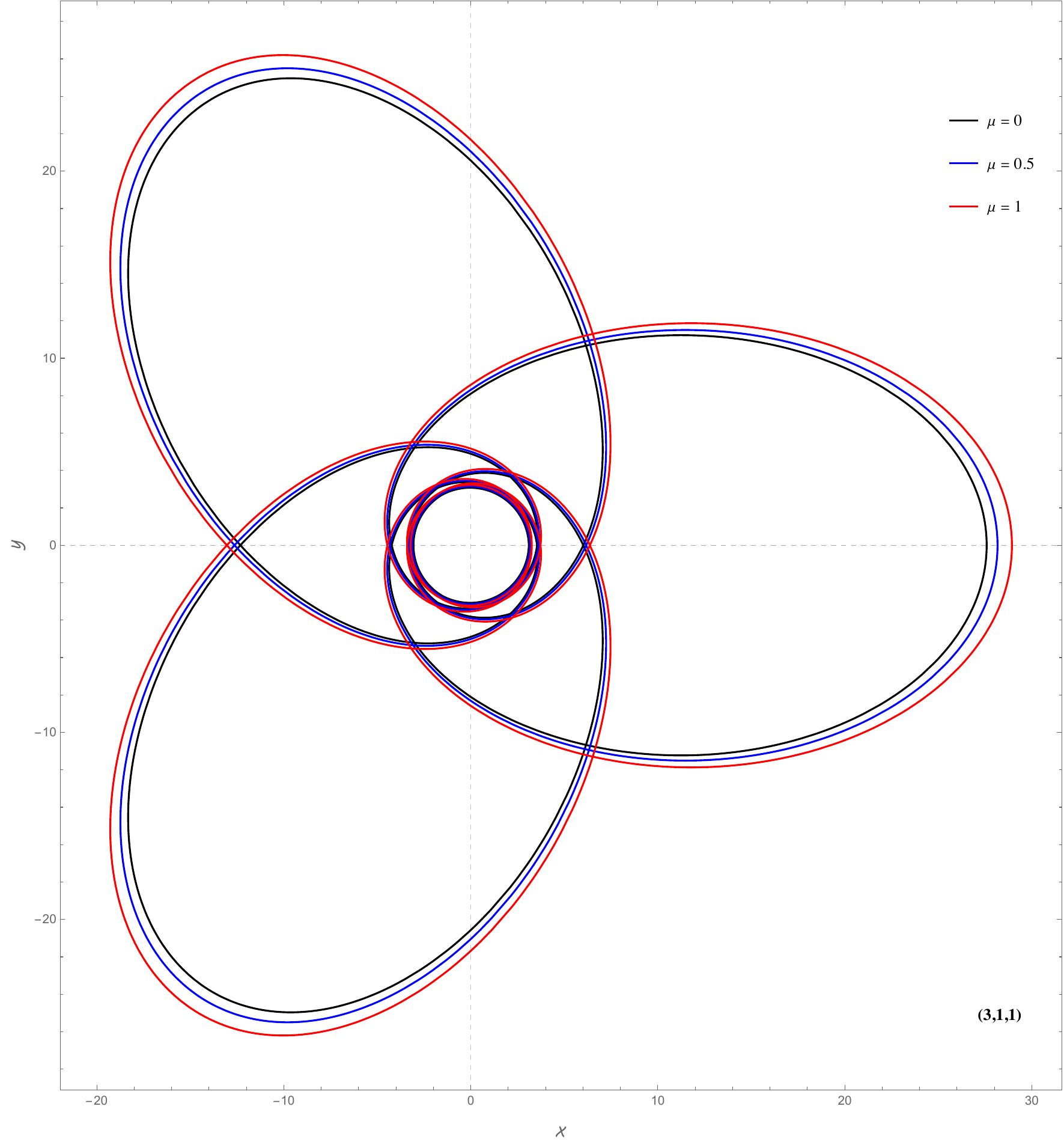}
    \end{minipage}

    \caption{Closed equatorial orbits for $a=0.6$ and $e=0.8$.
    From top to bottom, the rows show the $(1,0,0)$, $(2,0,1)$, and $(3,1,1)$
    families.  The left column shows the prograde branch ($l>0$, solid) and,
    where it exists, the retrograde branch ($l<0$, dashed) at
    $(B,\mu)=(0.05,0.1)$.  The middle and right columns show the
    prograde $B$ scan at $\mu=1$ and the prograde $\mu$ scan at $B=0.15$,
    respectively.  Solid and dashed curves in the middle column denote
    $B>0$ and $B<0$; colors distinguish $\mu$ in the right column.
    Coordinates are $(r\cos\Phi,r\sin\Phi)$, with
    $\Phi=\varphi/\mathcal C$.}
    \label{closedorbits}
\end{figure*}

Figure~\ref{closedorbits} relates the classification $(z,w,v)$ to orbital
morphology. In the left column, the retrograde $(2,0,1)$ and $(3,1,1)$ orbits
lie at larger radii and require larger $|l|$ and $\episode$ than their
prograde counterparts; no finite retrograde $(1,0,0)$ orbit is found in the
scanned domain. The finite $(1,0,0)$ orbit at $\mu=0$ shows that the deformed
geometry alone can cancel the relativistic periapsis advance.

The middle and right columns probe the direct dipole coupling. At fixed
$|B|$ and $\mu$, the transformation $B\to-B$ leaves the geometry unchanged
but reverses $U$; the $B<0$ orbits require larger $l$ and $\episode$, while
their change in size depends on the orbit family. At fixed $B$, increasing
$\mu$ lowers the required $l$ and $\episode$ but changes the radial extent
only modestly. The parameters are
collected in Table~\ref{tab:closed-orbit-parameters} and in
Tables~\ref{tab:field-scan-parameters} and
\ref{tab:dipole-scan-parameters} of
Appendix~\ref{app:closed-orbit-data}.

\begin{table*}[!htbp]
    \centering
    \scriptsize
    \setlength{\tabcolsep}{10pt}
    \renewcommand{\arraystretch}{1.5}
    \begin{tabular}{ccccccc}
        \toprule
        $(z,w,v)$ & Branch & $\lambda$ & $r_{\rm p}$ & $r_{\rm a}$ & $l$ & $\episode$ \\
        \midrule
        $(1,0,0)$ & prograde   & 38.185512 & 21.214173 & 190.927559 &  6.687308 & 1.034065 \\
        $(2,0,1)$ & prograde   &  7.786266 &  4.325703 &  38.931330 &  3.393044 & 0.981276 \\
        $(2,0,1)$ & retrograde & 15.142138 &  8.412299 &  75.710692 & -4.813857 & 1.018281 \\
        $(3,1,1)$ & prograde   &  5.510800 &  3.061556 &  27.554000 &  3.134272 & 0.971317 \\
        $(3,1,1)$ & retrograde & 10.404167 &  5.780093 &  52.020834 & -4.471693 & 1.009217 \\
        \bottomrule
    \end{tabular}

    \caption{Prograde and retrograde orbital parameters for the left
    column of Fig.~\ref{closedorbits}, with $a=0.6$, $e=0.8$, $B=0.05$, and
    $\mu=0.1$.}
    \label{tab:closed-orbit-parameters}
\end{table*}

\subsection{Circular orbits}
\begin{figure*}[!htbp]
    \centering
    \begin{subfigure}[b]{0.47\linewidth}
        \centering
        \includegraphics[width=0.85\linewidth]{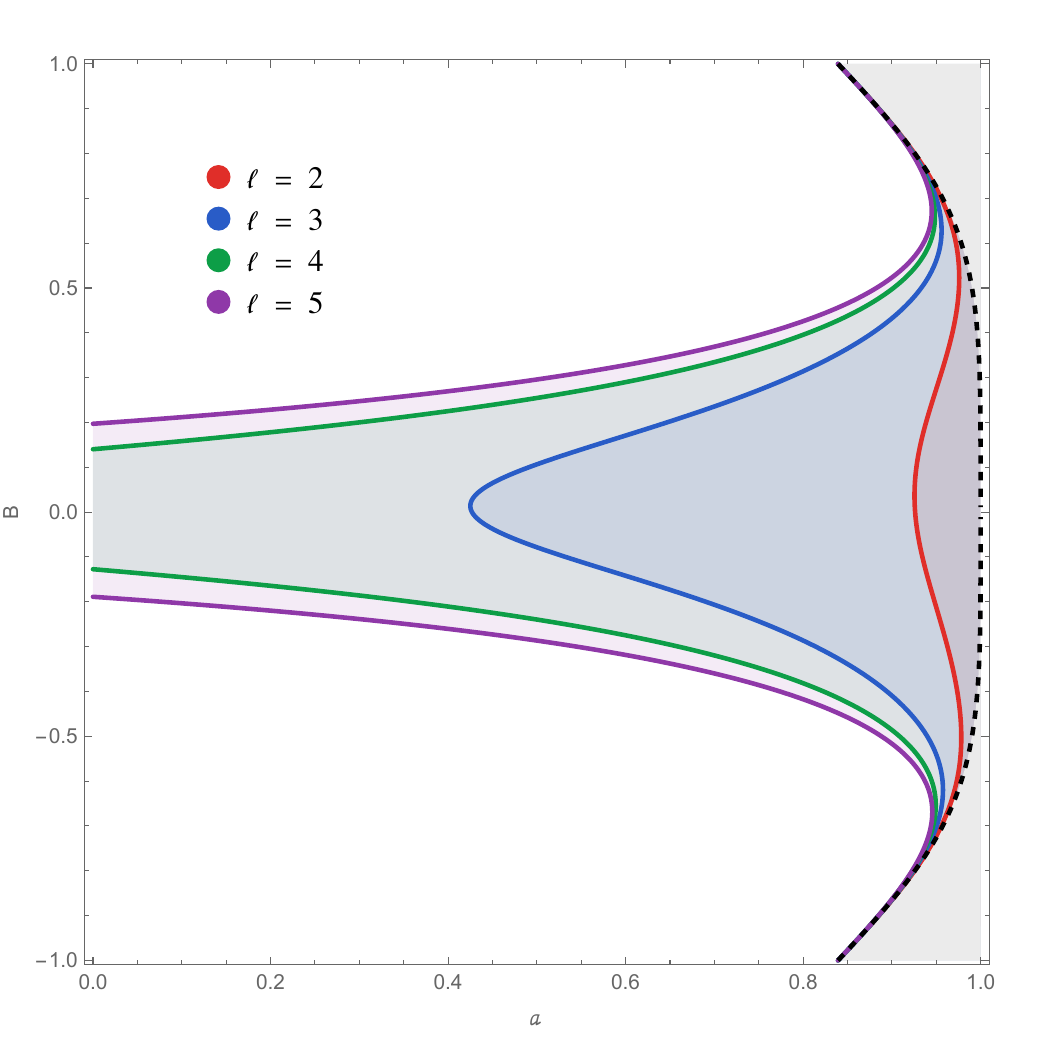}
        \caption{$\mu=0.1$ and $l=2,3,4,5$.}
        \label{fig:cocl-individual}
    \end{subfigure}
    \hspace{0.01\linewidth}
    \begin{subfigure}[b]{0.47\linewidth}
        \centering
        \includegraphics[width=0.85\linewidth]{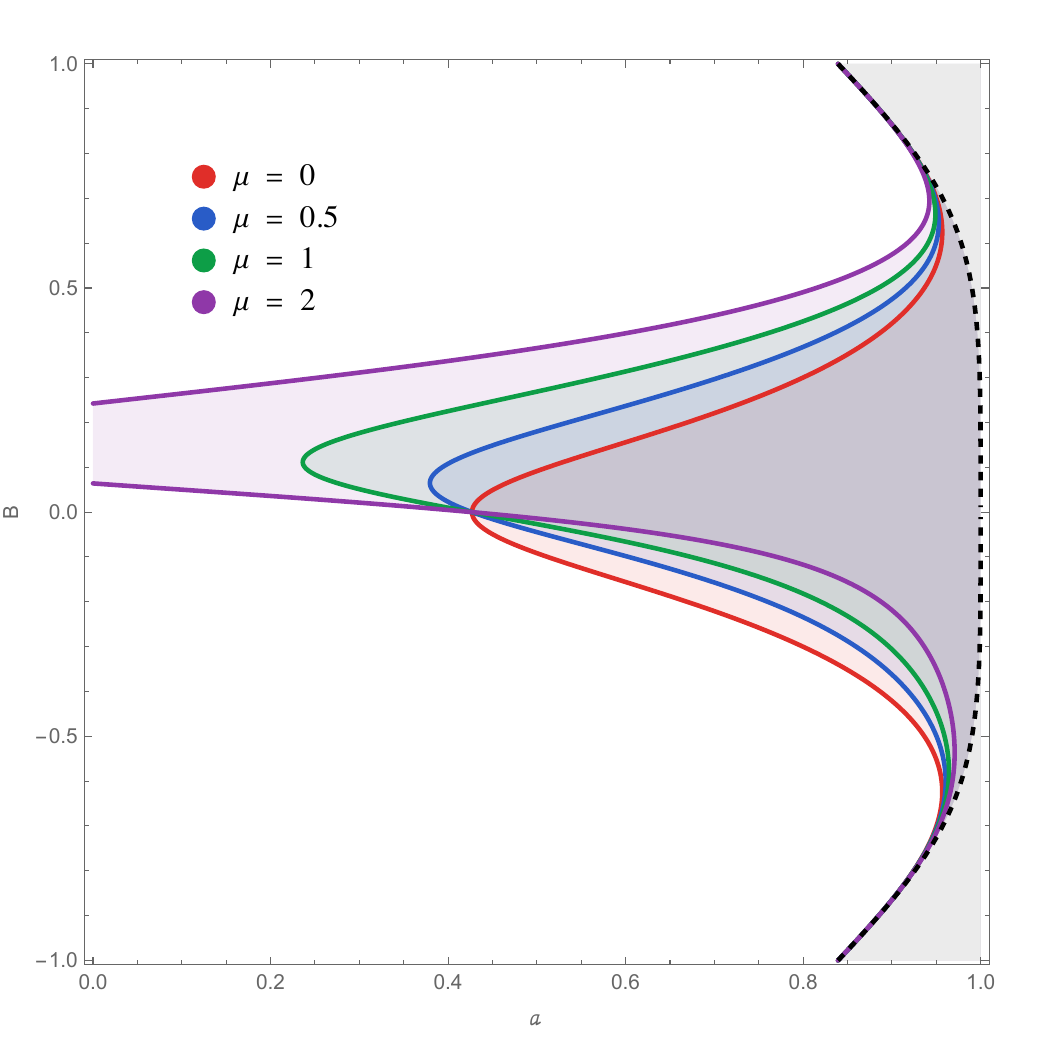}
        \caption{$l=3$ and $\mu=0,0.5,1,2$.}
        \label{fig:cocmu-individual}
    \end{subfigure}

    \caption{Numerically determined circular-orbit domains in the $(a,B)$
    parameter plane. Panel (a) shows $l=2,3,4,5$ at fixed $\mu=0.1$, while
    panel (b) shows $\mu=0,0.5,1,2$ at fixed $l=3$. The colored curves bound
    the domains satisfying Eq.~(\ref{orbits}). The black dashed curves mark
    the extremal horizon boundary $\mathcal D_{\rm H}=0$, or
    $B=\pm B_{\rm ext}(a)$; only the part with $\mathcal D_{\rm H}\geq0$
    corresponds to black-hole configurations.}
    \label{a-B}
\end{figure*}

Closed trajectories generally retain a finite radial
oscillation. Circular motion is obtained in the limiting case in which the
two turning points coalesce at a constant radius $r_0$. A circular orbit at $r=r_0$ satisfies
\begin{align}
    V_{\text{eff}}(r_0) = \episode, \quad \left. \frac{dV_{\text{eff}}}{dr} \right|_{r=r_0} = 0.
    \label{orbits}
\end{align}
Their numerical solutions define the
circular-orbit domains in Fig.~\ref{a-B}, the influences of the various parameters on the circular-orbit domains can be observed. As $l$ increases, the boundary extends toward larger $|B|$ at
low and moderate spin, while the curves approach one another in the rapidly
rotating regime. At fixed $l=3$, increasing $\mu$ enlarges the positive-$B$
domain but reduces its extent on the negative-$B$ side, consistent with the
sign dependence of $\beta=\mu B^{\hat\theta}/m$.

Fig.~\ref{a-B} identifies where circular motion is possible but
does not specify the conserved quantities at a prescribed radius. We denote the specific angular momentum and specific energy of a
circular orbit at radius $r_0$ by $l_c(r_0)$ and
$\episode_c(r_0)$, respectively.

\begin{figure*}[!htbp]
    \centering
    \begin{subfigure}[b]{0.47\linewidth}
        \centering
        \includegraphics[width=0.85\linewidth]{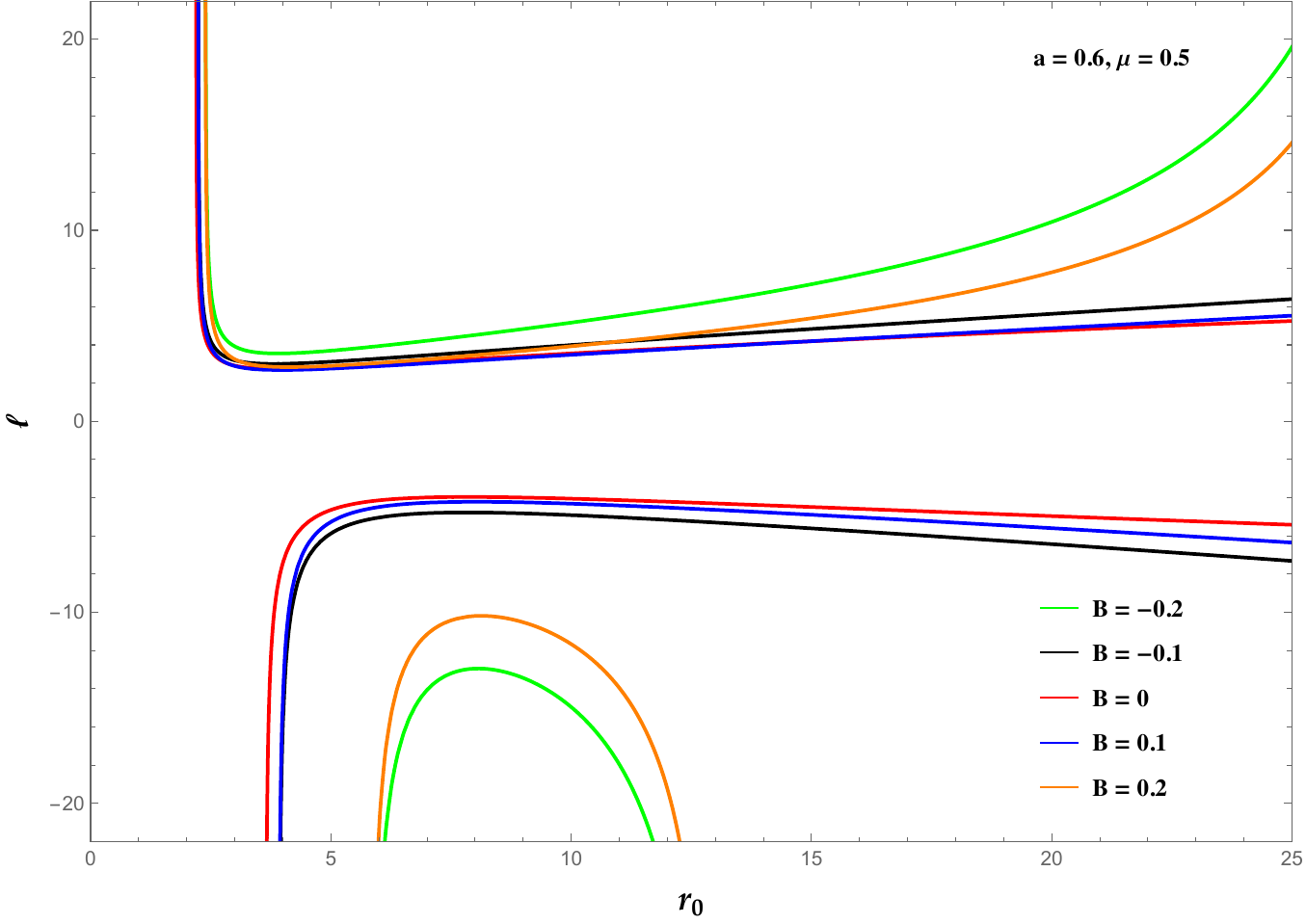}
        \caption{Varying $B$ at fixed $\mu=0.5$.}
        \label{fig:circular-l-B}
    \end{subfigure}
    \hspace{0.01\linewidth}
    \begin{subfigure}[b]{0.47\linewidth}
        \centering
        \includegraphics[width=0.85\linewidth]{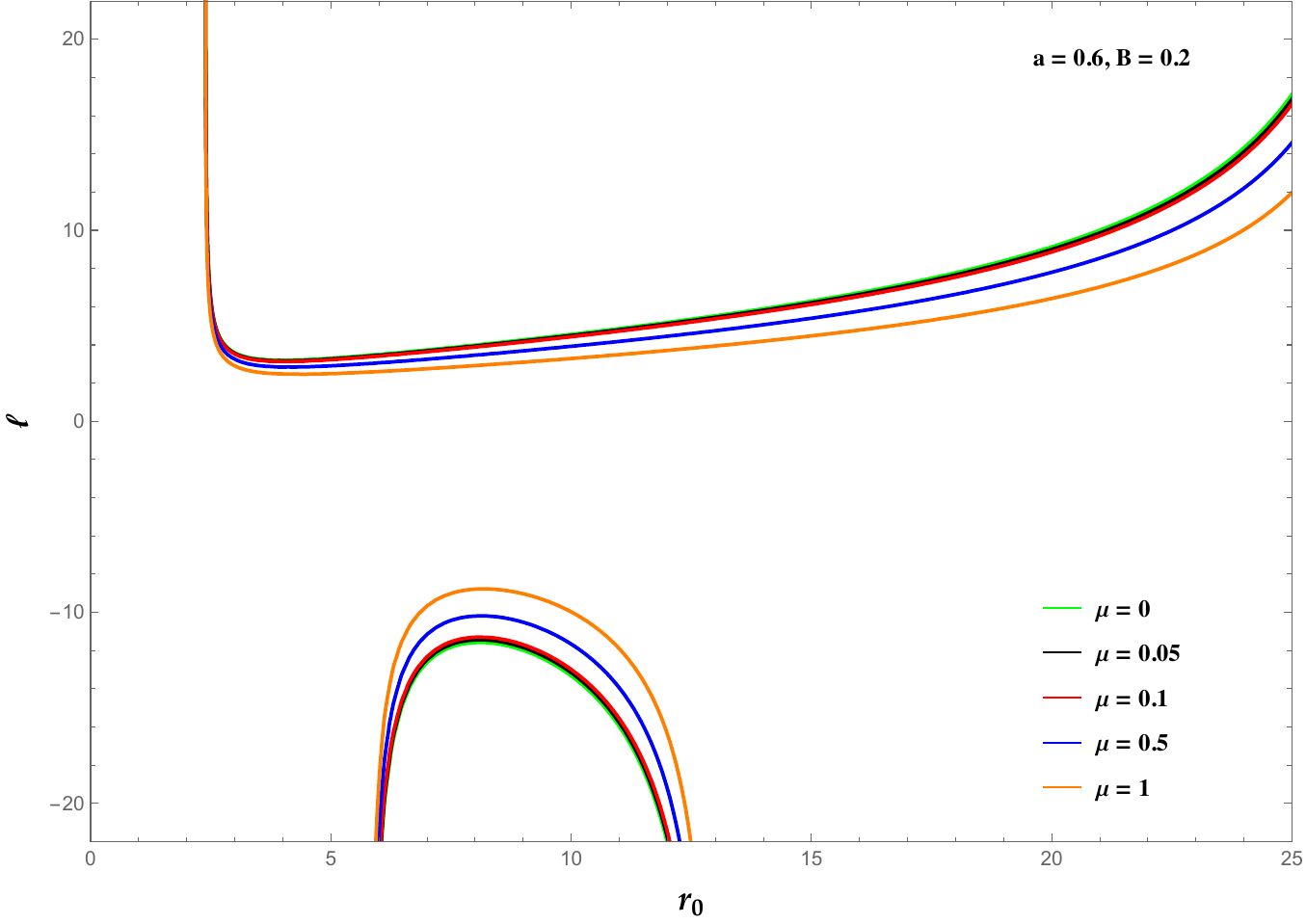}
        \caption{Varying $\mu$ at fixed $B=0.2$.}
        \label{fig:circular-l-mu}
    \end{subfigure}

    \caption{Specific angular momentum for equatorial circular orbits
    versus $r_0$ at $a=0.6$. Panels (a) and (b) show the dependence on $B$
    and $\mu$, respectively. The upper and lower branches correspond to
    $l>0$ and $l<0$.}
    \label{fig:circular-lr}
\end{figure*}

The resulting angular-momentum branches at $a=0.6$ are shown in
Fig.~\ref{fig:circular-lr}; only real solutions within the plotted range are
displayed. The $l_c>0$ and $l_c<0$ branches are asymmetric because frame
dragging introduces the mixed term $2g^{t\varphi}\episode l$, which changes
sign under $l\to-l$; their degeneracy is restored in the nonrotating limit.
In Fig.~\ref{fig:circular-l-B}, the negative-$B$ curves generally lie farther
from $l=0$ than their positive-$B$ counterparts, and the $l_c<0$ solutions
for $|B|=0.2$ exist only over a finite radial interval. This $B\to-B$
asymmetry reflects the sign-sensitive dipole interaction. Figure~\ref{fig:circular-l-mu} isolates the direct coupling
because varying $\mu$ at fixed $B=0.2$ leaves the geometry unchanged.
Increasing $\mu$ moves both branches toward $l_c=0$ while shifting their
radial extrema more modestly. Thus, the coupling mainly reduces the angular
momentum required for circular motion at a given radius.

\begin{figure*}[!htbp]
    \centering
    \begin{subfigure}[t]{0.47\linewidth}
        \vspace{0pt}
        \centering
        \includegraphics[width=0.85\linewidth]{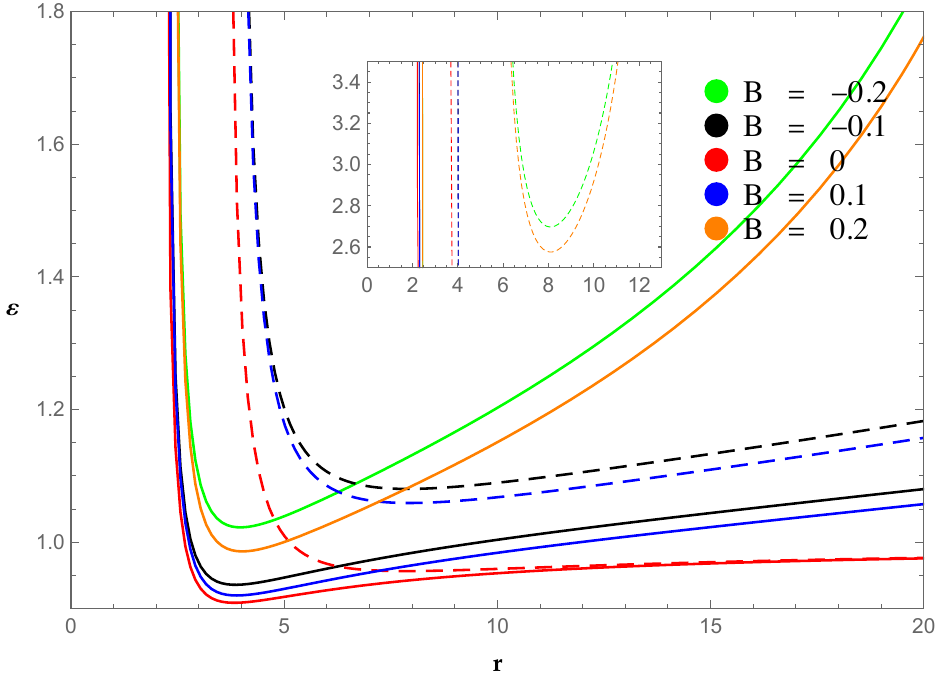}
        \caption{Varying $B$ at fixed $\mu=0.1$.}
        \label{fig:circular-energy-B}
    \end{subfigure}
    \hspace{0.01\linewidth}
    \begin{subfigure}[t]{0.47\linewidth}
        \vspace{0pt}
        \centering
        \includegraphics[width=0.85\linewidth]{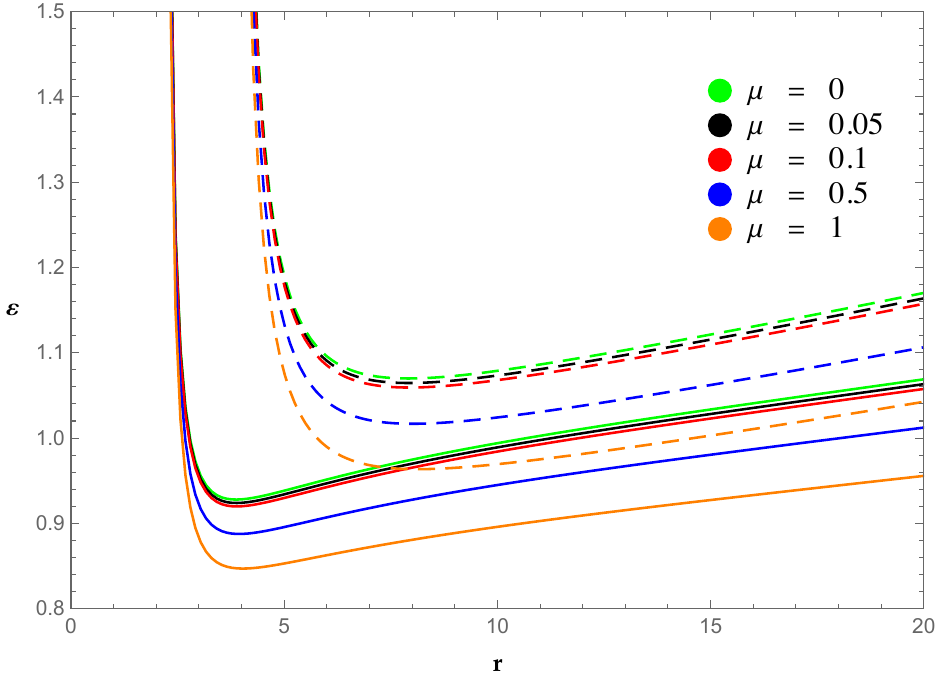}
        \caption{Varying $\mu$ at fixed $B=0.1$.}
        \label{fig:circular-energy-mu}
    \end{subfigure}

    \caption{Specific energy $\episode_c$ of equatorial circular orbits
    versus $r_0$ at $a=0.6$. Solid and dashed curves correspond to the
    $l_c>0$ and $l_c<0$ branches, respectively. Panels (a) and (b) show the
    dependence on $B$ and $\mu$, respectively.
    The inset in panel (a) resolves the high-energy minima of the
    $l_c<0$, $B=\pm0.2$ branches.}
    \label{fig:circular-energy}
\end{figure*}
Substituting each real branch $l_c(r_0)$ into $V_{\rm eff}$ gives
the energy curves $\episode_c(r_0)$ in Fig.~\ref{fig:circular-energy}. These
curves satisfy the circularity conditions only; radial stability is not yet
imposed. Frame dragging also separates the two energy branches: for the
parameters shown, the $l_c<0$ branch generally has a higher branch minimum at
a larger radius than the $l_c>0$ branch. This separation is already present at
$B=0$, showing that it is not caused by magnetic interaction; the two $B=0$
branches approach one another as $r_0$ increases. For $B\neq0$, the curves
need not approach $\episode_c=1$ because the Kerr--BR geometry is not
 asymptotically flat. For the displayed parameters,
 Fig.~\ref{fig:circular-energy-B} shows that nonzero $|B|$ raises the branch
 minima relative to Kerr, while the difference between $B$ and $-B$ reflects
 the orientation-sensitive coupling. In Fig.~\ref{fig:circular-energy-mu},
 increasing $\mu$ lowers both energy branches and their minima, whereas the
 minima shift less in radius. Together with Fig.~\ref{fig:circular-l-mu}, this
 shows that the direct coupling primarily reduces the energy and $|l_c|$
 required for circular motion.

\subsection{ISCO}

A circular orbit is
radially stable when $\partial_r^2V_{\rm eff}>0$ and
unstable when $\partial_r^2V_{\rm eff}<0$.  The innermost stable circular
orbit (ISCO) is the marginal orbit separating these two cases and satisfies
\begin{align}
    V_{\rm eff}(r_{\text{ISCO}}) = \episode, \quad
    \left. \frac{dV_{\rm eff}}{dr} \right|_{r=r_{\text{ISCO}}} = 0,
    \notag\\
    \left. \frac{d^2 V_{\rm eff}}{dr^2} \right|_{r=r_{\text{ISCO}}} = 0.
    \label{ISCO-conditions}
\end{align}
Here $V_{\rm eff}$ is differentiated only with respect to $r$.
We solve the last two equations for $(r_{\rm ISCO},l_{\rm ISCO})$, set
$\episode_{\rm ISCO}=V_{\rm eff}(r_{\rm ISCO})$.
We retain the real, future-directed solution outside $r_+$
 for which $\partial_r^2V_{\rm eff}$ changes locally from
negative on the inner side to positive on the outer side.

\begin{figure*}[!t]
    \centering
    \includegraphics[width=\linewidth]{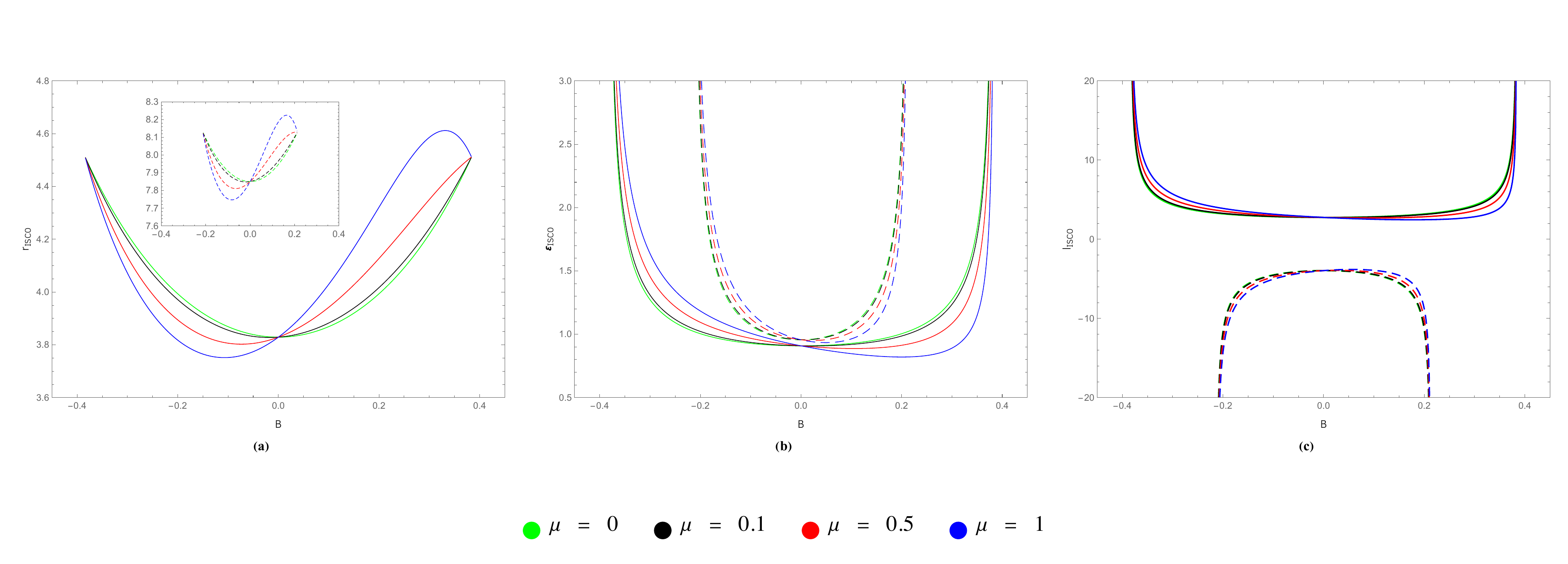}
    \caption{ISCO quantities versus $B$ for $a=0.6$ and
    $\mu=0,0.1,0.5,1$: $r_{\rm ISCO}$, $\episode_{\rm ISCO}$, and
    $l_{\rm ISCO}$ (left to right). Solid (dashed) curves denote
    $l_{\rm ISCO}>0$ ($l_{\rm ISCO}<0$); colors label $\mu$. The inset shows
    the retrograde radius.}
    \label{fig:isco}
\end{figure*}
Fig.~\ref{fig:isco} shows the dependence of the ISCO quantities on
$B$ for $a=0.6$. Frame dragging distinguishes the two angular-momentum
branches even at $B=0$: the retrograde ISCO lies farther from the
horizon and is therefore shown in the inset of the left panel. When
$\mu=0$, the curves are invariant under $B\to-B$, consistent with the
$B^2$ dependence of the metric, whereas a nonzero dipole moment breaks
this symmetry through the direct dipole--field interaction. As $|B|$ increases along the displayed curves, the difference between
the two angular-momentum branches becomes more pronounced. For
$l_{\rm ISCO}>0$, the unbounded growth of $\episode_{\rm ISCO}$ and
$|l_{\rm ISCO}|$ occurs near a larger critical value of $|B|$, while
$r_{\rm ISCO}$ approaches a smaller finite value. For
$l_{\rm ISCO}<0$, the corresponding growth occurs near a smaller
critical value of $|B|$, and $r_{\rm ISCO}$ approaches a larger finite
value. Thus, on both branches, the specific energy and the magnitude of
the specific angular momentum diverge while the ISCO radius remains
finite.

To understand this behavior, we divide Eq.~(\ref{Veff}) by $|l|$ and
take $l\to\pm\infty$, obtaining
\begin{align}
    \lim_{l\to\pm\infty}\frac{V_{\rm eff}}{|l|}
    ={}&\pm\frac{g^{t\varphi}}{g^{tt}}
    +\frac{\sqrt{(g^{t\varphi})^2-g^{tt}g^{\varphi\varphi}}}
    {-g^{tt}} .
    \label{ISCO-large-l-limit}
\end{align}
The upper and lower signs correspond to $l>0$ and $l<0$, respectively.
For finite $1+\beta$, the effective-mass term is suppressed by
$(1+\beta)^2/l^2$ after the overall $|l|$ scaling is extracted and
therefore drops out at leading order. The resulting radial dynamics
reduces to the corresponding null-particle form. Consequently, the
timelike ISCO approaches a circular null orbit, and its orbital speed
measured by a ZAMO tends to the speed of light. For a circular orbit,
$\episode_{\rm ISCO}=V_{\rm eff}(r_{\rm ISCO})$, so the same limiting
expression determines the ratio
$\episode_{\rm ISCO}/|l_{\rm ISCO}|$. This ratio approaches a finite,
nonzero value at the limiting radius. Hence
$\episode_{\rm ISCO}\propto|l_{\rm ISCO}|$, and the divergence of
$|l_{\rm ISCO}|$ is necessarily accompanied by a linear divergence of
$\episode_{\rm ISCO}$. The formal endpoints follow from the large-$|l|$ limits of the
circularity and marginal-stability conditions. Equivalently, the first
and second radial derivatives of the right-hand side of
Eq.~(\ref{ISCO-large-l-limit}) must vanish. For the
$l_{\rm ISCO}>0$ branch, these conditions give
\begin{align}
    |B_{\rm crit}^{(+)}|=0.383920,
    \qquad r_{\rm crit}^{(+)}=4.51149,
    \label{ISCO-critical-positive}
\end{align}
whereas for the $l_{\rm ISCO}<0$ branch they give
\begin{align}
    |B_{\rm crit}^{(-)}|=0.213070,
    \qquad r_{\rm crit}^{(-)}=8.12903.
    \label{ISCO-critical-negative}
\end{align}

Because the limiting expression is determined solely by the background
metric and is even in $B$, each angular-momentum branch has two
symmetry-related limiting points at $B=\pm|B_{\rm crit}|$ with the same
$r_{\rm crit}$. For each branch, the curves corresponding to the fixed
finite values of $\mu$ shown in Fig.~\ref{fig:isco} therefore
asymptotically approach the same formal endpoint in the
$(|B|,r_{\rm ISCO})$ plane. The background geometry thus determines the
common endpoint, even though the direct dipole--field interaction affects
the ISCO at finite $|l_{\rm ISCO}|$. In particular, on the $\mu=1$, $B>0$, $l_{\rm ISCO}>0$ branch,
$r_{\rm ISCO}$ lies above $r_{\rm crit}^{(+)}$, producing the overshoot
visible in Fig.~\ref{fig:isco}. As the limiting point is approached, the
dipole contribution becomes progressively weaker relative to the
angular-momentum terms, and $r_{\rm ISCO}$ approaches
$r_{\rm crit}^{(+)}$ from above.

\begin{figure*}[!t]
    \centering
    \includegraphics[width=\linewidth]{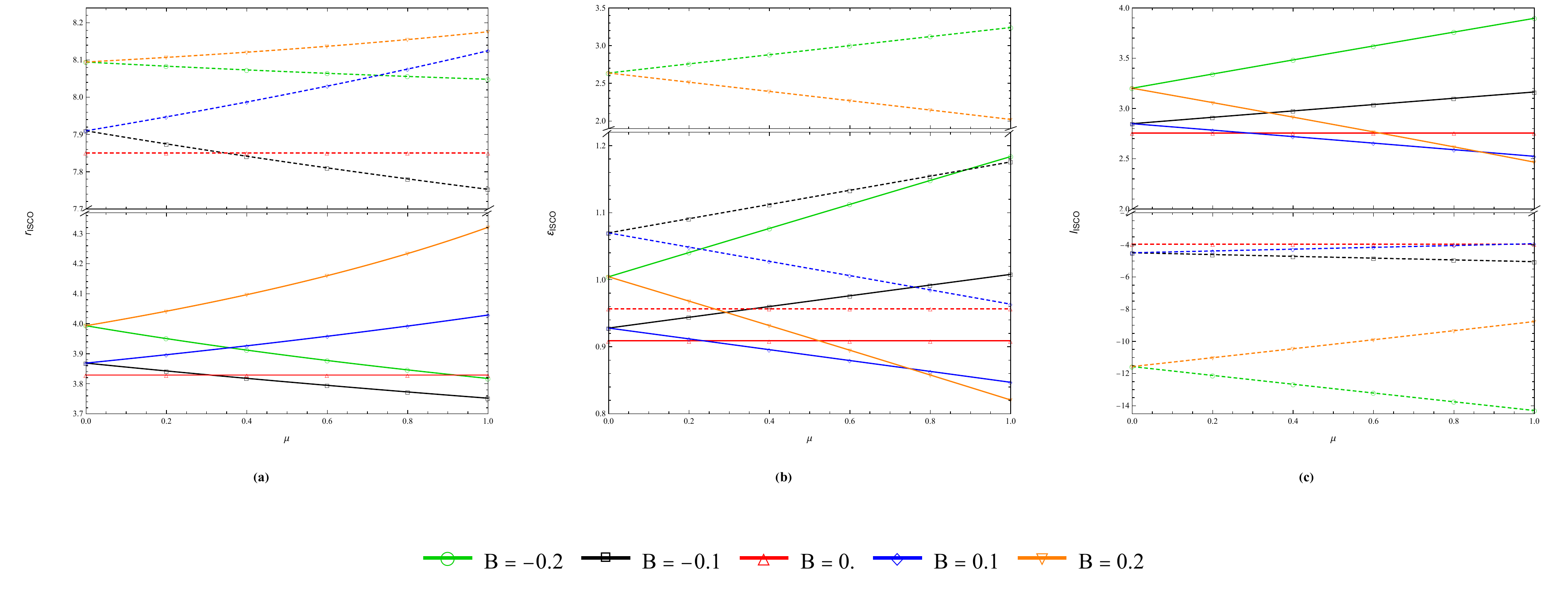}
    \caption{ISCO quantities versus $\mu$ for $a=0.6$ and
    $B=-0.2,-0.1,0,0.1,0.2$: $r_{\rm ISCO}$, $\episode_{\rm ISCO}$, and
    $l_{\rm ISCO}$ (left to right). Solid (dashed) curves denote
    $l_{\rm ISCO}>0$ ($l_{\rm ISCO}<0$); colors and markers label $B$. Broken
    vertical axes omit unused ranges, not physical discontinuities.}
    \label{fig:isco-mu}
\end{figure*}

For $0\leq\mu\leq1$, Fig.~\ref{fig:isco-mu} shows that increasing $\mu$
moves both ISCO branches outward and lowers $\episode_{\rm ISCO}$ and
$|l_{\rm ISCO}|$ for $B>0$, while the three trends reverse for $B<0$.
At $\mu=0$, the curves for $B=\pm|B|$ coincide pairwise because the
dipole coupling vanishes and the metric is even in $B$. Since varying
$\mu$ at fixed $B$ leaves the background geometry unchanged, their
separation for $\mu>0$ arises from the sign-sensitive dipole coupling.
At $B=0$, the coupling vanishes, so all three ISCO quantities are
independent of $\mu$.

\section{BSW mechanism in Kerr--Bertotti--Robinson spacetime}
\label{sec:BSW}

\subsection{Local collision kinematics and effective
center-of-mass energy}

We now consider the Ba\~nados--Silk--West (BSW) mechanism
\cite{Banados:2009pr,Harada:2014vka} for two
electrically neutral magnetized particles, labelled by
$i=1,2$, moving on the equatorial plane. We use the same reduced scalar
interaction as in Sec.~II,
$\beta_i=\mu_i B^{\hat\theta}/m_i$. The canonical momenta of the particles from Eq.~(\ref{Lagrangian}) are
\begin{align}
    p_i^\mu =m_i(1+\beta_i)\dot x_i^\mu .
    \label{BSW-conjugate-momentum}
\end{align}
Throughout this section, we assume $1+\beta_i>0$, corresponding
to the positive-effective-mass branch. Equivalently, the interaction energy
must satisfy $U_i>-m_i$.

For a stationary and axisymmetric metric, the equatorial line element can be
written in the ZAMO form
\begin{align}
    ds^2=-N^2dt^2+g_{\varphi\varphi}
    \left(d\varphi-\omega dt\right)^2+g_{rr}dr^2 ,
    \label{ADMform}
\end{align}
where $N$ is the lapse function and $\omega=-g_{t\varphi}/g_{\varphi\varphi}$
is the frame-dragging angular velocity. In the Kerr--BR spacetime these
quantities were given in Eq.~(\ref{ZAMO}). Since the event horizon is located
at $Q=0$, the horizon angular velocity is
\begin{align}
    \Omega_{\rm H}\equiv \omega(r_+)
    =\frac{a}{r_+^2+a^2}.
    \label{OmegaH}
\end{align}

We retain the original regular azimuthal coordinate $\varphi$ and define, for particles $i=1,2$, the specific conserved quantities by
$\episode_i\equiv-p_{it}/m_i$ and $l_i\equiv p_{i\varphi}/m_i$. If instead
the standard-period angle $\Phi=\varphi/\mathcal C$ is used, then
$l_i\to\mathcal C l_i$, $\omega\to\omega/\mathcal C$, and
$g_{\varphi\varphi}\to\mathcal C^2g_{\varphi\varphi}$. The combinations
entering the radial motion and collision energy are therefore unchanged, so
the following analysis is independent of this azimuthal normalization.

For each particle we define the standard near-horizon combination
\begin{align}
    X_i\equiv \episode_i-\omega l_i .
    \label{XiDef}
\end{align}
At any exterior point, contracting the conjugate momentum in
Eq.~(\ref{BSW-conjugate-momentum}) with the ZAMO four-velocity gives the
locally measured energy, $-p_{i\mu}u^\mu=m_iX_i/N$. Solving the definitions
of $\episode_i$ and $l_i$ then gives the temporal and azimuthal components
of the four-velocity:
\begin{align}
    \dot t_i&=\frac{X_i}{(1+\beta_i)N^2},
    \label{tdotBSW}\\
    \dot\varphi_i&=\frac{\omega X_i}{(1+\beta_i)N^2}
    +\frac{l_i}{(1+\beta_i)g_{\varphi\varphi}} .
    \label{phidotBSW}
\end{align}

Equation~(\ref{tdotBSW}) makes the future-directed condition
explicit. Outside the horizon $N^2>0$, and on the
positive-effective-mass branch $1+\beta_i>0$; hence $\dot t_i>0$ is
equivalent to $X_i(r)>0$ along the trajectory. Taking the continuous exterior
limit $r\to r_+$, where $\omega\to\Omega_{\rm H}$, gives
\begin{align}
    X_i(r_+)=\episode_i-\Omega_{\rm H}l_i\ge0 .
\end{align}
A particle with $X_i(r_+)>0$ is usual, whereas a critical
particle saturates this bound. For $a\ne0$, we take particle 1 to be critical,
which fixes its specific angular momentum as
\begin{align}
    X_1(r_+)=0,
    \qquad
    l_1=l_1^*\equiv\frac{\episode_1}{\Omega_{\rm H}}
    =\frac{\episode_1(r_+^2+a^2)}{a} .
    \label{criticalL}
\end{align}
Here $l_1^*$ denotes the critical specific angular momentum determined by
$X_1(r_+)=0$.
For $a=0$, $\Omega_{\rm H}=0$ and $X_1(r_+)=\episode_1$, so no
finite $l_1$ can satisfy the rotational criticality condition for
$\episode_1>0$. Substituting these velocity components into the normalization
condition suggests introducing the radial function
\begin{align}
    Z_i^2\equiv
    X_i^2
    -N^2\left[(1+\beta_i)^2
    +\frac{l_i^2}{g_{\varphi\varphi}}\right],
    \label{ZiDef}
\end{align}
The radial velocity is then
\begin{align}
    \dot r_i=\sigma_i
    \frac{Z_i}{(1+\beta_i)N\sqrt{g_{rr}}},
    \label{radialBSW1}
\end{align}
where $\sigma_i=+1$ and $-1$ denote outgoing and ingoing
motion, respectively, and $Z_i$ is the nonnegative square root. At any
exterior radius, local radial motion is possible only if
\begin{align}
Z_i^2(r)\ge0 .
\label{local-radial-admissibility}
\end{align}
A zero of $Z_i^2$ marks a radial turning point. Equation~(\ref{local-radial-admissibility}) is a
local condition. Connection to a specified outer region further requires
$Z_i^2(r)\ge0$, $X_i>0$, and $1+\beta_i>0$ to hold continuously along the
entire trajectory.

We now let the two particles collide at the same equatorial
event $r=r_*$. their local
effective center-of-mass energy is defined from the total conjugate momentum
by
\begin{align}
    \left(E_{\rm cm}^{\rm eff}\right)^2
    =-\left(p_1^\mu+p_2^\mu\right)
    \left(p_{1\mu}+p_{2\mu}\right).
\end{align}
Using Eqs.~(\ref{tdotBSW}), (\ref{phidotBSW}), and (\ref{ZiDef}) gives
\begin{widetext}
\begin{align}
    \left(E_{\rm cm}^{\rm eff}\right)^2
    ={}&m_1^2(1+\beta_1)^2+m_2^2(1+\beta_2)^2
    +2m_1m_2
    \left[
    \frac{X_1X_2-\sigma_1\sigma_2Z_1Z_2}{N^2}
    -\frac{l_1l_2}{g_{\varphi\varphi}}
    \right]\notag\\
    ={}&m_1^2(1+\beta_1)^2+m_2^2(1+\beta_2)^2
    +2m_1m_2(1+\beta_1)(1+\beta_2)\gamma_{12}.
    \label{EcmBSW}
\end{align}
\end{widetext}
Here,
$\gamma_{12}\equiv-g_{\mu\nu}\dot x_1^\mu\dot x_2^\nu$ is the
relative Lorentz factor of the two particles. We further define the local
ZAMO Lorentz factor of particle $i$ by
\begin{align}
\gamma_i^{\rm Z}
&\equiv-u_\mu\dot x_i^\mu
=N\dot t_i
=\frac{X_i}{(1+\beta_i)N}.
\label{local-ZAMO-Lorentz-factor}
\end{align}

We focus on the standard BSW channel in which both particles
are ingoing, $\sigma_1=\sigma_2=-1$. Here and below,
$\Theta(N^p)$ denotes a quantity whose leading near-horizon order is $N^p$
as $N\to0$. A usual particle has $X_i(r_+)>0$, and hence
$\gamma_i^{\rm Z}=\Theta(N^{-1})$; it becomes ultrarelativistic relative to
the ZAMO. For two usual ingoing particles, both $X_i$ approach finite positive
horizon values, so they acquire the same leading inward boost. Indeed,
\begin{align}
    Z_i={}&X_i-\frac{N^2}{2X_i}
    \left[(1+\beta_i)^2
    +\frac{l_i^2}{g_{\varphi\varphi}}\right]
    +\mathcal{O}(N^4),
\end{align}
substitution into Eq.~(\ref{EcmBSW}) shows that the leading
$X_1X_2$ contribution cancels against that from $Z_1Z_2$. The remaining
numerator is $\Theta(N^2)$, and division by $N^2$ therefore gives
$\gamma_{12}=\Theta(1)$. Thus the divergence of the individual
$\gamma_i^{\rm Z}$ does not imply a divergent collision energy: both
particles approach the speed of light relative to the ZAMO in the same
inward direction, while their relative Lorentz factor remains finite. This cancellation relies on both particles being usual, with finite positive
$X_i(r_+)$. If particle 1 is critical, $X_1(r_+)=0$, while particle 2 remains
usual, $X_2(r_+)>0$, the usual-particle expansion above, which assumes
$N^2/X_i^2\to0$, is no longer valid for particle 1. Its ZAMO boost has a
different near-horizon scaling from that of particle 2, so the common-boost
cancellation is no longer guaranteed and $\gamma_{12}$ may therefore diverge. Criticality alone,
however, does not produce a collision: the relation $X_1(r_+)=0$ fixes
$l_1$, but particle 1 must also satisfy
Eq.~(\ref{local-radial-admissibility}) on a future-directed ingoing branch. For completeness, if $\sigma_1\sigma_2=-1$, two usual particles instead have
$\gamma_{12}=\Theta(N^{-2})$ and
\begin{align}
    \left(E_{\rm cm}^{\rm eff}\right)^2
    =\Theta(N^{-2}).
\end{align}

This head-on divergence is not the
standard ingoing BSW process and requires a separate physical mechanism for
producing an outgoing particle near the future horizon.  We do not pursue
that channel here.

\subsection{Critical collisions around extremal Kerr--BR black holes}

We first test whether a critical trajectory remains radially
admissible as the collision approaches the horizon. On the extremal branch
$\mathcal D_{\rm H}=M^2I_2-a^2I_1^2=0$, the rotation and field parameters are
linked, and the metric functions satisfy
\begin{align}
    r_+=\frac{M}{I_1},
    \qquad
    a^2=I_2r_+^2,
    \qquad
    \Delta=I_2(r-r_+)^2 .
    \label{extremal-KBR-identities}
\end{align}
On the equatorial plane,
$N^2=\Delta r^2/R$ and
$g_{\varphi\varphi}=R/[(1+B^2r^2)r^2]$.  For a critical particle with
$l_1=l_1^*=\episode_1(r_+^2+a^2)/a$, the near-horizon expansions are
\begin{align}
    X_1={}&
    \frac{2\episode_1r_+}{r_+^2+a^2}(r-r_+)
    +\mathcal{O}\left((r-r_+)^2\right),\notag\\
    \beta_1={}&\mathcal{O}(r-r_+),\notag\\
    Z_1^2={}&
    \frac{(r-r_+)^2}
    {I_2r_+^2(1+I_2)^2}
    \left[(4I_2-1)\episode_1^2-I_2^2\right]
    +\mathcal{O}\left((r-r_+)^3\right).
    \label{extremalReach}
\end{align}
Consequently, the leading local reachability condition is
\begin{align}
    (4I_2-1)\episode_1^2\ge I_2^2 .
    \label{extremal-reachability-condition}
\end{align}

The sign of the leading coefficient in $Z_1^2$, rather than
criticality alone, decides whether
a locally allowed critical branch exists sufficiently close to
the horizon. A strict inequality in
Eq.~(\ref{extremal-reachability-condition}) gives a locally allowed
near-horizon branch; at equality the next radial order must be examined.
Future-directedness additionally requires $\episode_1>0$. Hence a
finite-energy critical branch has the local necessary condition $I_2>1/4$,
equivalent on the extremal branch to $|a|/M>4/5$, while the Kerr limit
$I_2=1$ recovers $\episode_1^2\ge1/3$.

Let $r_*>r_+$ denote the exterior collision radius. Along
the ingoing critical trajectory, the extremal BSW limit is obtained by taking
$r_*\to r_+$. At every finite collision radius in the interior of this
near-horizon allowed branch, particle 1 remains ingoing,
$\dot r_1(r_*)<0$; along this branch, $\dot r_1\to0$ as
$r_*\to r_+$.

The Lorentz factors now identify the physical source of the
divergence. On the strictly allowed critical branch,
$X_1=\Theta(N)$, $Z_1=\Theta(N)$, and $\beta_1=\mathcal{O}(N)$, so
$\gamma_1^{\rm Z}=\Theta(1)$. The critical particle therefore has a finite
ZAMO-measured speed. By contrast, the usual particle obeys
$X_2\to X_{2{\rm H}}\equiv X_2(r_+)>0$ and hence
$\gamma_2^{\rm Z}=\Theta(N^{-1})$; its ZAMO-measured speed approaches the
speed of light.

The different individual boosts suggest an unbounded relative
motion, but its precise order follows by combining the radial function in
Eq.~(\ref{ZiDef}) with the collision invariant in
Eq.~(\ref{EcmBSW}). Since
$(1+\beta_1)^2+l_1^2/g_{\varphi\varphi}$ has a finite positive horizon limit
and $X_1+Z_1=\Theta(N)$, Eq.~(\ref{ZiDef}) gives
\begin{align}
X_1-Z_1&=\Theta(N),\qquad X_2-Z_2=\Theta(N^2),\notag\\
X_1X_2-Z_1Z_2
&=X_2(X_1-Z_1)+Z_1(X_2-Z_2)
=\Theta(N).
\label{critical-usual-radial-scaling}
\end{align}
For fixed finite $\mu_i/m_i$,
Eqs.~(\ref{Btheta-equatorial-compact}) and
(\ref{extremal-KBR-identities}) give
$B^{\hat\theta}=-BI_2N$ and
$\beta_i=-(\mu_i/m_i)BI_2N$, so
$1+\beta_i\to1$. Equation~(\ref{EcmBSW}) then gives
$\gamma_{12}=\Theta(N^{-1})$. Consequently,
\begin{align}
E_{\rm cm}^{\rm eff}
=\Theta(N^{-1/2})
\propto(r_*-r_+)^{-1/2}.
\label{BSW-divergence-rate}
\end{align}

\begin{figure*}[!t]
    \centering
    \includegraphics[width=\linewidth]{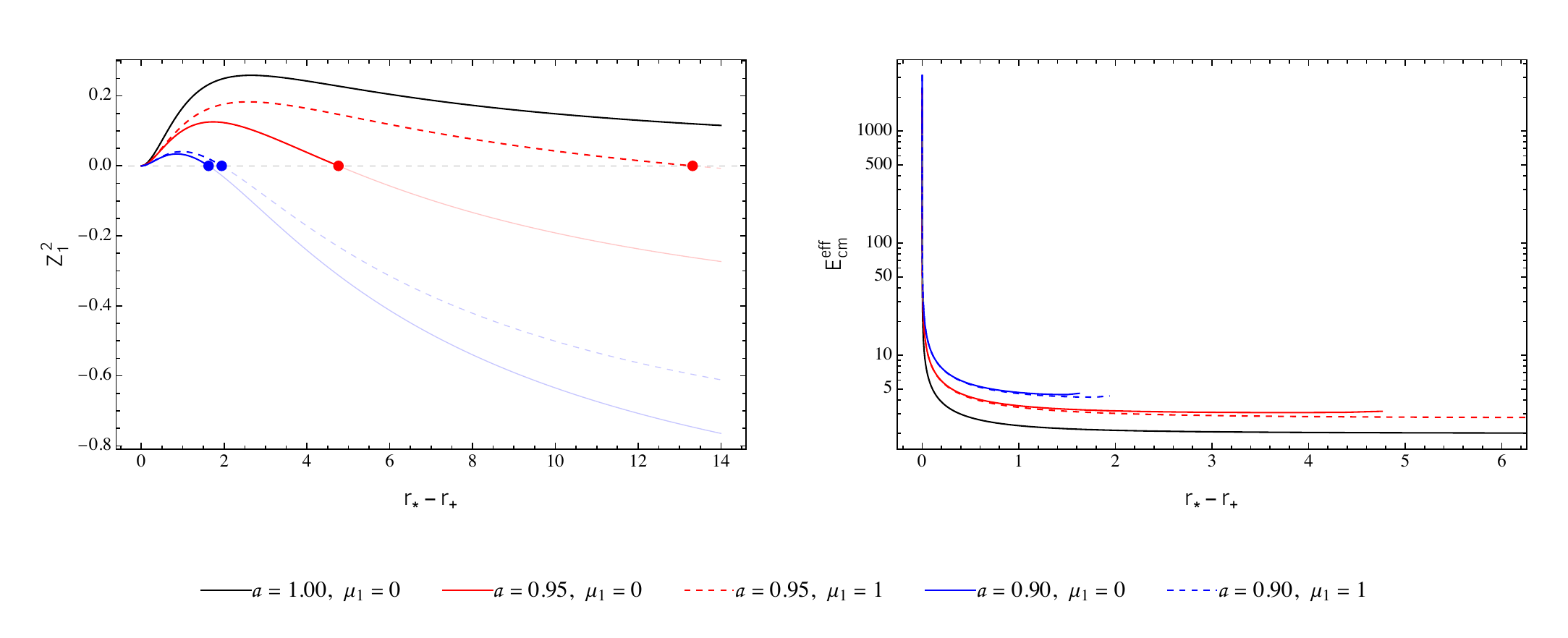}
    \caption{Radial function $Z_1^2(r_*)$ (left) and effective
    center-of-mass energy $E_{\rm cm}^{\rm eff}(r_*)$ (right) for an exactly
    critical particle colliding with a usual ingoing particle near extremal
    Kerr--BR horizons. We set $M=m_1=m_2=1$,
    $\episode_1=\episode_2=1$, and $l_2=\mu_2=0$. Each curve lies on the
    positive extremal branch $\mathcal D_{\rm H}=0$, with
    $B=B_{\rm ext}(a)$ and
    $l_1=l_1^*=\episode_1/\Omega_{\rm H}$. Dark and faint
    segments denote $Z_1^2\ge0$ and $Z_1^2<0$, respectively, and filled
    markers denote finite-radius roots of $Z_1^2$. The right panel uses a
    logarithmic vertical scale and follows
    $E_{\rm cm}^{\rm eff}\propto(r_*-r_+)^{-1/2}$ near the horizon.}
    \label{fig:bsw-extremal}
\end{figure*}

As shown in Fig.~\ref{fig:bsw-extremal}, the double-zero extremal horizon gives
$N\propto r_*-r_+$, every finite $r_*>r_+$ yields a finite collision energy;
the unbounded value refers only to the limiting family $r_*\to r_+$. The
divergence in Eq.~(\ref{BSW-divergence-rate}) is therefore produced by the
unbounded relative Lorentz factor between the critical and usual particles. The background field $B$ remains in the leading coefficients through $I_2$,
$r_+$, $\Omega_{\rm H}$, and $l_1^*$. It can change the local
radial-admissibility condition, the prefactor of $\gamma_{12}$ and
$E_{\rm cm}^{\rm eff}$, and even whether the critical branch exists. It does
not change the exponent because every regular extremal configuration
satisfying the strict radial condition retains the double-zero structure
$\Delta\propto(r-r_+)^2$ and hence $N\propto r-r_+$. Since extremality fixes
$B=B_{\rm ext}(a)$, the differently colored curves in
Fig.~\ref{fig:bsw-extremal} compare distinct extremal geometries. The left panel displays the radial effect. Each
dark segment with $Z_1^2\ge0$ is the radially allowed part of the exactly
critical branch connected to the horizon, and its marked finite-radius zero
is the outer boundary of that segment rather than the location of the
divergence. Within each fixed-geometry pair, the solid $\mu_1=0$ and dashed
$\mu_1=1$ curves isolate the direct dipole interaction. For the displayed
$B>0$ configurations, $B^{\hat\theta}=-BI_2N<0$, so positive $\mu_1$ gives
$\beta_1<0$ and reduces the $(1+\beta_1)^2$ contribution to the radial
barrier in Eq.~(\ref{ZiDef}). This moves the outer turning point outward and
extends the horizon-connected allowed segment, as shown by the dashed curves. The right panel samples the same allowed branches at progressively smaller
lapse. Since $\beta_i=\mathcal{O}(N)$, the dressed factors approach unity and
$E_{\rm cm}^{\rm eff}\propto\gamma_{12}^{1/2}$. A fixed finite $\mu_1$
changes the finite-radius profile but not the leading near-horizon exponent,
whereas different colors carry the geometry-dependent prefactors. The
limiting $(a,B,\mu_1)=(1,0,0)$ curve provides the standard Kerr BSW check. At
finite radius, the vertical separation of the $E_{\rm cm}^{\rm eff}$ curves
contains both the relative Lorentz factor $\gamma_{12}$ and the dressed
factors $1+\beta_i$ and should not be attributed to either contribution
alone. Finally, the marked root establishes only a local horizon-connected
segment; reachability from infinity requires the complete radial path to
remain allowed.

\subsection{Critical-particle collisions around nonextremal Kerr--BR black holes}

\begin{figure*}[!t]
    \centering
    \includegraphics[width=\linewidth]{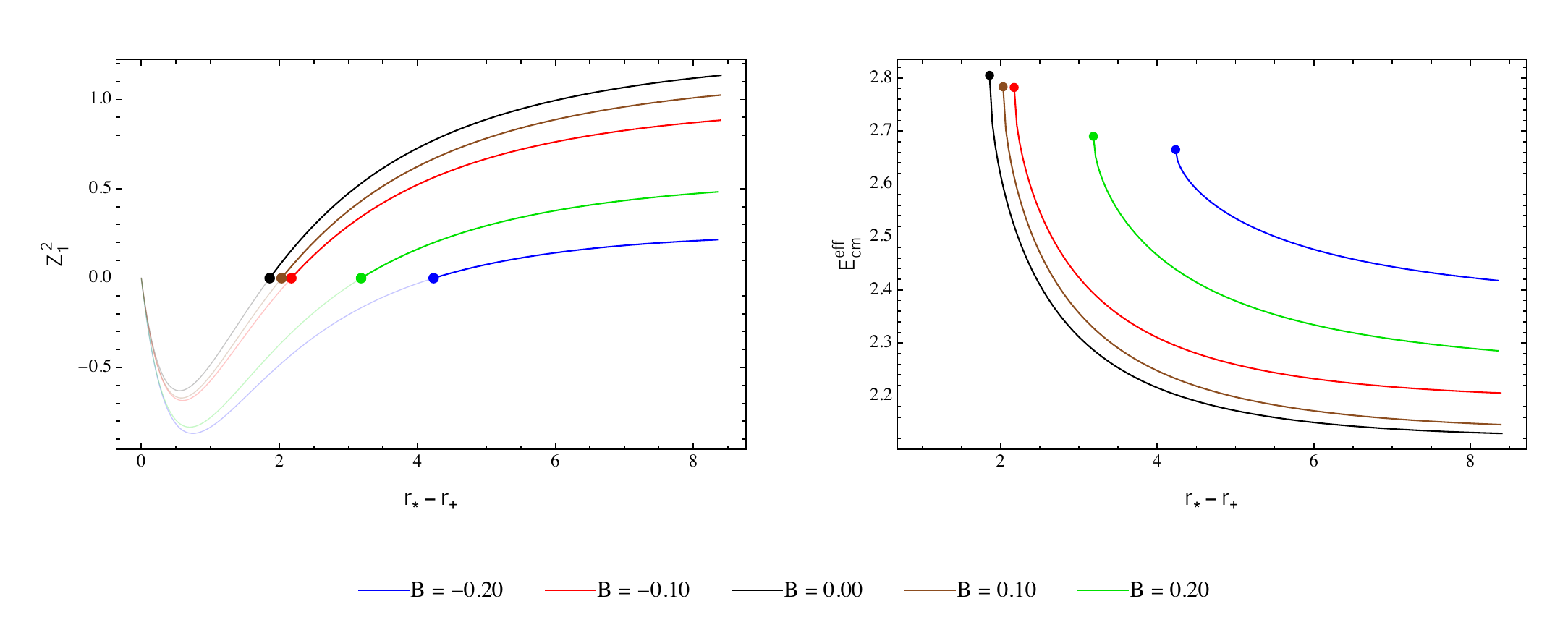}
    \caption{$Z_1^2(r_*)$ (left) and
    $E_{\rm cm}^{\rm eff}(r_*)$ (right), illustrating the magnetic-field
    dependence for a nonextremal black hole. We set $M=m_1=m_2=1$, $a=0.8$,
    $\episode_1=1.5$,
    $\episode_2=1$, $\mu_1=0.5$, and $l_2=\mu_2=0$. For each $B$, the
    critical angular momentum
    $l_1=l_1^*=\episode_1/\Omega_{\rm H}$ and the collision point $r_*$ are
    recalculated. The collision point satisfies $\dot r_1(r_*)=0$; the
    markers in the right panel give the corresponding finite energies.
    Dark and faint segments denote $Z_1^2\ge0$ and $Z_1^2<0$, respectively.}
    \label{fig:bsw-nonext-B}
\end{figure*}

\begin{figure*}[!t]
    \centering
    \includegraphics[width=\linewidth]{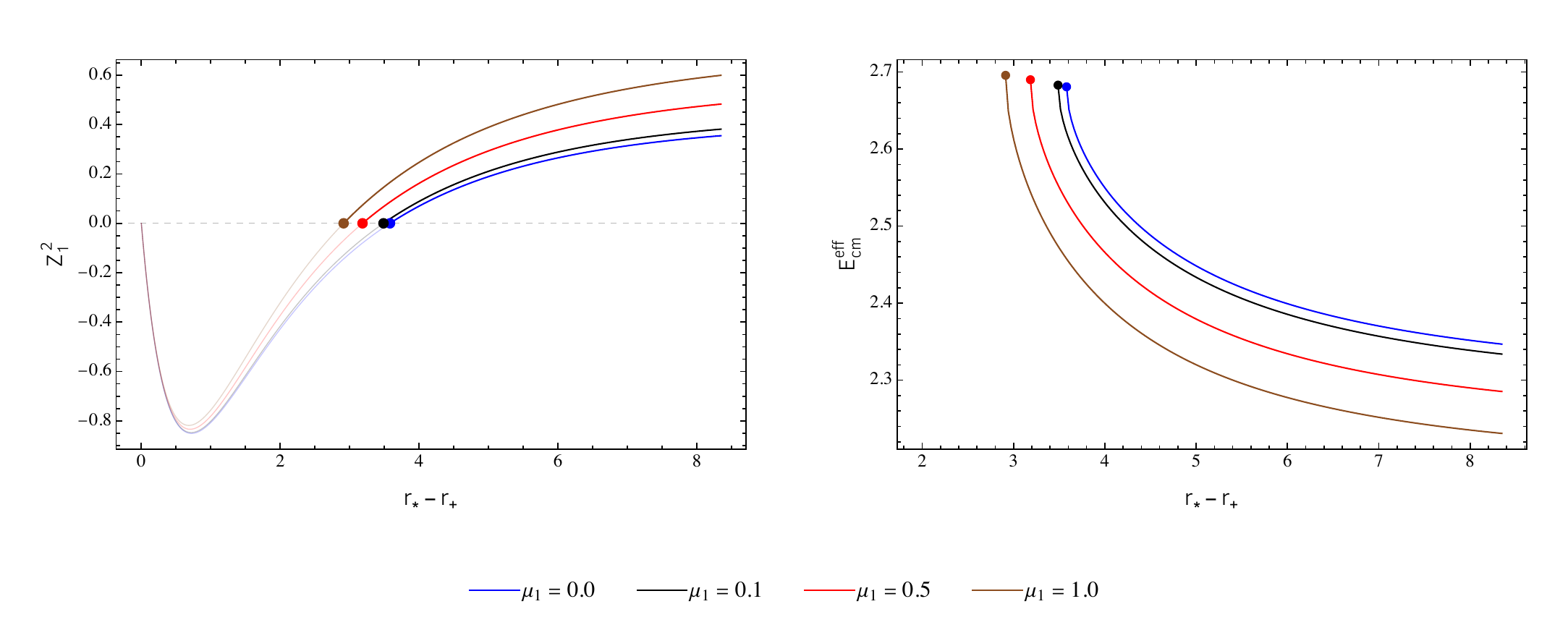}
    \caption{$Z_1^2(r_*)$ (left) and
    $E_{\rm cm}^{\rm eff}(r_*)$ (right), illustrating the magnetic-moment
    dependence for a nonextremal black hole. We set $M=m_1=m_2=1$, $a=0.8$,
    $B=0.2$,
    $\episode_1=1.5$, $\episode_2=1$, and $l_2=\mu_2=0$. Particle 1 is
    exactly critical and approaches the collision point along the ingoing
    branch, with $\dot r_1(r_*)=0$.}
    \label{fig:bsw-nonext-mu}
\end{figure*}

The extremal divergence relies on the double zero of $N^2$
(equivalently, of $\Delta$). At a nonextremal horizon, the different
near-horizon orders reverse the balance in $Z_1^2$ and prevent an exactly
critical timelike particle from entering the arbitrarily small-lapse region.
Specifically, $N^2=\Theta(r-r_+)$, whereas an exactly critical particle has
$X_1=\mathcal{O}(r-r_+)$. Since $\beta_1\to0$ and
$g_{\varphi\varphi}$ has a finite positive horizon limit,
\begin{align}
    Z_1^2={}&X_1^2-N^2
    \left[(1+\beta_1)^2
    +\frac{l_1^2}{g_{\varphi\varphi}}\right]<0,
    \qquad r\to r_+ .
    \label{nonextremalReach}
\end{align}
The same obstruction is visible directly in the ZAMO Lorentz
factor. Here $N=\Theta((r-r_+)^{1/2})$, while
$X_1=\mathcal{O}(r-r_+)=\mathcal{O}(N^2)$, so the formal critical scaling would give
$\gamma_1^{\rm Z}=\mathcal{O}(N)\to0$. This is impossible for a future-directed
timelike particle, for which $\gamma_1^{\rm Z}\ge1$. Equation
(\ref{nonextremalReach}) is the radial manifestation of this kinematic
contradiction: the negative lapse term dominates $X_1^2$ and creates a
forbidden near-horizon region. Thus an exactly critical trajectory is excluded from a neighborhood outside a
nonextremal horizon. In the left panels of
Figs.~\ref{fig:bsw-nonext-B} and \ref{fig:bsw-nonext-mu}, we select this root as the collision point
shown in the right panels, so that
\begin{align}
    r_*>r_+,\qquad
    Z_1^2(r_*)=0,\qquad
    \dot r_1(r_*)=0 .
    \label{nonextremal-collision-point}
\end{align}

If an exterior allowed branch exists, it is cut off at the
finite turning point $r_*>r_+$, where $N(r_*)>0$. For the parameters used in Figs.~\ref{fig:bsw-nonext-B} and
\ref{fig:bsw-nonext-mu}, $l_2=\beta_2=0$, and the relative factor at the
turning point becomes
\begin{align}
\gamma_{12}(r_*)
=\frac{X_2}{N(r_*)}
\left[
1+\frac{(l_1^*)^2}
{(1+\beta_1)^2g_{\varphi\varphi}}
\right]_{r_*}^{1/2}.
\label{turning-point-relative-Lorentz}
\end{align}
On the positive-effective-mass branch, $N(r_*)>0$ and all
remaining factors in Eq.~(\ref{turning-point-relative-Lorentz}) are finite.
Thus $\gamma_{12}(r_*)$ is finite, and Eq.~(\ref{EcmBSW}) gives a finite
$E_{\rm cm}^{\rm eff}(r_*)$.
The contrast with the extremal case is
qualitative: an allowed extremal critical branch can sample arbitrarily
small lapse and develop an unbounded $\gamma_{12}$, whereas the exactly
critical nonextremal branch is intercepted at finite lapse by the radial
barrier,the divergence at the horizon has been ruled out analytically.

Fig.~\ref{fig:bsw-nonext-B} shows how $B$ changes the
finite-lapse cutoff. In the left panel, every curve is negative sufficiently
close to the horizon, as required by Eq.~(\ref{nonextremalReach}). Each
marker is the inner boundary of the exterior ingoing allowed branch: the
exactly critical particle reaches that turning point but cannot continue
into the smaller-lapse region. Relative to $B=0$, both signs of the displayed
field move this boundary outward. Changing $|B|$ modifies the horizon
geometry and $l_1^*$, whereas reversing $B$ changes the orientation-odd part
of the direct dipole--field interaction. At equal $|B|$, the two signs have
the same $r_+$ and $l_1^*$ because the geometry is even in $B$. Their
different turning points therefore expose the sign-sensitive interaction:
for the retained $\mu_1>0$, negative $B$ gives $\beta_1>0$, raises the
$(1+\beta_1)^2$ radial barrier, and intercepts the critical branch farther
from the horizon than positive $B$. The right panel separates fixed-position and endpoint effects. At equal
$|B|$ and the same plotted $r-r_+$, the negative-$B$ curve lies above the
positive-$B$ curve. This is an ordering of $E_{\rm cm}^{\rm eff}$, not of
$\gamma_{12}$ alone, because both the relative factor and the dressed masses
enter the ordinate. At the marked endpoints the ordering is reversed:
negative $B$ places the collision farther out, at a larger attainable lapse,
and gives a slightly smaller endpoint energy than positive $B$. The $B=0$
branch reaches the smallest lapse and has the largest endpoint energy.
These trends show that the endpoint energy is controlled mainly by how far
the allowed branch can penetrate toward small $N$. They cannot be attributed
to that factor alone, however, because $\Omega_{\rm H}$, $l_1^*$, the local
metric functions, and $\beta_1$ vary simultaneously.

Fig.~\ref{fig:bsw-nonext-mu} isolates the direct
dipole--field interaction because the geometry, $r_+$, $\Omega_{\rm H}$, and
$l_1^*$ remain fixed. In the left panel, $B^{\hat\theta}<0$ for $B=0.2$, so
increasing positive $\mu_1$ makes $\beta_1$ more negative, lowers the
$(1+\beta_1)^2$ radial barrier, and increases $Z_1^2$ at a fixed radius. The
inner boundary of the exterior allowed branch consequently moves toward the
horizon. The critical particle can then reach a smaller lapse, although the
nonextremal forbidden neighborhood never disappears. The right panel displays two competing consequences. Within the overlap of
the physical domains, the larger values of $\mu_1$ correspond to lower values of
$E_{\rm cm}^{\rm eff}$. At a common
radius, the direct coupling changes both the dressed mass
$m_1(1+\beta_1)$ and the relative factor $\gamma_{12}$, so this vertical
ordering should not be assigned to either contribution separately. The endpoint markers instead compare collisions occurring at different
radii. At their respective endpoints, increasing $\mu_1$ both moves the
collision to a smaller $N(r_*)$ and changes the finite dipole-dependent factor
in Eq.~(\ref{turning-point-relative-Lorentz}). Together these effects increase
the attainable relative boost and slightly outweigh the accompanying
reduction of the dressed-mass
contribution, producing only a modest net increase in the endpoint
$E_{\rm cm}^{\rm eff}$.

\section{Summary and Discussion}

We studied the equatorial motion and near-horizon collisions of
electrically neutral particles with intrinsic magnetic dipole moments in the
exact Kerr--BR spacetime. In the $\xi=0$ sector, the dipoles were fixed normal
to the equatorial plane in the ZAMO frame, and the scalar interaction energy
$U=\mu B^{\hat\theta}$ was adopted following
Ref.~\cite{Preti:2004Polarized}. We restricted the motion to the
future-directed branch $1+\beta>0$. The field then enters in two distinct
ways: $B$ deforms the exact background geometry, whereas
$\beta=\mu B^{\hat\theta}/m$ represents the sign-sensitive direct coupling.

The radial effective potential connects these effects to the orbital
structure. Within the parameter ranges considered, increasing $|B|$ shifts
the turning points and circular-orbit domains and can eliminate a finite
potential well together with its bound-orbit family. At $\mu=0$, the dynamics
is invariant under $B\to-B$ because the geometry is even in $B$; a nonzero
dipole moment breaks this symmetry. Closed orbits were classified using the
regularized azimuthal angle $\Phi=\varphi/\mathcal C$. For $a=0.6$ and $e=0.8$,
the backreacted geometry can offset the Kerr periapsis advance and support a
finite $(1,0,0)$ orbit even at $\mu=0$. This demonstrates a qualitative
geometric effect that cannot be attributed to direct dipole coupling.

The ISCO analysis shows the same separation. Finite $\mu$ shifts the ISCO
radius, energy, and angular momentum in an orientation-dependent manner, but
the formal large-$|l_{\rm ISCO}|$ endpoints are fixed by the background
geometry. For $a=0.6$, the $l_{\rm ISCO}>0$ and $l_{\rm ISCO}<0$ branches
approach $(|B|,r_{\rm ISCO})\simeq(0.384,4.51)$ and $(0.213,8.13)$,
respectively. At either endpoint,
$\episode_{\rm ISCO}/|l_{\rm ISCO}|$ approaches a finite nonzero limit, so
$\episode_{\rm ISCO}\propto|l_{\rm ISCO}|$, while both quantities diverge and
$r_{\rm ISCO}$ remains finite. The leading radial dynamics becomes null, and
the timelike ISCO approaches the corresponding circular null orbit.

For collisions, $E_{\rm cm}^{\rm eff}$ is the local invariant constructed
from the conjugate momenta of the scalar-coupling model. On a locally
admissible extremal branch, the critical particle has
$\gamma_1^{\rm Z}=\Theta(1)$, whereas the usual ingoing particle has
$\gamma_2^{\rm Z}=\Theta(N^{-1})$. Their relative Lorentz factor therefore
scales as $\gamma_{12}=\Theta(N^{-1})$. Because
$B^{\hat\theta}\to0$ and $1+\beta_i\to1$ for fixed finite $\mu_i/m_i$,
\begin{align}
E_{\rm cm}^{\rm eff}
=\Theta(N^{-1/2})
\propto(r_*-r_+)^{-1/2}.
\end{align}
The divergence is thus driven by the relative boost between the critical and
usual particles. Finite dipole coupling shifts finite-radius barriers but
does not change the leading reachability condition or divergence exponent.
For a nonextremal horizon, by contrast, exact criticality would imply
$\gamma_1^{\rm Z}\to0$, which is incompatible with future-directed timelike
motion. The critical particle is therefore excluded from the near-horizon
region, and any exterior allowed branch ends at a finite turning point where
$E_{\rm cm}^{\rm eff}$ remains finite. This statement applies to exact
criticality; near-critical trajectories require a separate analysis.

All physical trajectories must satisfy $Z_i^2\ge0$, $X_i>0$, and
$1+\beta_i>0$ along a connected radial path. For the nonzero-$B$ examples
studied here, the exactly critical particle cannot reach the extremal horizon
by direct infall from $r\to\infty$, although this obstruction is
parameter dependent. The divergent extremal result is therefore local and
does not determine the energy carried by collision products to an outer
observation region.

Several extensions follow naturally. A systematic study of near-critical
trajectories and their possible production mechanisms is needed to determine
whether the locally admissible extremal branch can be populated dynamically.
The kinematics and escape conditions of collision products in the collisional
Penrose process must then be included to connect the local effective
center-of-mass energy to the energy carried into an outer observation region~
\cite{Piran:1977CollisionalPenrose,Zaslavskii:2012BSWPenrose,
Bejger:2012CollisionalPenrose,Schnittman:2014RevisedLimit}. Extending the model
to generic dipole orientations and nonequatorial motion, together with fully
three-dimensional spin--dipole dynamics, would test the robustness of the
equatorial results. Self-force, collision backreaction, and finite-size effects
should also be included to assess the limits of the test-particle
approximation. Finally, comparison with localized or spatially varying
magnetic-field configurations would help distinguish generic magnetic effects
from features specific to the Kerr--BR background.

\onecolumngrid
\appendix
\section{Electromagnetic-field components in the ZAMO frame}
\label{app:magnetic-field}

For $\xi=0$, the real potential is stationary and axisymmetric
and has only the components $A_t$ and $A_\varphi$. Using
Eqs.~(\ref{electromagnetic}), (\ref{ZAMO}), and
(\ref{polar-tetrad-vector}), the potentially nonzero ZAMO-frame field
components are
\begin{align}
    E^{\hat r}
    &=
    -\frac{\Omega^2\sqrt{R}}{\rho^2\sqrt{P}}
    \left(
    \partial_r A_t+\omega\,\partial_r A_\varphi
    \right),\notag\\
    E^{\hat\theta}
    &=
    \frac{\Omega^2\sqrt{R}}{\rho^2\sqrt{Q}}
    \left(
    \partial_\theta A_t+\omega\,\partial_\theta A_\varphi
    \right),\notag\\
    B^{\hat r}
    &=
    \Omega^2\sqrt{\frac{P}{R}}\,
    \csc\theta\,\partial_\theta A_\varphi,\notag\\
    B^{\hat\theta}
    &=
    \Omega^2\sqrt{\frac{Q}{R}}\,
    \csc\theta\,\partial_r A_\varphi.
    \label{app:all-field-components}
\end{align}
The azimuthal components vanish,
$E^{\hat\varphi}=B^{\hat\varphi}=0$. The choice $\xi=0$ fixes the duality
sector but does not, for $M\neq0$, imply a purely magnetic field in the
ZAMO frame. For completeness, direct substitution of $A_t$ and
$A_\varphi$ gives

\begin{align}
    E^{\hat\theta}
    ={}&
    \frac{aB\Omega\sin\theta\cos\theta}
    {2\rho^4\sqrt{QR}}
    \Bigg\{
    R
    \Bigg[
    -2\left(2\Delta-r\partial_r\Delta\right)
    \notag\\
    &\quad
    +\left[
    2r^2+\left(2\Delta-r\partial_r\Delta\right)\cos^2\theta
    \right]
    \left(
    \frac{2a^2}{\rho^2}
    -\frac{B^2\Delta}{\Omega^2}
    \right)
    \Bigg]
    \notag\\
    &\quad
    +\left[P(r^2+a^2)-Q\right]
    \Bigg[
    2\rho^2\Delta
    -2r(r^2+a^2)\partial_r\Delta
    -4a^2\Delta\cos(2\theta)
    \notag\\
    &\qquad
    +\Bigg\{
    r(r^2+a^2)
    \left[
    2r-(\partial_r\Delta)\cos^2\theta
    \right]
    +2a^2\Delta\cos^2\theta\sin^2\theta
    \Bigg\}
    \notag\\
    &\qquad\qquad\times
    \left(
    -\frac{2a^2}{\rho^2}
    +\frac{B^2\Delta}{\Omega^2}
    \right)
    \Bigg]
    \Bigg\},
    \label{app:Etheta-full}
\end{align}

\begin{align}
    E^{\hat r}
    ={}&
    \frac{aB}{4\rho^4\Omega\sqrt{PR}}
    \Bigg\{
    -(r^2+a^2)P\csc\theta
    \Bigg[
    -B^2\rho^2\Delta\cos\theta\sin(2\theta)
    \left[2r-(\partial_r\Delta)\cos^2\theta\right]
    \notag\\
    &\quad
    +2\Omega^2
    \Bigg(
    -2r\Delta\cos\theta\sin(2\theta)
    +(a^2\cos^2\theta-r^2)\sin\theta
    \left[2r-(\partial_r\Delta)\cos^2\theta\right]
    \notag\\
    &\qquad\qquad
    +\rho^2(\partial_r\Delta)\cos\theta\sin(2\theta)
    \Bigg)
    \Bigg]
    \notag\\
    &\quad
    +r\rho^2Q
    \Bigg\{
    B^2\left[2r-(\partial_r\Delta)\cos^2\theta\right]^2
    -4\Omega^2
    \left(
    \sin^2\theta
    +\frac{B^2M^2I_2}{I_1^2}\cos^2\theta
    \right)
    \Bigg\}
    \Bigg\},
    \label{app:Er-full}
\end{align}

\begin{align}
    B^{\hat r}
    ={}&
    \frac{B\sqrt{P}\cos\theta}
    {4\rho^4\Omega\sqrt{R}}
    \Bigg\{
    B^2\rho^2\Delta
    \Bigg[
    a^2\Delta\sin^2(2\theta)
    +2r(r^2+a^2)
    \left[2r-(\partial_r\Delta)\cos^2\theta\right]
    \Bigg]
    \notag\\
    &\quad
    +\Omega^2
    \Bigg[
    \left(
    4r^2(r^2+a^2)-a^4\sin^2(2\theta)
    \right)\Delta
    -4a^2\rho^2\Delta\cos(2\theta)
    \notag\\
    &\qquad\qquad
    -4r(r^2+a^2)
    \Bigg(
    a^2\left[2r-(\partial_r\Delta)\cos^2\theta\right]
    +\rho^2(\partial_r\Delta)
    \Bigg)
    \Bigg]
    \Bigg\}.
    \label{app:Br-full}
\end{align}

\begin{align}
    B^{\hat{\theta}}
    ={}&
    \frac{B\csc\theta}{8\rho^4\Omega}
    \sqrt{\frac{Q}{R}}
    \Bigg\{
    B^2\rho^2
    \left[2r-(\partial_r\Delta)\cos^2\theta\right]\notag\\
    &\quad\times
    \Bigg[
    2r(a^2+r^2)
    \left[2r-(\partial_r\Delta)\cos^2\theta\right]
    +a^2\Delta\sin^2(2\theta)
    \Bigg]\notag\\
    &\quad
    -4\Omega^2\Bigg\{
    a^2\left[2r-(\partial_r\Delta)\cos^2\theta\right]
    (a^2\cos^2\theta-r^2)\sin^2\theta
    -a^2r\Delta\sin^2(2\theta)\notag\\
    &\qquad\qquad
    +\rho^2\Bigg[
    2r(a^2+r^2)
    \left(
    \sin^2\theta
    +\frac{B^2M^2I_2}{I_1^2}\cos^2\theta
    \right)
    +\frac{a^2}{2}(\partial_r\Delta)\sin^2(2\theta)
    \Bigg]
    \Bigg\}
    \Bigg\}.
    \label{app:Btheta-full}
\end{align}

On the equatorial plane, reflection symmetry gives
$E^{\hat\theta}=B^{\hat r}=0$, while the azimuthal components already
vanish identically. The remaining components reduce to
\begin{align}
\left.E^{\hat r}\right|_{\theta=\pi/2}
&=
\frac{aB\sqrt{1+B^2r^2}\,
\left(r^2+a^2-\Delta\right)}
{r\sqrt{R_{\rm eq}}},
\label{app:Er-equator}\\
\left.B^{\hat\theta}\right|_{\theta=\pi/2}
&=
-B I_2 r\sqrt{\frac{\Delta}{R_{\rm eq}}},
\label{app:Btheta-equator}
\end{align}
where
$R_{\rm eq}=(r^2+a^2)^2-a^2(1+B^2r^2)\Delta$.
These expressions hold for arbitrary $M$. In particular,
$E^{\hat r}$ need not vanish for $M\neq0$, and the second relation
reproduces Eq.~(\ref{Btheta-equatorial-compact}).

For the Bertotti--Robinson limit $M=0$, the metric functions become
$\Delta=r^2+a^2$, $P=1-a^2B^2\cos^2\theta$, and
$\Omega^2=1+B^2r^2-B^2(r^2+a^2)\cos^2\theta$. Moreover,
$A_t=aB/\Omega$, $A_\varphi=I_2/(B\Omega)-1/B$, and
$\omega=-aB^2/I_2$. It follows that
$\partial_iA_t+\omega\partial_iA_\varphi=0$ for $i=r,\theta$, and hence
\begin{align}
E^{\hat r}=E^{\hat\theta}=E^{\hat\varphi}=0.
\label{app:electric-M0}
\end{align}
For $I_2>0$, the magnetic components are
\begin{align}
B^{\hat r}
={}&-B\cos\theta
\left[
\frac{
I_2(r^2+a^2)(1-a^2B^2\cos^2\theta)
}{
(r^2+a^2\cos^2\theta)
\left[1+B^2r^2-B^2(r^2+a^2)\cos^2\theta\right]
}
\right]^{1/2},
\notag\\
B^{\hat\theta}
={}&-Br\sin\theta
\left[
\frac{
I_2(1+B^2r^2)
}{
(r^2+a^2\cos^2\theta)
\left[1+B^2r^2-B^2(r^2+a^2)\cos^2\theta\right]
}
\right]^{1/2},
\notag\\
B^{\hat\varphi}
={}&0.
\label{app:magnetic-M0}
\end{align}
They satisfy
$(B^{\hat r})^2+(B^{\hat\theta})^2=B^2I_2$, so the magnetic-field
magnitude is uniform. At $\theta=\pi/2$,
$B^{\hat r}=B^{\hat\varphi}=0$ and
$B^{\hat\theta}=-B\sqrt{I_2}$.

\section{Closed-orbit parameters for the field and dipole scans}
\label{app:closed-orbit-data}

The tables in this appendix are organized by the varied parameter.  Table~
\ref{tab:field-scan-parameters} gives the prograde $B$ scan at $\mu=1$, whereas
Table~\ref{tab:dipole-scan-parameters} gives the prograde $\mu$ scan at
$B=0.15$.  Together with the orientation data in Table~
\ref{tab:closed-orbit-parameters}, they provide the parameters of every curve
in Fig.~\ref{closedorbits}.

\begingroup
\captionsetup{hypcap=false}
\noindent
\begin{minipage}[t]{0.485\textwidth}
    \vspace{0pt}
    \centering
    \scriptsize
    \setlength{\tabcolsep}{3pt}
    \renewcommand{\arraystretch}{1.05}
    \resizebox{\linewidth}{!}{%
    \begin{tabular}{ccccccc}
        \toprule
        $(z,w,v)$ & $B$ & $\lambda$ & $r_{\rm p}$ & $r_{\rm a}$ & $l$ & $\episode$ \\
        \midrule
        $(1,0,0)$ & $-0.15$ & 42.997299 & 23.887389 & 214.986497 & 24.489244 & 3.494314 \\
        $(1,0,0)$ & $-0.05$ & 38.186991 & 21.214995 & 190.934954 &  7.228981 & 1.093392 \\
        $(1,0,0)$ & $\phantom{-}0.05$ & 38.174136 & 21.207853 & 190.870681 &  6.239599 & 0.985507 \\
        $(1,0,0)$ & $\phantom{-}0.15$ & 42.114431 & 23.396906 & 210.572156 & 14.269731 & 2.096776 \\
        \addlinespace[2pt]
        $(2,0,1)$ & $-0.15$ & 7.105353 & 3.947419 & 35.526767 & 4.362745 & 1.221481 \\
        $(2,0,1)$ & $-0.05$ & 7.648705 & 4.249281 & 38.243526 & 3.622550 & 1.034049 \\
        $(2,0,1)$ & $\phantom{-}0.05$ & 7.917813 & 4.398785 & 39.589066 & 3.203227 & 0.938060 \\
        $(2,0,1)$ & $\phantom{-}0.15$ & 7.700617 & 4.278121 & 38.503085 & 3.000460 & 0.900381 \\
        \addlinespace[2pt]
        $(3,1,1)$ & $-0.15$ & 5.352156 & 2.973420 & 26.760781 & 4.052661 & 1.193832 \\
        $(3,1,1)$ & $-0.05$ & 5.425389 & 3.014105 & 27.126945 & 3.343699 & 1.022914 \\
        $(3,1,1)$ & $\phantom{-}0.05$ & 5.593245 & 3.107359 & 27.966227 & 2.961134 & 0.929063 \\
        $(3,1,1)$ & $\phantom{-}0.15$ & 5.787698 & 3.215388 & 28.938488 & 2.804539 & 0.884831 \\
        \bottomrule
    \end{tabular}%
    }
    \captionof{table}{Prograde $B$-scan parameters with $a=0.6$, $e=0.8$, and
    $\mu=1$.}
    \label{tab:field-scan-parameters}
\end{minipage}\hfill%
\begin{minipage}[t]{0.485\textwidth}
    \vspace{0pt}
    \centering
    \scriptsize
    \setlength{\tabcolsep}{3pt}
    \renewcommand{\arraystretch}{1.05}
    \resizebox{\linewidth}{!}{%
    \begin{tabular}{ccccccc}
        \toprule
        $(z,w,v)$ & $\mu$ & $\lambda$ & $r_{\rm p}$ & $r_{\rm a}$ & $l$ & $\episode$ \\
        \midrule
        $(1,0,0)$ & $0$   & 42.757666 & 23.754259 & 213.788331 & 19.592780 & 2.820724 \\
        $(1,0,0)$ & $0.5$ & 42.521474 & 23.623041 & 212.607371 & 17.003113 & 2.467009 \\
        $(1,0,0)$ & $1$   & 42.114431 & 23.396906 & 210.572156 & 14.269731 & 2.096776 \\
        \addlinespace[2pt]
        $(2,0,1)$ & $0$   & 7.336419 & 4.075788 & 36.682096 & 3.692699 & 1.061503 \\
        $(2,0,1)$ & $0.5$ & 7.494822 & 4.163790 & 37.474112 & 3.350212 & 0.981119 \\
        $(2,0,1)$ & $1$   & 7.700617 & 4.278121 & 38.503085 & 3.000460 & 0.900381 \\
        \addlinespace[2pt]
        $(3,1,1)$ & $0$   & 5.517490 & 3.065272 & 27.587450 & 3.438336 & 1.039826 \\
        $(3,1,1)$ & $0.5$ & 5.633770 & 3.129872 & 28.168852 & 3.124648 & 0.962484 \\
        $(3,1,1)$ & $1$   & 5.787698 & 3.215388 & 28.938488 & 2.804539 & 0.884831 \\
        \bottomrule
    \end{tabular}%
    }
    \captionof{table}{Prograde $\mu$-scan parameters with $a=0.6$, $e=0.8$, and
    $B=0.15$.}
    \label{tab:dipole-scan-parameters}
\end{minipage}
\par
\endgroup

\section{Parameter sets for the BSW analysis}
\label{app:bsw-parameters}
\begingroup
    \captionsetup{hypcap=false}
    \noindent
    \begin{minipage}[t]{0.485\textwidth}
        \vspace{0pt}
        \centering
        \scriptsize
        \setlength{\tabcolsep}{2.5pt}
        \renewcommand{\arraystretch}{1.05}

        \resizebox{\linewidth}{!}{%
        \begin{tabular}{@{}cccccc@{}}
        \toprule
        $a$ & $B$ & $r_+$ & $l_1^*$ & $r_*$
            & $E_{\rm cm}^{\rm eff}(r_*)$ \\
        \midrule

        \multicolumn{6}{@{}l}{\textit{Extremal collisions}} \\
        \cmidrule(lr){1-6}
        1.00 & 0        & 1.000000 & 2.000000
             & $r_*\to r_+$
             & \text{Divergent} \\
        0.95 & 0.726163 & 1.312250 & 2.762631
             & $r_*\to r_+$
             & \text{Divergent} \\
        0.90 & 0.865766 & 1.435890 & 3.190866
             & $r_*\to r_+$
             & \text{Divergent} \\

        \midrule
        \multicolumn{6}{@{}l}{\textit{Nonextremal: $B$ dependence}} \\
        \cmidrule(lr){1-6}
        0.80 & $-0.20$ & 1.653003 & 6.323285
             & 5.889212 & 2.664866 \\
        0.80 & $-0.10$ & 1.612910 & 6.077773
             & 3.786893 & 2.782116 \\
        0.80 & $\phantom{-}0.00$ & 1.600000 & 6.000000
             & 3.459644 & 2.804983 \\
        0.80 & $\phantom{-}0.10$ & 1.612910 & 6.077773
             & 3.645381 & 2.783299 \\
        0.80 & $\phantom{-}0.20$ & 1.653003 & 6.323285
             & 4.838373 & 2.689944 \\

        \midrule
        \multicolumn{6}{@{}l}{\textit{Nonextremal: $\mu_1$ dependence
        ($a=0.8$, $B=0.2$)}} \\
        \cmidrule(lr){1-6}
        \multicolumn{2}{c}{$\mu_1=0\phantom{.1}$} & 1.653003 & 6.323285
             & 5.233807 & 2.680763 \\
        \multicolumn{2}{c}{$\mu_1=0.1$} & 1.653003 & 6.323285
             & 5.140512 & 2.682985 \\
        \multicolumn{2}{c}{$\mu_1=0.5$} & 1.653003 & 6.323285
             & 4.838373 & 2.689944 \\
        \multicolumn{2}{c}{$\mu_1=1\phantom{.1}$} & 1.653003 & 6.323285
             & 4.566107 & 2.695685 \\
        \bottomrule
        \end{tabular}%
        }
    \end{minipage}\hfill%
    \begin{minipage}[t]{0.485\textwidth}
        \vspace{0pt}
        \captionof{table}{Parameter sets used in Fig.~\ref{fig:bsw-extremal},
    Fig.~\ref{fig:bsw-nonext-B}, and Fig.~\ref{fig:bsw-nonext-mu}. Common to all
    panels are $M=m_1=m_2=1$, $\episode_2=1$, and $l_2=\mu_2=0$; both
    particles approach the collision on ingoing branches. Particle 1 is
    exactly critical, with
    $l_1=l_1^*\equiv\episode_1/\Omega_{\rm H}$. The extremal configurations
    use $\episode_1=1$ and $B=B_{\rm ext}(a)$;
    for these cases $r_*\to r_+$ and
    $E_{\rm cm}^{\rm eff}$ diverges. The two nonextremal parameter families use
    $a=0.8$ and $\episode_1=1.5$. Their finite collision positions satisfy
    $\dot r_1(r_*)=0$, and the tabulated energies are the corresponding
    endpoint values on the ingoing branch.
    }
        \label{tab:bsw-parameters}
    \end{minipage}
\par
\endgroup

\begin{acknowledgments}
This work is supported by the National Natural Science Foundation of China (NNSFC) under Grant No 12075207.
\end{acknowledgments}

\twocolumngrid

\bibliography{ref}

\begin{thebibliography}{55}%
\makeatletter
\providecommand \@ifxundefined [1]{%
 \@ifx{#1\undefined}
}%
\providecommand \@ifnum [1]{%
 \ifnum #1\expandafter \@firstoftwo
 \else \expandafter \@secondoftwo
 \fi
}%
\providecommand \@ifx [1]{%
 \ifx #1\expandafter \@firstoftwo
 \else \expandafter \@secondoftwo
 \fi
}%
\providecommand \natexlab [1]{#1}%
\providecommand \enquote  [1]{``#1''}%
\providecommand \bibnamefont  [1]{#1}%
\providecommand \bibfnamefont [1]{#1}%
\providecommand \citenamefont [1]{#1}%
\providecommand \href@noop [0]{\@secondoftwo}%
\providecommand \href [0]{\begingroup \@sanitize@url \@href}%
\providecommand \@href[1]{\@@startlink{#1}\@@href}%
\providecommand \@@href[1]{\endgroup#1\@@endlink}%
\providecommand \@sanitize@url [0]{\catcode `\\12\catcode `\$12\catcode
  `\&12\catcode `\#12\catcode `\^12\catcode `\_12\catcode `\%12\relax}%
\providecommand \@@startlink[1]{}%
\providecommand \@@endlink[0]{}%
\providecommand \url  [0]{\begingroup\@sanitize@url \@url }%
\providecommand \@url [1]{\endgroup\@href {#1}{\urlprefix }}%
\providecommand \urlprefix  [0]{URL }%
\providecommand \Eprint [0]{\href }%
\providecommand \doibase [0]{https://doi.org/}%
\providecommand \selectlanguage [0]{\@gobble}%
\providecommand \bibinfo  [0]{\@secondoftwo}%
\providecommand \bibfield  [0]{\@secondoftwo}%
\providecommand \translation [1]{[#1]}%
\providecommand \BibitemOpen [0]{}%
\providecommand \bibitemStop [0]{}%
\providecommand \bibitemNoStop [0]{.\EOS\space}%
\providecommand \EOS [0]{\spacefactor3000\relax}%
\providecommand \BibitemShut  [1]{\csname bibitem#1\endcsname}%
\let\auto@bib@innerbib\@empty
\bibitem [{\citenamefont {Stuchlik}\ \emph {et~al.}(2020)\citenamefont
  {Stuchlik}, \citenamefont {Kolos}, \citenamefont {Kovar}, \citenamefont
  {Slany},\ and\ \citenamefont {Tursunov}}]{Stuchlik:2020Cosmic}%
  \BibitemOpen
  \bibfield  {author} {\bibinfo {author} {\bibfnamefont {Z.}~\bibnamefont
  {Stuchlik}}, \bibinfo {author} {\bibfnamefont {M.}~\bibnamefont {Kolos}},
  \bibinfo {author} {\bibfnamefont {J.}~\bibnamefont {Kovar}}, \bibinfo
  {author} {\bibfnamefont {P.}~\bibnamefont {Slany}},\ and\ \bibinfo {author}
  {\bibfnamefont {A.}~\bibnamefont {Tursunov}},\ }\bibfield  {title} {\bibinfo
  {title} {{Influence of Cosmic Repulsion and Magnetic Fields on Accretion
  Disks Rotating around Kerr Black Holes}},\ }\href
  {https://doi.org/10.3390/universe6020026} {\bibfield  {journal} {\bibinfo
  {journal} {Universe}\ }\textbf {\bibinfo {volume} {6}},\ \bibinfo {pages}
  {26} (\bibinfo {year} {2020})}\BibitemShut {NoStop}%
\bibitem [{\citenamefont {Blandford}\ and\ \citenamefont
  {Znajek}(1977)}]{Blandford:1977BZ}%
  \BibitemOpen
  \bibfield  {author} {\bibinfo {author} {\bibfnamefont {R.~D.}\ \bibnamefont
  {Blandford}}\ and\ \bibinfo {author} {\bibfnamefont {R.~L.}\ \bibnamefont
  {Znajek}},\ }\bibfield  {title} {\bibinfo {title} {{Electromagnetic
  extraction of energy from Kerr black holes}},\ }\href
  {https://doi.org/10.1093/mnras/179.3.433} {\bibfield  {journal} {\bibinfo
  {journal} {Mon. Not. Roy. Astron. Soc.}\ }\textbf {\bibinfo {volume} {179}},\
  \bibinfo {pages} {433} (\bibinfo {year} {1977})}\BibitemShut {NoStop}%
\bibitem [{\citenamefont {Wald}(1974)}]{Wald:1974UniformField}%
  \BibitemOpen
  \bibfield  {author} {\bibinfo {author} {\bibfnamefont {R.~M.}\ \bibnamefont
  {Wald}},\ }\bibfield  {title} {\bibinfo {title} {{Black hole in a uniform
  magnetic field}},\ }\href {https://doi.org/10.1103/PhysRevD.10.1680}
  {\bibfield  {journal} {\bibinfo  {journal} {Phys. Rev. D}\ }\textbf {\bibinfo
  {volume} {10}},\ \bibinfo {pages} {1680} (\bibinfo {year}
  {1974})}\BibitemShut {NoStop}%
\bibitem [{\citenamefont {Melvin}(1964)}]{Melvin:1964Geons}%
  \BibitemOpen
  \bibfield  {author} {\bibinfo {author} {\bibfnamefont {M.~A.}\ \bibnamefont
  {Melvin}},\ }\bibfield  {title} {\bibinfo {title} {{Pure magnetic and
  electric geons}},\ }\href {https://doi.org/10.1016/0031-9163(64)90801-7}
  {\bibfield  {journal} {\bibinfo  {journal} {Phys. Lett.}\ }\textbf {\bibinfo
  {volume} {8}},\ \bibinfo {pages} {65} (\bibinfo {year} {1964})}\BibitemShut
  {NoStop}%
\bibitem [{\citenamefont {Ernst}(1976)}]{Ernst:1976MagneticUniverse}%
  \BibitemOpen
  \bibfield  {author} {\bibinfo {author} {\bibfnamefont {F.~J.}\ \bibnamefont
  {Ernst}},\ }\bibfield  {title} {\bibinfo {title} {{Black holes in a magnetic
  universe}},\ }\href {https://doi.org/10.1063/1.522781} {\bibfield  {journal}
  {\bibinfo  {journal} {J. Math. Phys.}\ }\textbf {\bibinfo {volume} {17}},\
  \bibinfo {pages} {54} (\bibinfo {year} {1976})}\BibitemShut {NoStop}%
\bibitem [{\citenamefont {Ernst}\ and\ \citenamefont
  {Wild}(1976)}]{ErnstWild:1976KerrMagnetic}%
  \BibitemOpen
  \bibfield  {author} {\bibinfo {author} {\bibfnamefont {F.~J.}\ \bibnamefont
  {Ernst}}\ and\ \bibinfo {author} {\bibfnamefont {W.~J.}\ \bibnamefont
  {Wild}},\ }\bibfield  {title} {\bibinfo {title} {{{Kerr} black holes in a
  magnetic universe}},\ }\href {https://doi.org/10.1063/1.522875} {\bibfield
  {journal} {\bibinfo  {journal} {J. Math. Phys.}\ }\textbf {\bibinfo {volume}
  {17}},\ \bibinfo {pages} {182} (\bibinfo {year} {1976})}\BibitemShut
  {NoStop}%
\bibitem [{\citenamefont {Gibbons}\ \emph {et~al.}(2013)\citenamefont
  {Gibbons}, \citenamefont {Mujtaba},\ and\ \citenamefont
  {Pope}}]{Gibbons:2013Ergoregions}%
  \BibitemOpen
  \bibfield  {author} {\bibinfo {author} {\bibfnamefont {G.~W.}\ \bibnamefont
  {Gibbons}}, \bibinfo {author} {\bibfnamefont {A.~H.}\ \bibnamefont
  {Mujtaba}},\ and\ \bibinfo {author} {\bibfnamefont {C.~N.}\ \bibnamefont
  {Pope}},\ }\bibfield  {title} {\bibinfo {title} {{Ergoregions in Magnetised
  Black Hole Spacetimes}},\ }\href
  {https://doi.org/10.1088/0264-9381/30/12/125008} {\bibfield  {journal}
  {\bibinfo  {journal} {Class. Quant. Grav.}\ }\textbf {\bibinfo {volume}
  {30}},\ \bibinfo {pages} {125008} (\bibinfo {year} {2013})},\ \Eprint
  {https://arxiv.org/abs/1301.3927} {arXiv:1301.3927 [gr-qc]} \BibitemShut
  {NoStop}%
\bibitem [{\citenamefont {Bertotti}(1959)}]{Bertotti:1959Uniform}%
  \BibitemOpen
  \bibfield  {author} {\bibinfo {author} {\bibfnamefont {B.}~\bibnamefont
  {Bertotti}},\ }\bibfield  {title} {\bibinfo {title} {{Uniform Electromagnetic
  Field in the Theory of General Relativity}},\ }\href
  {https://doi.org/10.1103/PhysRev.116.1331} {\bibfield  {journal} {\bibinfo
  {journal} {Phys. Rev.}\ }\textbf {\bibinfo {volume} {116}},\ \bibinfo {pages}
  {1331} (\bibinfo {year} {1959})}\BibitemShut {NoStop}%
\bibitem [{\citenamefont {Robinson}(1959)}]{Robinson:1959MaxwellEinstein}%
  \BibitemOpen
  \bibfield  {author} {\bibinfo {author} {\bibfnamefont {I.}~\bibnamefont
  {Robinson}},\ }\bibfield  {title} {\bibinfo {title} {{A Solution of the
  Maxwell--Einstein Equations}},\ }\href@noop {} {\bibfield  {journal}
  {\bibinfo  {journal} {Bull. Acad. Polon. Sci. Ser. Sci. Math. Astron. Phys.}\
  }\textbf {\bibinfo {volume} {7}},\ \bibinfo {pages} {351} (\bibinfo {year}
  {1959})}\BibitemShut {NoStop}%
\bibitem [{\citenamefont {Podolsky}\ and\ \citenamefont
  {Ovcharenko}(2025)}]{Podolsky:2025KerrBR}%
  \BibitemOpen
  \bibfield  {author} {\bibinfo {author} {\bibfnamefont {J.}~\bibnamefont
  {Podolsky}}\ and\ \bibinfo {author} {\bibfnamefont {H.}~\bibnamefont
  {Ovcharenko}},\ }\bibfield  {title} {\bibinfo {title} {{Kerr black hole in a
  uniform Bertotti-Robinson magnetic field: An exact solution}},\ }\href
  {https://doi.org/10.1103/rfgv-ybz5} {\bibfield  {journal} {\bibinfo
  {journal} {Phys. Rev. Lett.}\ }\textbf {\bibinfo {volume} {135}},\ \bibinfo
  {pages} {181401} (\bibinfo {year} {2025})},\ \Eprint
  {https://arxiv.org/abs/2507.05199} {arXiv:2507.05199 [gr-qc]} \BibitemShut
  {NoStop}%
\bibitem [{\citenamefont {Ovcharenko}\ and\ \citenamefont
  {Podolsky}(2026)}]{Ovcharenko:2026KerrNewmanBR}%
  \BibitemOpen
  \bibfield  {author} {\bibinfo {author} {\bibfnamefont {H.}~\bibnamefont
  {Ovcharenko}}\ and\ \bibinfo {author} {\bibfnamefont {J.}~\bibnamefont
  {Podolsky}},\ }\href {https://doi.org/10.48550/arXiv.2608.21672} {\bibinfo
  {title} {{Kerr-Newman and genuine Kerr black holes in the Bertotti-Robinson
  magnetic field}}} (\bibinfo {year} {2026}),\ \Eprint
  {https://arxiv.org/abs/2608.21672} {arXiv:2608.21672 [gr-qc]} \BibitemShut
  {NoStop}%
\bibitem [{\citenamefont {Ovcharenko}\ and\ \citenamefont
  {Podolsky}(2025)}]{Ovcharenko:2025Nonaligned}%
  \BibitemOpen
  \bibfield  {author} {\bibinfo {author} {\bibfnamefont {H.}~\bibnamefont
  {Ovcharenko}}\ and\ \bibinfo {author} {\bibfnamefont {J.}~\bibnamefont
  {Podolsky}},\ }\bibfield  {title} {\bibinfo {title} {{New class of rotating
  charged black holes with nonaligned electromagnetic field}},\ }\href
  {https://doi.org/10.1103/8wkz-th6v} {\bibfield  {journal} {\bibinfo
  {journal} {Phys. Rev. D}\ }\textbf {\bibinfo {volume} {112}},\ \bibinfo
  {pages} {064076} (\bibinfo {year} {2025})},\ \Eprint
  {https://arxiv.org/abs/2508.04850} {arXiv:2508.04850 [gr-qc]} \BibitemShut
  {NoStop}%
\bibitem [{\citenamefont {Wang}\ \emph
  {et~al.}(2026{\natexlab{a}})\citenamefont {Wang}, \citenamefont {Hou},
  \citenamefont {Wan}, \citenamefont {Guo},\ and\ \citenamefont
  {Chen}}]{Wang:2026KerrBRGeodesics}%
  \BibitemOpen
  \bibfield  {author} {\bibinfo {author} {\bibfnamefont {X.}~\bibnamefont
  {Wang}}, \bibinfo {author} {\bibfnamefont {Y.}~\bibnamefont {Hou}}, \bibinfo
  {author} {\bibfnamefont {X.}~\bibnamefont {Wan}}, \bibinfo {author}
  {\bibfnamefont {M.}~\bibnamefont {Guo}},\ and\ \bibinfo {author}
  {\bibfnamefont {B.}~\bibnamefont {Chen}},\ }\bibfield  {title} {\bibinfo
  {title} {{Geodesics and shadows in the Kerr-Bertotti-Robinson black hole
  spacetime}},\ }\href {https://doi.org/10.1088/1475-7516/2026/02/050}
  {\bibfield  {journal} {\bibinfo  {journal} {JCAP}\ }\textbf {\bibinfo
  {volume} {2026}}\bibfield  {number} {\bibinfo  {number} { (02)},\ \bibinfo
  {pages} {050}},\ }\Eprint {https://arxiv.org/abs/2507.22494}
  {arXiv:2507.22494 [gr-qc]} \BibitemShut {NoStop}%
\bibitem [{\citenamefont {Wang}(2025)}]{Wang:2025KerrBRISCO}%
  \BibitemOpen
  \bibfield  {author} {\bibinfo {author} {\bibfnamefont {T.}~\bibnamefont
  {Wang}},\ }\href {https://doi.org/10.48550/arXiv.2508.04684} {\bibinfo
  {title} {{Innermost stable circular orbit of Kerr-Bertotti-Robinson black
  holes and inspirals from it: Exact solutions}}} (\bibinfo {year} {2025}),\
  \Eprint {https://arxiv.org/abs/2508.04684} {arXiv:2508.04684 [gr-qc]}
  \BibitemShut {NoStop}%
\bibitem [{\citenamefont {Mustafa}\ \emph {et~al.}(2026)\citenamefont
  {Mustafa}, \citenamefont {Donmez}, \citenamefont {Gogoi}, \citenamefont
  {Ghosh}, \citenamefont {Hussain},\ and\ \citenamefont
  {Yuan}}]{Mustafa:2026KerrBRDynamics}%
  \BibitemOpen
  \bibfield  {author} {\bibinfo {author} {\bibfnamefont {G.}~\bibnamefont
  {Mustafa}}, \bibinfo {author} {\bibfnamefont {O.}~\bibnamefont {Donmez}},
  \bibinfo {author} {\bibfnamefont {D.~J.}\ \bibnamefont {Gogoi}}, \bibinfo
  {author} {\bibfnamefont {S.~G.}\ \bibnamefont {Ghosh}}, \bibinfo {author}
  {\bibfnamefont {I.}~\bibnamefont {Hussain}},\ and\ \bibinfo {author}
  {\bibfnamefont {C.}~\bibnamefont {Yuan}},\ }\bibfield  {title} {\bibinfo
  {title} {{Dynamics, Ringdown, and Accretion-Driven Multiple Quasi-Periodic
  Oscillations of Kerr-Bertotti-Robinson Black Holes}},\ }\href
  {https://doi.org/10.1088/1475-7516/2026/07/065} {\bibfield  {journal}
  {\bibinfo  {journal} {JCAP}\ }\textbf {\bibinfo {volume} {2026}}\bibfield
  {number} {\bibinfo  {number} { (07)},\ \bibinfo {pages} {065}},\ }\Eprint
  {https://arxiv.org/abs/2602.08911} {arXiv:2602.08911 [gr-qc]} \BibitemShut
  {NoStop}%
\bibitem [{\citenamefont {Wang}\ \emph
  {et~al.}(2026{\natexlab{b}})\citenamefont {Wang}, \citenamefont {Meng},\ and\
  \citenamefont {Wei}}]{Wang:2026KBRSpherical}%
  \BibitemOpen
  \bibfield  {author} {\bibinfo {author} {\bibfnamefont {C.-H.}\ \bibnamefont
  {Wang}}, \bibinfo {author} {\bibfnamefont {X.-C.}\ \bibnamefont {Meng}},\
  and\ \bibinfo {author} {\bibfnamefont {S.-W.}\ \bibnamefont {Wei}},\ }\href
  {https://doi.org/10.48550/arXiv.2602.03161} {\bibinfo {title} {{Magnetic
  field effects on spherical orbit in Kerr-Bertotti-Robinson spacetime:
  constraints from jet precession of M87*}}} (\bibinfo {year}
  {2026}{\natexlab{b}}),\ \Eprint {https://arxiv.org/abs/2602.03161}
  {arXiv:2602.03161 [gr-qc]} \BibitemShut {NoStop}%
\bibitem [{\citenamefont {Lu}\ and\ \citenamefont
  {Wu}(2026)}]{Lu:2026KBRIntegrability}%
  \BibitemOpen
  \bibfield  {author} {\bibinfo {author} {\bibfnamefont {J.}~\bibnamefont
  {Lu}}\ and\ \bibinfo {author} {\bibfnamefont {X.}~\bibnamefont {Wu}},\
  }\bibfield  {title} {\bibinfo {title} {{Third type of spacetime with the
  coexistence of integrability and non-integrability}},\ }\href
  {https://doi.org/10.1140/epjc/s10052-026-15482-w} {\bibfield  {journal}
  {\bibinfo  {journal} {Eur. Phys. J. C}\ }\textbf {\bibinfo {volume} {86}},\
  \bibinfo {pages} {256} (\bibinfo {year} {2026})},\ \Eprint
  {https://arxiv.org/abs/2603.12674} {arXiv:2603.12674 [gr-qc]} \BibitemShut
  {NoStop}%
\bibitem [{\citenamefont {Zeng}\ \emph {et~al.}(2025)\citenamefont {Zeng},
  \citenamefont {Yang},\ and\ \citenamefont {Yu}}]{Zeng:2025KerrBROptics}%
  \BibitemOpen
  \bibfield  {author} {\bibinfo {author} {\bibfnamefont {X.-X.}\ \bibnamefont
  {Zeng}}, \bibinfo {author} {\bibfnamefont {C.-Y.}\ \bibnamefont {Yang}},\
  and\ \bibinfo {author} {\bibfnamefont {H.}~\bibnamefont {Yu}},\ }\bibfield
  {title} {\bibinfo {title} {{Optical characteristics of the
  Kerr-Bertotti-Robinson black hole}},\ }\href
  {https://doi.org/10.1140/epjc/s10052-025-14989-y} {\bibfield  {journal}
  {\bibinfo  {journal} {Eur. Phys. J. C}\ }\textbf {\bibinfo {volume} {85}},\
  \bibinfo {pages} {1242} (\bibinfo {year} {2025})},\ \Eprint
  {https://arxiv.org/abs/2508.03020} {arXiv:2508.03020 [gr-qc]} \BibitemShut
  {NoStop}%
\bibitem [{\citenamefont {Wan}\ \emph {et~al.}(2026)\citenamefont {Wan},
  \citenamefont {Zhang}, \citenamefont {Wei}, \citenamefont {Hou},\ and\
  \citenamefont {Chen}}]{Wan:2026KBRPhotonRings}%
  \BibitemOpen
  \bibfield  {author} {\bibinfo {author} {\bibfnamefont {X.}~\bibnamefont
  {Wan}}, \bibinfo {author} {\bibfnamefont {Z.}~\bibnamefont {Zhang}}, \bibinfo
  {author} {\bibfnamefont {F.-S.}\ \bibnamefont {Wei}}, \bibinfo {author}
  {\bibfnamefont {Y.}~\bibnamefont {Hou}},\ and\ \bibinfo {author}
  {\bibfnamefont {B.}~\bibnamefont {Chen}},\ }\bibfield  {title} {\bibinfo
  {title} {{Critical behavior of photon rings in Kerr-Bertotti-Robinson
  spacetime}},\ }\href {https://doi.org/10.1103/fyky-bkbg} {\bibfield
  {journal} {\bibinfo  {journal} {Phys. Rev. D}\ }\textbf {\bibinfo {volume}
  {114}},\ \bibinfo {pages} {044005} (\bibinfo {year} {2026})},\ \Eprint
  {https://arxiv.org/abs/2603.25049} {arXiv:2603.25049 [gr-qc]} \BibitemShut
  {NoStop}%
\bibitem [{\citenamefont {Li}\ \emph {et~al.}(2026)\citenamefont {Li},
  \citenamefont {Yan},\ and\ \citenamefont {Yue}}]{Li:2026KerrBRGW}%
  \BibitemOpen
  \bibfield  {author} {\bibinfo {author} {\bibfnamefont {X.-Q.}\ \bibnamefont
  {Li}}, \bibinfo {author} {\bibfnamefont {H.-P.}\ \bibnamefont {Yan}},\ and\
  \bibinfo {author} {\bibfnamefont {X.-J.}\ \bibnamefont {Yue}},\ }\bibfield
  {title} {\bibinfo {title} {{Gravitational-wave imprints of
  Kerr-Bertotti-Robinson black holes: frequency blue-shift and waveform
  dephasing}},\ }\href {https://doi.org/10.1140/epjc/s10052-026-15441-5}
  {\bibfield  {journal} {\bibinfo  {journal} {Eur. Phys. J. C}\ }\textbf
  {\bibinfo {volume} {86}},\ \bibinfo {pages} {176} (\bibinfo {year} {2026})},\
  \Eprint {https://arxiv.org/abs/2512.02921} {arXiv:2512.02921 [gr-qc]}
  \BibitemShut {NoStop}%
\bibitem [{\citenamefont {Zeng}\ and\ \citenamefont
  {Wang}(2025)}]{Zeng:2025KerrBRReconnection}%
  \BibitemOpen
  \bibfield  {author} {\bibinfo {author} {\bibfnamefont {X.-X.}\ \bibnamefont
  {Zeng}}\ and\ \bibinfo {author} {\bibfnamefont {K.}~\bibnamefont {Wang}},\
  }\bibfield  {title} {\bibinfo {title} {{Energy extraction from the
  Kerr-Bertotti-Robinson black hole via magnetic reconnection in a circular and
  a plunging plasma}},\ }\href {https://doi.org/10.1103/vc96-snjm} {\bibfield
  {journal} {\bibinfo  {journal} {Phys. Rev. D}\ }\textbf {\bibinfo {volume}
  {112}},\ \bibinfo {pages} {064032} (\bibinfo {year} {2025})},\ \Eprint
  {https://arxiv.org/abs/2507.21777} {arXiv:2507.21777 [gr-qc]} \BibitemShut
  {NoStop}%
\bibitem [{\citenamefont {Mirkhaydarov}\ \emph {et~al.}(2026)\citenamefont
  {Mirkhaydarov}, \citenamefont {Xamidov}, \citenamefont {Sheoran},
  \citenamefont {Shaymatov},\ and\ \citenamefont
  {Nandan}}]{Mirkhaydarov:2026KerrBRMPP}%
  \BibitemOpen
  \bibfield  {author} {\bibinfo {author} {\bibfnamefont {M.}~\bibnamefont
  {Mirkhaydarov}}, \bibinfo {author} {\bibfnamefont {T.}~\bibnamefont
  {Xamidov}}, \bibinfo {author} {\bibfnamefont {P.}~\bibnamefont {Sheoran}},
  \bibinfo {author} {\bibfnamefont {S.}~\bibnamefont {Shaymatov}},\ and\
  \bibinfo {author} {\bibfnamefont {H.}~\bibnamefont {Nandan}},\ }\bibfield
  {title} {\bibinfo {title} {{Non-Monotonic Enhancement of the Magnetic Penrose
  Process in Kerr-Bertotti-Robinson Spacetime and its Implication for Electron
  Acceleration}},\ }\href {https://doi.org/10.1088/1475-7516/2026/08/035}
  {\bibfield  {journal} {\bibinfo  {journal} {JCAP}\ }\textbf {\bibinfo
  {volume} {2026}}\bibfield  {number} {\bibinfo  {number} { (08)},\ \bibinfo
  {pages} {035}},\ }\Eprint {https://arxiv.org/abs/2601.09919}
  {arXiv:2601.09919 [gr-qc]} \BibitemShut {NoStop}%
\bibitem [{\citenamefont {Xamidov}\ \emph {et~al.}(2026)\citenamefont
  {Xamidov}, \citenamefont {Shaymatov}, \citenamefont {Wu},\ and\ \citenamefont
  {Zhu}}]{Xamidov:2026SchwarzschildBRPeriodic}%
  \BibitemOpen
  \bibfield  {author} {\bibinfo {author} {\bibfnamefont {T.}~\bibnamefont
  {Xamidov}}, \bibinfo {author} {\bibfnamefont {S.}~\bibnamefont {Shaymatov}},
  \bibinfo {author} {\bibfnamefont {Q.}~\bibnamefont {Wu}},\ and\ \bibinfo
  {author} {\bibfnamefont {T.}~\bibnamefont {Zhu}},\ }\href
  {https://doi.org/10.48550/arXiv.2602.09453} {\bibinfo {title} {{Gravitational
  wave signatures from periodic orbits around a Schwarzschild-Bertotti-Robinson
  black hole}}} (\bibinfo {year} {2026}),\ \Eprint
  {https://arxiv.org/abs/2602.09453} {arXiv:2602.09453 [gr-qc]} \BibitemShut
  {NoStop}%
\bibitem [{\citenamefont {Xu}\ \emph {et~al.}(2026)\citenamefont {Xu},
  \citenamefont {Uktamov}, \citenamefont {Ntelis}, \citenamefont
  {Abdujabbarov}, \citenamefont {Ahmedov},\ and\ \citenamefont
  {Yuan}}]{Xu:2026SBRChaos}%
  \BibitemOpen
  \bibfield  {author} {\bibinfo {author} {\bibfnamefont {Y.}~\bibnamefont
  {Xu}}, \bibinfo {author} {\bibfnamefont {U.}~\bibnamefont {Uktamov}},
  \bibinfo {author} {\bibfnamefont {P.}~\bibnamefont {Ntelis}}, \bibinfo
  {author} {\bibfnamefont {A.}~\bibnamefont {Abdujabbarov}}, \bibinfo {author}
  {\bibfnamefont {B.}~\bibnamefont {Ahmedov}},\ and\ \bibinfo {author}
  {\bibfnamefont {C.}~\bibnamefont {Yuan}},\ }\href
  {https://doi.org/10.48550/arXiv.2603.19797} {\bibinfo {title} {{Chaotic
  motion and power spectral density in Schwarzschild Bertotti-Robinson black
  hole spacetime}}} (\bibinfo {year} {2026}),\ \Eprint
  {https://arxiv.org/abs/2603.19797} {arXiv:2603.19797 [gr-qc]} \BibitemShut
  {NoStop}%
\bibitem [{\citenamefont {Zhang}\ and\ \citenamefont
  {Wei}(2026)}]{Zhang:2026KerrBRSpin}%
  \BibitemOpen
  \bibfield  {author} {\bibinfo {author} {\bibfnamefont {Y.-K.}\ \bibnamefont
  {Zhang}}\ and\ \bibinfo {author} {\bibfnamefont {S.-W.}\ \bibnamefont
  {Wei}},\ }\bibfield  {title} {\bibinfo {title} {{Effects of magnetic fields
  on spinning test particles orbiting Kerr-Bertotti-Robinson black holes}},\
  }\href {https://doi.org/10.1103/dmh3-ht32} {\bibfield  {journal} {\bibinfo
  {journal} {Phys. Rev. D}\ }\textbf {\bibinfo {volume} {113}},\ \bibinfo
  {pages} {104024} (\bibinfo {year} {2026})},\ \Eprint
  {https://arxiv.org/abs/2510.07914} {arXiv:2510.07914 [gr-qc]} \BibitemShut
  {NoStop}%
\bibitem [{\citenamefont
  {Siahaan}(2026{\natexlab{a}})}]{Siahaan:2026KerrBRCFT}%
  \BibitemOpen
  \bibfield  {author} {\bibinfo {author} {\bibfnamefont {H.~M.}\ \bibnamefont
  {Siahaan}},\ }\bibfield  {title} {\bibinfo {title} {{Kerr-Bertotti-Robinson
  spacetime and the Kerr/CFT correspondence}},\ }\href
  {https://doi.org/10.1016/j.nuclphysb.2026.117514} {\bibfield  {journal}
  {\bibinfo  {journal} {Nucl. Phys. B}\ }\textbf {\bibinfo {volume} {1028}},\
  \bibinfo {pages} {117514} (\bibinfo {year} {2026}{\natexlab{a}})},\ \Eprint
  {https://arxiv.org/abs/2512.12533} {arXiv:2512.12533 [gr-qc]} \BibitemShut
  {NoStop}%
\bibitem [{\citenamefont
  {Siahaan}(2026{\natexlab{b}})}]{Siahaan:2026KBRMeissner}%
  \BibitemOpen
  \bibfield  {author} {\bibinfo {author} {\bibfnamefont {H.~M.}\ \bibnamefont
  {Siahaan}},\ }\href {https://doi.org/10.48550/arXiv.2603.00653} {\bibinfo
  {title} {{Meissner Effect in Kerr--Bertotti--Robinson Spacetime}}} (\bibinfo
  {year} {2026}{\natexlab{b}}),\ \Eprint {https://arxiv.org/abs/2603.00653}
  {arXiv:2603.00653 [gr-qc]} \BibitemShut {NoStop}%
\bibitem [{\citenamefont {Hu}\ \emph {et~al.}(2026)\citenamefont {Hu},
  \citenamefont {Cai},\ and\ \citenamefont
  {Wang}}]{Hu:2026KerrBRThermodynamics}%
  \BibitemOpen
  \bibfield  {author} {\bibinfo {author} {\bibfnamefont {L.}~\bibnamefont
  {Hu}}, \bibinfo {author} {\bibfnamefont {R.-G.}\ \bibnamefont {Cai}},\ and\
  \bibinfo {author} {\bibfnamefont {S.-J.}\ \bibnamefont {Wang}},\ }\bibfield
  {title} {\bibinfo {title} {{Thermodynamics of Kerr-Bertotti-Robinson black
  hole}},\ }\href {https://doi.org/10.1103/dybn-n2vx} {\bibfield  {journal}
  {\bibinfo  {journal} {Phys. Rev. D}\ }\textbf {\bibinfo {volume} {114}},\
  \bibinfo {pages} {L041501} (\bibinfo {year} {2026})},\ \Eprint
  {https://arxiv.org/abs/2603.18821} {arXiv:2603.18821 [gr-qc]} \BibitemShut
  {NoStop}%
\bibitem [{\citenamefont {Zhou}\ \emph {et~al.}(2026)\citenamefont {Zhou},
  \citenamefont {Wu}, \citenamefont {Ji}, \citenamefont {Fu}, \citenamefont
  {Cao},\ and\ \citenamefont {Cai}}]{Zhou:2026KerrBRNotes}%
  \BibitemOpen
  \bibfield  {author} {\bibinfo {author} {\bibfnamefont {Y.-S.}\ \bibnamefont
  {Zhou}}, \bibinfo {author} {\bibfnamefont {L.-B.}\ \bibnamefont {Wu}},
  \bibinfo {author} {\bibfnamefont {M.-F.}\ \bibnamefont {Ji}}, \bibinfo
  {author} {\bibfnamefont {W.-T.}\ \bibnamefont {Fu}}, \bibinfo {author}
  {\bibfnamefont {L.-M.}\ \bibnamefont {Cao}},\ and\ \bibinfo {author}
  {\bibfnamefont {R.-G.}\ \bibnamefont {Cai}},\ }\href
  {https://doi.org/10.48550/arXiv.2608.19136} {\bibinfo {title} {{Notes on
  Kerr-Bertotti-Robinson Spacetime}}} (\bibinfo {year} {2026}),\ \Eprint
  {https://arxiv.org/abs/2608.19136} {arXiv:2608.19136 [gr-qc]} \BibitemShut
  {NoStop}%
\bibitem [{\citenamefont {Preti}(2004)}]{Preti:2004Polarized}%
  \BibitemOpen
  \bibfield  {author} {\bibinfo {author} {\bibfnamefont {G.}~\bibnamefont
  {Preti}},\ }\bibfield  {title} {\bibinfo {title} {{General relativistic
  dynamics of polarized particles in electromagnetic fields}},\ }\href
  {https://doi.org/10.1103/PhysRevD.70.024012} {\bibfield  {journal} {\bibinfo
  {journal} {Phys. Rev. D}\ }\textbf {\bibinfo {volume} {70}},\ \bibinfo
  {pages} {024012} (\bibinfo {year} {2004})}\BibitemShut {NoStop}%
\bibitem [{\citenamefont {De~Felice}\ and\ \citenamefont
  {Sorge}(2003)}]{DeFelice:2003Schwarzschild}%
  \BibitemOpen
  \bibfield  {author} {\bibinfo {author} {\bibfnamefont {F.}~\bibnamefont
  {De~Felice}}\ and\ \bibinfo {author} {\bibfnamefont {F.}~\bibnamefont
  {Sorge}},\ }\bibfield  {title} {\bibinfo {title} {{Magnetized orbits around a
  Schwarzschild black hole}},\ }\href
  {https://doi.org/10.1088/0264-9381/20/3/306} {\bibfield  {journal} {\bibinfo
  {journal} {Class. Quantum Grav.}\ }\textbf {\bibinfo {volume} {20}},\
  \bibinfo {pages} {469} (\bibinfo {year} {2003})}\BibitemShut {NoStop}%
\bibitem [{\citenamefont {De~Felice}\ \emph {et~al.}(2004)\citenamefont
  {De~Felice}, \citenamefont {Sorge},\ and\ \citenamefont
  {Zilio}}]{DeFelice:2004Kerr}%
  \BibitemOpen
  \bibfield  {author} {\bibinfo {author} {\bibfnamefont {F.}~\bibnamefont
  {De~Felice}}, \bibinfo {author} {\bibfnamefont {F.}~\bibnamefont {Sorge}},\
  and\ \bibinfo {author} {\bibfnamefont {S.}~\bibnamefont {Zilio}},\ }\bibfield
   {title} {\bibinfo {title} {{Magnetized orbits around a Kerr black hole}},\
  }\href {https://doi.org/10.1088/0264-9381/21/4/016} {\bibfield  {journal}
  {\bibinfo  {journal} {Class. Quantum Grav.}\ }\textbf {\bibinfo {volume}
  {21}},\ \bibinfo {pages} {961} (\bibinfo {year} {2004})}\BibitemShut
  {NoStop}%
\bibitem [{\citenamefont {Abdujabbarov}\ \emph {et~al.}(2014)\citenamefont
  {Abdujabbarov}, \citenamefont {Ahmedov}, \citenamefont {Rahimov},\ and\
  \citenamefont {Salikhbaev}}]{Abdujabbarov:2014MagnetizedAcceleration}%
  \BibitemOpen
  \bibfield  {author} {\bibinfo {author} {\bibfnamefont {A.}~\bibnamefont
  {Abdujabbarov}}, \bibinfo {author} {\bibfnamefont {B.}~\bibnamefont
  {Ahmedov}}, \bibinfo {author} {\bibfnamefont {O.}~\bibnamefont {Rahimov}},\
  and\ \bibinfo {author} {\bibfnamefont {U.}~\bibnamefont {Salikhbaev}},\
  }\bibfield  {title} {\bibinfo {title} {{Magnetized particle motion and
  acceleration around a {Schwarzschild} black hole in a magnetic field}},\
  }\href {https://doi.org/10.1088/0031-8949/89/8/084008} {\bibfield  {journal}
  {\bibinfo  {journal} {Phys. Scr.}\ }\textbf {\bibinfo {volume} {89}},\
  \bibinfo {pages} {084008} (\bibinfo {year} {2014})}\BibitemShut {NoStop}%
\bibitem [{\citenamefont {Uktamov}\ \emph {et~al.}(2024)\citenamefont
  {Uktamov}, \citenamefont {Fathi}, \citenamefont {Rayimbaev},\ and\
  \citenamefont {Abdujabbarov}}]{Uktamov:2024MagnetizedSchwarzschild}%
  \BibitemOpen
  \bibfield  {author} {\bibinfo {author} {\bibfnamefont {U.}~\bibnamefont
  {Uktamov}}, \bibinfo {author} {\bibfnamefont {M.}~\bibnamefont {Fathi}},
  \bibinfo {author} {\bibfnamefont {J.}~\bibnamefont {Rayimbaev}},\ and\
  \bibinfo {author} {\bibfnamefont {A.}~\bibnamefont {Abdujabbarov}},\
  }\bibfield  {title} {\bibinfo {title} {{Orbits of particles with magnetic
  dipole moment around magnetized Schwarzschild black holes: Applications to S2
  star orbit}},\ }\href {https://doi.org/10.1103/PhysRevD.110.084084}
  {\bibfield  {journal} {\bibinfo  {journal} {Phys. Rev. D}\ }\textbf {\bibinfo
  {volume} {110}},\ \bibinfo {pages} {084084} (\bibinfo {year} {2024})},\
  \Eprint {https://arxiv.org/abs/2406.03371} {arXiv:2406.03371 [gr-qc]}
  \BibitemShut {NoStop}%
\bibitem [{\citenamefont {Narzilloev}\ \emph {et~al.}(2021)\citenamefont
  {Narzilloev}, \citenamefont {Rayimbaev}, \citenamefont {Abdujabbarov},
  \citenamefont {Ahmedov},\ and\ \citenamefont
  {Bambi}}]{Narzilloev:2021QuasiSchwarzschild}%
  \BibitemOpen
  \bibfield  {author} {\bibinfo {author} {\bibfnamefont {B.}~\bibnamefont
  {Narzilloev}}, \bibinfo {author} {\bibfnamefont {J.}~\bibnamefont
  {Rayimbaev}}, \bibinfo {author} {\bibfnamefont {A.}~\bibnamefont
  {Abdujabbarov}}, \bibinfo {author} {\bibfnamefont {B.}~\bibnamefont
  {Ahmedov}},\ and\ \bibinfo {author} {\bibfnamefont {C.}~\bibnamefont
  {Bambi}},\ }\bibfield  {title} {\bibinfo {title} {{Dynamics of charged
  particles and magnetic dipoles around magnetized quasi-{Schwarzschild} black
  holes}},\ }\href {https://doi.org/10.1140/epjc/s10052-021-09074-z} {\bibfield
   {journal} {\bibinfo  {journal} {Eur. Phys. J. C}\ }\textbf {\bibinfo
  {volume} {81}},\ \bibinfo {pages} {269} (\bibinfo {year} {2021})},\ \Eprint
  {https://arxiv.org/abs/2103.11090} {arXiv:2103.11090 [gr-qc]} \BibitemShut
  {NoStop}%
\bibitem [{\citenamefont {Glampedakis}\ and\ \citenamefont
  {Kennefick}(2002)}]{Glampedakis:2002ZoomWhirl}%
  \BibitemOpen
  \bibfield  {author} {\bibinfo {author} {\bibfnamefont {K.}~\bibnamefont
  {Glampedakis}}\ and\ \bibinfo {author} {\bibfnamefont {D.}~\bibnamefont
  {Kennefick}},\ }\bibfield  {title} {\bibinfo {title} {{Zoom and whirl:
  Eccentric equatorial orbits around spinning black holes and their evolution
  under gravitational radiation reaction}},\ }\href
  {https://doi.org/10.1103/PhysRevD.66.044002} {\bibfield  {journal} {\bibinfo
  {journal} {Phys. Rev. D}\ }\textbf {\bibinfo {volume} {66}},\ \bibinfo
  {pages} {044002} (\bibinfo {year} {2002})},\ \Eprint
  {https://arxiv.org/abs/gr-qc/0203086} {arXiv:gr-qc/0203086 [gr-qc]}
  \BibitemShut {NoStop}%
\bibitem [{\citenamefont {Levin}\ and\ \citenamefont
  {Perez-Giz}(2008)}]{Levin:2008Periodic}%
  \BibitemOpen
  \bibfield  {author} {\bibinfo {author} {\bibfnamefont {J.}~\bibnamefont
  {Levin}}\ and\ \bibinfo {author} {\bibfnamefont {G.}~\bibnamefont
  {Perez-Giz}},\ }\bibfield  {title} {\bibinfo {title} {{A periodic table for
  black hole orbits}},\ }\href {https://doi.org/10.1103/PhysRevD.77.103005}
  {\bibfield  {journal} {\bibinfo  {journal} {Phys. Rev. D}\ }\textbf {\bibinfo
  {volume} {77}},\ \bibinfo {pages} {103005} (\bibinfo {year}
  {2008})}\BibitemShut {NoStop}%
\bibitem [{\citenamefont {Jumaniyozov}\ \emph {et~al.}(2026)\citenamefont
  {Jumaniyozov}, \citenamefont {Rayimbaev}, \citenamefont {Yuan}, \citenamefont
  {Palvanov}, \citenamefont {Khaknazarova},\ and\ \citenamefont
  {Javed}}]{Jumaniyozov:2026MagnetizedKerrPeriodic}%
  \BibitemOpen
  \bibfield  {author} {\bibinfo {author} {\bibfnamefont {S.}~\bibnamefont
  {Jumaniyozov}}, \bibinfo {author} {\bibfnamefont {J.}~\bibnamefont
  {Rayimbaev}}, \bibinfo {author} {\bibfnamefont {C.}~\bibnamefont {Yuan}},
  \bibinfo {author} {\bibfnamefont {S.}~\bibnamefont {Palvanov}}, \bibinfo
  {author} {\bibfnamefont {K.}~\bibnamefont {Khaknazarova}},\ and\ \bibinfo
  {author} {\bibfnamefont {F.}~\bibnamefont {Javed}},\ }\href
  {https://doi.org/10.48550/arXiv.2608.11048} {\bibinfo {title} {{Periodic
  orbits and gravitational wave signatures from magnetic dipoles around
  magnetized Kerr black holes}}} (\bibinfo {year} {2026}),\ \Eprint
  {https://arxiv.org/abs/2608.11048} {arXiv:2608.11048 [gr-qc]} \BibitemShut
  {NoStop}%
\bibitem [{\citenamefont {Ba{\~n}ados}\ \emph {et~al.}(2009)\citenamefont
  {Ba{\~n}ados}, \citenamefont {Silk},\ and\ \citenamefont
  {West}}]{Banados:2009pr}%
  \BibitemOpen
  \bibfield  {author} {\bibinfo {author} {\bibfnamefont {M.}~\bibnamefont
  {Ba{\~n}ados}}, \bibinfo {author} {\bibfnamefont {J.}~\bibnamefont {Silk}},\
  and\ \bibinfo {author} {\bibfnamefont {S.~M.}\ \bibnamefont {West}},\
  }\bibfield  {title} {\bibinfo {title} {{Kerr Black Holes as Particle
  Accelerators to Arbitrarily High Energy}},\ }\href
  {https://doi.org/10.1103/PhysRevLett.103.111102} {\bibfield  {journal}
  {\bibinfo  {journal} {Phys. Rev. Lett.}\ }\textbf {\bibinfo {volume} {103}},\
  \bibinfo {pages} {111102} (\bibinfo {year} {2009})},\ \Eprint
  {https://arxiv.org/abs/0909.0169} {arXiv:0909.0169 [hep-ph]} \BibitemShut
  {NoStop}%
\bibitem [{\citenamefont {Berti}\ \emph {et~al.}(2009)\citenamefont {Berti},
  \citenamefont {Cardoso}, \citenamefont {Gualtieri}, \citenamefont
  {Pretorius},\ and\ \citenamefont {Sperhake}}]{Berti:2009BSWComment}%
  \BibitemOpen
  \bibfield  {author} {\bibinfo {author} {\bibfnamefont {E.}~\bibnamefont
  {Berti}}, \bibinfo {author} {\bibfnamefont {V.}~\bibnamefont {Cardoso}},
  \bibinfo {author} {\bibfnamefont {L.}~\bibnamefont {Gualtieri}}, \bibinfo
  {author} {\bibfnamefont {F.}~\bibnamefont {Pretorius}},\ and\ \bibinfo
  {author} {\bibfnamefont {U.}~\bibnamefont {Sperhake}},\ }\bibfield  {title}
  {\bibinfo {title} {{Comment on `Kerr Black Holes as Particle Accelerators to
  Arbitrarily High Energy'}},\ }\href
  {https://doi.org/10.1103/PhysRevLett.103.239001} {\bibfield  {journal}
  {\bibinfo  {journal} {Phys. Rev. Lett.}\ }\textbf {\bibinfo {volume} {103}},\
  \bibinfo {pages} {239001} (\bibinfo {year} {2009})},\ \Eprint
  {https://arxiv.org/abs/0911.2243} {arXiv:0911.2243 [gr-qc]} \BibitemShut
  {NoStop}%
\bibitem [{\citenamefont {Jacobson}\ and\ \citenamefont
  {Sotiriou}(2010)}]{Jacobson:2009BSWLimits}%
  \BibitemOpen
  \bibfield  {author} {\bibinfo {author} {\bibfnamefont {T.}~\bibnamefont
  {Jacobson}}\ and\ \bibinfo {author} {\bibfnamefont {T.~P.}\ \bibnamefont
  {Sotiriou}},\ }\bibfield  {title} {\bibinfo {title} {{Spinning Black Holes as
  Particle Accelerators}},\ }\href
  {https://doi.org/10.1103/PhysRevLett.104.021101} {\bibfield  {journal}
  {\bibinfo  {journal} {Phys. Rev. Lett.}\ }\textbf {\bibinfo {volume} {104}},\
  \bibinfo {pages} {021101} (\bibinfo {year} {2010})},\ \Eprint
  {https://arxiv.org/abs/0911.3363} {arXiv:0911.3363 [gr-qc]} \BibitemShut
  {NoStop}%
\bibitem [{\citenamefont {Harada}\ and\ \citenamefont
  {Kimura}(2011)}]{Harada:2011GeneralGeodesic}%
  \BibitemOpen
  \bibfield  {author} {\bibinfo {author} {\bibfnamefont {T.}~\bibnamefont
  {Harada}}\ and\ \bibinfo {author} {\bibfnamefont {M.}~\bibnamefont
  {Kimura}},\ }\bibfield  {title} {\bibinfo {title} {{Collision of two general
  geodesic particles around a Kerr black hole}},\ }\href
  {https://doi.org/10.1103/PhysRevD.83.084041} {\bibfield  {journal} {\bibinfo
  {journal} {Phys. Rev. D}\ }\textbf {\bibinfo {volume} {83}},\ \bibinfo
  {pages} {084041} (\bibinfo {year} {2011})},\ \Eprint
  {https://arxiv.org/abs/1102.3316} {arXiv:1102.3316 [gr-qc]} \BibitemShut
  {NoStop}%
\bibitem [{\citenamefont {Harada}\ and\ \citenamefont
  {Kimura}(2014)}]{Harada:2014vka}%
  \BibitemOpen
  \bibfield  {author} {\bibinfo {author} {\bibfnamefont {T.}~\bibnamefont
  {Harada}}\ and\ \bibinfo {author} {\bibfnamefont {M.}~\bibnamefont
  {Kimura}},\ }\bibfield  {title} {\bibinfo {title} {{Black holes as particle
  accelerators: a brief review}},\ }\href
  {https://doi.org/10.1088/0264-9381/31/24/243001} {\bibfield  {journal}
  {\bibinfo  {journal} {Class. Quant. Grav.}\ }\textbf {\bibinfo {volume}
  {31}},\ \bibinfo {pages} {243001} (\bibinfo {year} {2014})},\ \Eprint
  {https://arxiv.org/abs/1409.7502} {arXiv:1409.7502 [gr-qc]} \BibitemShut
  {NoStop}%
\bibitem [{\citenamefont {Zaslavskii}(2010)}]{Zaslavskii:2010Universal}%
  \BibitemOpen
  \bibfield  {author} {\bibinfo {author} {\bibfnamefont {O.~B.}\ \bibnamefont
  {Zaslavskii}},\ }\bibfield  {title} {\bibinfo {title} {{Acceleration of
  particles as a universal property of rotating black holes}},\ }\href
  {https://doi.org/10.1103/PhysRevD.82.083004} {\bibfield  {journal} {\bibinfo
  {journal} {Phys. Rev. D}\ }\textbf {\bibinfo {volume} {82}},\ \bibinfo
  {pages} {083004} (\bibinfo {year} {2010})},\ \Eprint
  {https://arxiv.org/abs/1007.3678} {arXiv:1007.3678 [gr-qc]} \BibitemShut
  {NoStop}%
\bibitem [{\citenamefont {Hejda}\ and\ \citenamefont
  {Bicak}(2017)}]{Hejda:2017KinematicRestrictions}%
  \BibitemOpen
  \bibfield  {author} {\bibinfo {author} {\bibfnamefont {F.}~\bibnamefont
  {Hejda}}\ and\ \bibinfo {author} {\bibfnamefont {J.}~\bibnamefont {Bicak}},\
  }\bibfield  {title} {\bibinfo {title} {{Kinematic restrictions on particle
  collisions near extremal black holes: A unified picture}},\ }\href
  {https://doi.org/10.1103/PhysRevD.95.084055} {\bibfield  {journal} {\bibinfo
  {journal} {Phys. Rev. D}\ }\textbf {\bibinfo {volume} {95}},\ \bibinfo
  {pages} {084055} (\bibinfo {year} {2017})},\ \Eprint
  {https://arxiv.org/abs/1612.04959} {arXiv:1612.04959 [gr-qc]} \BibitemShut
  {NoStop}%
\bibitem [{\citenamefont {Ovcharenko}\ and\ \citenamefont
  {Zaslavskii}(2023)}]{Ovcharenko:2023FiniteForces}%
  \BibitemOpen
  \bibfield  {author} {\bibinfo {author} {\bibfnamefont {H.~V.}\ \bibnamefont
  {Ovcharenko}}\ and\ \bibinfo {author} {\bibfnamefont {O.~B.}\ \bibnamefont
  {Zaslavskii}},\ }\bibfield  {title} {\bibinfo {title} {{Ba{\~n}ados-Silk-West
  effect with finite forces near different types of horizons: General
  classification of scenarios}},\ }\href
  {https://doi.org/10.1103/PhysRevD.108.064029} {\bibfield  {journal} {\bibinfo
   {journal} {Phys. Rev. D}\ }\textbf {\bibinfo {volume} {108}},\ \bibinfo
  {pages} {064029} (\bibinfo {year} {2023})},\ \Eprint
  {https://arxiv.org/abs/2304.13087} {arXiv:2304.13087 [gr-qc]} \BibitemShut
  {NoStop}%
\bibitem [{\citenamefont
  {Zaslavskii}(2020)}]{Zaslavskii:2020NonextremalAccelerator}%
  \BibitemOpen
  \bibfield  {author} {\bibinfo {author} {\bibfnamefont {O.~B.}\ \bibnamefont
  {Zaslavskii}},\ }\bibfield  {title} {\bibinfo {title} {{Can a nonextremal
  black hole be a particle accelerator?}},\ }\href
  {https://doi.org/10.1103/PhysRevD.102.104004} {\bibfield  {journal} {\bibinfo
   {journal} {Phys. Rev. D}\ }\textbf {\bibinfo {volume} {102}},\ \bibinfo
  {pages} {104004} (\bibinfo {year} {2020})},\ \Eprint
  {https://arxiv.org/abs/2007.09413} {arXiv:2007.09413 [gr-qc]} \BibitemShut
  {NoStop}%
\bibitem [{\citenamefont {Frolov}(2012)}]{Frolov:2011ea}%
  \BibitemOpen
  \bibfield  {author} {\bibinfo {author} {\bibfnamefont {V.~P.}\ \bibnamefont
  {Frolov}},\ }\bibfield  {title} {\bibinfo {title} {{Weakly magnetized black
  holes as particle accelerators}},\ }\href
  {https://doi.org/10.1103/PhysRevD.85.024020} {\bibfield  {journal} {\bibinfo
  {journal} {Phys. Rev. D}\ }\textbf {\bibinfo {volume} {85}},\ \bibinfo
  {pages} {024020} (\bibinfo {year} {2012})},\ \Eprint
  {https://arxiv.org/abs/1110.6274} {arXiv:1110.6274 [gr-qc]} \BibitemShut
  {NoStop}%
\bibitem [{\citenamefont {Igata}\ \emph {et~al.}(2012)\citenamefont {Igata},
  \citenamefont {Harada},\ and\ \citenamefont {Kimura}}]{Igata:2012js}%
  \BibitemOpen
  \bibfield  {author} {\bibinfo {author} {\bibfnamefont {T.}~\bibnamefont
  {Igata}}, \bibinfo {author} {\bibfnamefont {T.}~\bibnamefont {Harada}},\ and\
  \bibinfo {author} {\bibfnamefont {M.}~\bibnamefont {Kimura}},\ }\bibfield
  {title} {\bibinfo {title} {{Effect of a Weak Electromagnetic Field on
  Particle Acceleration by a Rotating Black Hole}},\ }\href
  {https://doi.org/10.1103/PhysRevD.85.104028} {\bibfield  {journal} {\bibinfo
  {journal} {Phys. Rev. D}\ }\textbf {\bibinfo {volume} {85}},\ \bibinfo
  {pages} {104028} (\bibinfo {year} {2012})},\ \Eprint
  {https://arxiv.org/abs/1202.4859} {arXiv:1202.4859 [gr-qc]} \BibitemShut
  {NoStop}%
\bibitem [{\citenamefont {Zaslavskii}(2015)}]{Zaslavskii:2014eda}%
  \BibitemOpen
  \bibfield  {author} {\bibinfo {author} {\bibfnamefont {O.~B.}\ \bibnamefont
  {Zaslavskii}},\ }\bibfield  {title} {\bibinfo {title} {{Kinematics of
  ultra-high energy particle collisions near black holes in the magnetic
  field}},\ }\href {https://doi.org/10.1142/S0217732315500273} {\bibfield
  {journal} {\bibinfo  {journal} {Mod. Phys. Lett. A}\ }\textbf {\bibinfo
  {volume} {30}},\ \bibinfo {pages} {1550027} (\bibinfo {year} {2015})},\
  \Eprint {https://arxiv.org/abs/1407.5440} {arXiv:1407.5440 [gr-qc]}
  \BibitemShut {NoStop}%
\bibitem [{\citenamefont {Jumaniyozov}\ \emph {et~al.}(2024)\citenamefont
  {Jumaniyozov}, \citenamefont {Khan}, \citenamefont {Rayimbaev}, \citenamefont
  {Abdujabbarov},\ and\ \citenamefont
  {Ahmedov}}]{Jumaniyozov:2024MagnetizedKerr}%
  \BibitemOpen
  \bibfield  {author} {\bibinfo {author} {\bibfnamefont {S.}~\bibnamefont
  {Jumaniyozov}}, \bibinfo {author} {\bibfnamefont {S.~U.}\ \bibnamefont
  {Khan}}, \bibinfo {author} {\bibfnamefont {J.}~\bibnamefont {Rayimbaev}},
  \bibinfo {author} {\bibfnamefont {A.}~\bibnamefont {Abdujabbarov}},\ and\
  \bibinfo {author} {\bibfnamefont {B.}~\bibnamefont {Ahmedov}},\ }\bibfield
  {title} {\bibinfo {title} {{Collisions and dynamics of particles with
  magnetic dipole moment and electric charge near magnetized rotating Kerr
  black holes}},\ }\href {https://doi.org/10.1140/epjc/s10052-024-12605-z}
  {\bibfield  {journal} {\bibinfo  {journal} {Eur. Phys. J. C}\ }\textbf
  {\bibinfo {volume} {84}},\ \bibinfo {pages} {291} (\bibinfo {year}
  {2024})}\BibitemShut {NoStop}%
\bibitem [{\citenamefont {Piran}\ and\ \citenamefont
  {Shaham}(1977)}]{Piran:1977CollisionalPenrose}%
  \BibitemOpen
  \bibfield  {author} {\bibinfo {author} {\bibfnamefont {T.}~\bibnamefont
  {Piran}}\ and\ \bibinfo {author} {\bibfnamefont {J.}~\bibnamefont {Shaham}},\
  }\bibfield  {title} {\bibinfo {title} {{Upper bounds on collisional Penrose
  processes near rotating black-hole horizons}},\ }\href
  {https://doi.org/10.1103/PhysRevD.16.1615} {\bibfield  {journal} {\bibinfo
  {journal} {Phys. Rev. D}\ }\textbf {\bibinfo {volume} {16}},\ \bibinfo
  {pages} {1615} (\bibinfo {year} {1977})}\BibitemShut {NoStop}%
\bibitem [{\citenamefont {Zaslavskii}(2012)}]{Zaslavskii:2012BSWPenrose}%
  \BibitemOpen
  \bibfield  {author} {\bibinfo {author} {\bibfnamefont {O.~B.}\ \bibnamefont
  {Zaslavskii}},\ }\bibfield  {title} {\bibinfo {title} {{Energetics of
  particle collisions near dirty rotating extremal black holes:
  Ba{\~n}ados-Silk-West effect versus Penrose process}},\ }\href
  {https://doi.org/10.1103/PhysRevD.86.084030} {\bibfield  {journal} {\bibinfo
  {journal} {Phys. Rev. D}\ }\textbf {\bibinfo {volume} {86}},\ \bibinfo
  {pages} {084030} (\bibinfo {year} {2012})},\ \Eprint
  {https://arxiv.org/abs/1205.4410} {arXiv:1205.4410 [gr-qc]} \BibitemShut
  {NoStop}%
\bibitem [{\citenamefont {Bejger}\ \emph {et~al.}(2012)\citenamefont {Bejger},
  \citenamefont {Piran}, \citenamefont {Abramowicz},\ and\ \citenamefont
  {Hakanson}}]{Bejger:2012CollisionalPenrose}%
  \BibitemOpen
  \bibfield  {author} {\bibinfo {author} {\bibfnamefont {M.}~\bibnamefont
  {Bejger}}, \bibinfo {author} {\bibfnamefont {T.}~\bibnamefont {Piran}},
  \bibinfo {author} {\bibfnamefont {M.}~\bibnamefont {Abramowicz}},\ and\
  \bibinfo {author} {\bibfnamefont {F.}~\bibnamefont {Hakanson}},\ }\bibfield
  {title} {\bibinfo {title} {{Collisional Penrose process near the horizon of
  extreme Kerr black holes}},\ }\href
  {https://doi.org/10.1103/PhysRevLett.109.121101} {\bibfield  {journal}
  {\bibinfo  {journal} {Phys. Rev. Lett.}\ }\textbf {\bibinfo {volume} {109}},\
  \bibinfo {pages} {121101} (\bibinfo {year} {2012})},\ \Eprint
  {https://arxiv.org/abs/1205.4350} {arXiv:1205.4350 [astro-ph.HE]}
  \BibitemShut {NoStop}%
\bibitem [{\citenamefont {Schnittman}(2014)}]{Schnittman:2014RevisedLimit}%
  \BibitemOpen
  \bibfield  {author} {\bibinfo {author} {\bibfnamefont {J.~D.}\ \bibnamefont
  {Schnittman}},\ }\bibfield  {title} {\bibinfo {title} {{Revised upper limit
  to energy extraction from a Kerr black hole}},\ }\href
  {https://doi.org/10.1103/PhysRevLett.113.261102} {\bibfield  {journal}
  {\bibinfo  {journal} {Phys. Rev. Lett.}\ }\textbf {\bibinfo {volume} {113}},\
  \bibinfo {pages} {261102} (\bibinfo {year} {2014})},\ \Eprint
  {https://arxiv.org/abs/1410.6446} {arXiv:1410.6446 [astro-ph.HE]}
  \BibitemShut {NoStop}%
\end{thebibliography}%

\end{document}